\documentclass[review]{elsarticle}
\usepackage[T1]{fontenc}
\usepackage{lineno}
\usepackage{xcolor}
\usepackage{amsmath,amssymb}
\usepackage{siunitx}
\usepackage{tikz}
\usepackage{epstopdf}
\usepackage{booktabs}
\modulolinenumbers[5]
\journal{Computers \& Fluids}

\usepackage[a4paper, total={6in, 8in}]{geometry}
\newsavebox\mysavebox
\graphicspath{{./figures/}}
\begin{document}

%========================================================================================
% TIKZ
\usetikzlibrary{positioning}
\usetikzlibrary{chains,arrows,shapes}
\usetikzlibrary{decorations.pathmorphing}
\usetikzlibrary{decorations.markings}
\tikzset{snake it/.style={decorate, decoration=snake}}

\newcommand{\mysquare}[2]{\tikz{\node[draw=#1,fill=#2,rectangle,minimum width=0.2cm,minimum height=0.2cm,inner sep=0pt] at (0,0) {};}}
\newcommand{\mycirc}[2]  {\tikz{\node[draw=#1,fill=#2,circle,minimum width=0.2cm,minimum height=0.2cm,inner sep=0pt] at (0,0) {};}}
\newcommand{\mytriang}[2]{\tikz{\node[draw=#1,fill=#2,isosceles triangle,isosceles triangle stretches,shape border rotate=90,minimum width=0.2cm,minimum height=0.2cm,inner sep=0pt] at (0,0) {};}}
\newcommand{\mytriangleft}[2]{\tikz{\node[draw=#1,fill=#2,isosceles triangle,isosceles triangle stretches,shape border rotate=180,minimum width=0.2cm,minimum height=0.2cm,inner sep=0pt] at (0,0) {};}}
\newcommand{\mygrad}[2]{\tikz{\node[draw=#1,fill=#2,isosceles triangle,isosceles triangle stretches,shape border rotate=270,minimum width=0.2cm,minimum height=0.2cm,inner sep=0pt] at (0,0) {};}}
\newcommand{\mydiam}[2]  {\tikz{\node[draw=#1,fill=#2,rectangle,rotate=45,minimum width=0.16cm,minimum height=0.16cm,inner sep=0pt] at (0,0) {};}}
\newcommand{\mystar}[2]  {\tikz{\node[draw=#1,fill=#2,star, star points=5, star point ratio=2.5,scale=0.3] at (0,0) {};}}

\tikzstyle{longdash}=[dash pattern=on 6pt off 2pt]
\tikzstyle{longdash}=[dash pattern=on 6pt off 2pt]
\tikzstyle{dashdotdot}=[dash pattern=on 3pt off 2pt on \the\pgflinewidth off 2pt on \the\pgflinewidth off 2pt]
\newcommand{\lsolid}  [1]   {\raisebox{2pt}{\tikz{\draw[#1,solid     ,line width=1pt](0,0)--(07mm,0);}}}
\newcommand{\ldash}   [1]   {\raisebox{2pt}{\tikz{\draw[#1,dashed    ,line width=1pt](0,0)--(07mm,0);}}}
\newcommand{\ldott}   [1]   {\raisebox{2pt}{\tikz{\draw[#1,dotted    ,line width=1pt](0,0)--(07mm,0);}}}
\newcommand{\ldashdot}[1]   {\raisebox{2pt}{\tikz{\draw[#1,dashdotted,line width=1pt](0,0)--(07mm,0);}}}
\newcommand{\ldashlon}[1]   {\raisebox{2pt}{\tikz{\draw[#1,longdash  ,line width=1pt](0,0)--(07mm,0);}}}
\newcommand{\ldashdotdot}[1]{\raisebox{2pt}{\tikz{\draw[#1,dashdotdot,line width=1pt](0,0)--(07mm,0);}}}

\newcommand{\lsolidsquare}[1]{\raisebox{2pt}{\tikz{\draw[#1,solid,line width=1pt](0,0)--(10mm,0);}}
                             \hspace{-0.75cm}\mysquare{black}{white}\hspace{0.5cm}}
\newcommand{\lsolidcirc}[1]{\raisebox{2pt}{\tikz{\draw[#1,solid,line width=1pt](0,0)--(10mm,0);}}
                             \hspace{-0.75cm}\mycirc{black}{white}\hspace{0.5cm}}
\newcommand{\lsolidtriang}[1]{\raisebox{2pt}{\tikz{\draw[#1,solid,line width=1pt](0,0)--(10mm,0);}}
                             \hspace{-0.75cm}\mytriang{black}{white}\hspace{0.5cm}}
\newcommand{\ldashdotgrad}[1]{\raisebox{2pt}{\tikz{\draw[#1,dashdotted, line width=1pt]
                             (0,0)--(10mm,0);}}\hspace{-0.6cm}\mygrad{#1}{#1}\hspace{0.5cm}}
\newcommand{\ldashtriang}[1]{\raisebox{2pt}{\tikz{\draw[#1,dashed,line width=1pt](0,0)--(10mm,0);}}
                             \hspace{-0.75cm}\mytriang{#1}{#1}\hspace{0.5cm}}
\newcommand{\lsolidcircle}[1]{\raisebox{2pt}{\tikz{\draw[#1,solid,line width=1pt](0,0)--(10mm,0);}}
                             \hspace{-0.75cm}\mycirc{#1}{#1}\hspace{0.5cm}}

% ---------------------------------------------------------------------------
\newcommand{\markerone}{\raisebox{0.5pt}{\tikz{\node[draw,scale=0.4,circle,black,fill=none](){};}}}
\newcommand{\markertwo}{\raisebox{0pt}{\tikz{\node[draw,scale=0.3,regular polygon, regular polygon sides=3,orange,fill=none,rotate=270](){};}}}
\newcommand{\markerthree}{\raisebox{0.5pt}{\tikz{\node[draw,scale=0.3,regular polygon, regular polygon sides=3,fill=black!10!red,rotate=0](){};}}}
\newcommand{\markerfour}{\raisebox{0.5pt}{\tikz{\node[draw,scale=0.4,regular polygon, regular polygon sides=4,fill=none](){};}}}
\newcommand{\markerfive}{\raisebox{0pt}{\tikz{\node[draw,scale=0.4,diamond,blue,fill=none](){};}}}
\newcommand{\markersix}{\raisebox{0.6pt}{\tikz{\node[draw,scale=0.3,circle,fill=black!100!](){};}}}

\tikzstyle{decision} = [diamond, draw, fill=blue!20,
    text width=4.5em, text badly centered, node distance=3cm, inner sep=0pt]
\tikzstyle{start} = [rectangle, draw, fill=blue!20,
    text width=4.5em, text centered, rounded corners, minimum height=4em]
\tikzstyle{block} = [rectangle, draw, fill=blue!20,
    text width=4.5em, text centered, minimum height=4em]
\tikzstyle{line} = [draw, -latex']
%========================================================================================

\newcommand {\Caption} [1]{ \caption[~#1]{#1} }
\newcommand {\B}      [1] { \boldsymbol{#1} }
\newcommand {\DPS}        { \displaystyle }
\newcommand {\D}      [1] { \delta #1 }

\newcommand {\oline}  [1] { \overline{#1} }
\newcommand {\doline} [1] { \overline{\overline{#1}} }
\newcommand {\uline}  [1] { \underline{#1} }
\newcommand {\dline}  [1] { \overline{\overline{#1}} }
\newcommand {\duline} [1] { \underline{\underline{#1}} }
\newcommand {\dvec}   [1] { \vec{\vec{#1}} }
\newcommand {\wtilde} [1] { \widetilde{#1} }
\newcommand {\dtilde} [1] { \widetilde{\widetilde{#1}} }
\newcommand {\what}   [1] { \widehat{#1} }
\newcommand {\dhat}   [1] { \widehat{\widehat{#1}} }
\newcommand {\derp}   [2] { \partial #1 / \partial #2  }
\newcommand {\derpp}  [2] { \partial^2 #1 / \partial #2^2  }
\newcommand {\derpc}  [3] { \partial^2 #1 / \partial #2 \partial #3  }
\newcommand {\derpn}  [3] { \partial^#1 #2 / \partial #3^#1 }
\newcommand {\Derpnc} [6] { \frac { \partial^{#1} #2 } { \partial #3^{#4} \partial #5^{#6}} }
\newcommand {\derd}   [2] { d #1 / d #2  }
\newcommand {\derdd}  [2] { d^2 #1 / d #2^2  }
\newcommand {\derD}   [2] { D #1 / D #2  }
\newcommand {\Derp}   [2] { \frac { \partial #1 } { \partial #2 } }
\newcommand {\Derpp}  [2] { \frac { \partial^2 #1 } { \partial #2^2 } }
\newcommand {\Derpc}  [3] { \frac { \partial^2 #1 } { \partial #2 \partial #3 } }
\newcommand {\Derpn}  [3] { \frac { \partial^{#1} #2 } { \partial #3^{#1} } }
\newcommand {\Derd}   [2] { \frac { d #1 } { d #2 } }
\newcommand {\Derdd}  [2] { \frac { d^2 #1 } { d #2^2 } }
\newcommand {\Derdn}  [3] { \frac { d^{#1} #2 } { d #3^{#1} } }
\newcommand {\DerD}   [2] { \frac { D #1 } { D #2 } }
\newcommand {\pp}     [1] {  { \left( #1 \right) } }
\newcommand {\cro}    [1] {  { \left[ #1 \right] } }
\newcommand {\pip}    [1] {  { \left| #1 \right| } }
\newcommand {\acc}    [1] {  { \left\{ #1 \right\} } }
\newcommand {\bra}    [1] {  { \left\langle #1 \right\rangle } }
\newcommand {\vecteur}[1] { \left( \begin{array}{c} #1 \end{array} \right) }
\newcommand {\Remarque} [1]   {{\noindent\bf\uline{Remarque{#1}~:}}}

\newcommand {\CR}        [1] {{\color{brick}{#1}}}
\newcommand {\CB}        [1] {{\color{blue}{#1}}}
\newcommand {\CG}        [1] {{\color{green}{#1}}}
\newcommand {\CC}        [1] {{\color{cyan}{#1}}}

\sisetup{group-digits=false}
%\sisetup{table-parse-only}
%\newcolumntype{L}{>{\phantom{$-$}}l}
%\newcolumntype{C}[1]{>{\centering\arraybackslash}p{#1}}
\newcolumntype{D}{>{$\displaystyle} c <{$}}
\newcolumntype{L}[1]{>{\raggedright\let\newline\\\arraybackslash\hspace{0pt}}m{#1}}
\newcolumntype{C}[1]{>{\centering\let\newline\\\arraybackslash\hspace{0pt}}m{#1}}
\newcolumntype{R}[1]{>{\raggedleft\let\newline\\\arraybackslash\hspace{0pt}}m{#1}}
\newcommand*{\Scale}[2][4]{\scalebox{#1}{$#2$}}%
%% SYMBOLS %%
%%%%%%%%%%%%%%%%%%%%%%%%%%%% Commandes personnels %%%%%%%%%%%%%%%%%%%%%%%%%%%%%%%%%%%%%%%%%%%%%%%%%%%
\newcommand{\md}{\mathrm{d}}
\newcommand{\mdd}{\, \mathrm{d}}
\newcommand{\ovl}{\overline}
\newcommand{\p}{\partial}
\newcommand{\dd}[2][]{\frac{\partial#1}{\partial#2}}
\newcommand{\mf}[1]{\mathbf{#1}}
\newcommand{\ov}{\overline}
\newcommand{\wt}{\widetilde}
\newcommand{\rref}{\ensuremath{\text{ref}}}
\newcommand{\rms}{\ensuremath{\text{rms}}}
\newcommand{\sij}{\ensuremath{S_{ij}}}
\newcommand{\wij}{\ensuremath{W_{ij}}}
\newcommand{\aij}{\ensuremath{A_{ij}}}
\newcommand{\aji}{\ensuremath{A_{ji}}}
\newcommand{\oij}{\ensuremath{\Omega_{ij}}}
\newcommand{\rhorms}{\ensuremath{\rho_{\text{rms}}}}
\newcommand{\prms}{\ensuremath{p_{\text{rms}}}}
\newcommand{\crms}{\ensuremath{c_{\text{rms}}}}
\newcommand{\Trms}{\ensuremath{T_{\text{rms}}}}
\newcommand{\vrtrms}{\ensuremath{\omega_{\text{rms}}}}
\newcommand{\divrms}{\ensuremath{\theta_{\text{rms}}}}
\newcommand{\thetarms}{\ensuremath{\theta_{\text{rms}}}}
\newcommand{\mt}  {\ensuremath{M_{t_0}}}
\newcommand{\mbw} {\ensuremath{M_{B}}}
\newcommand{\rebw}{\ensuremath{Re_{B}}}
\newcommand{\rebb}{\ensuremath{Re_{Bb}}}
\newcommand{\row}{\ensuremath{\overline{\rho}_w}}
\newcommand{\tw}{\ensuremath{\overline{T}_w}}
\newcommand{\muw}{\ensuremath{\overline{\mu}_w}}
\newcommand{\tcl} {\ensuremath{\overline{T}_{cl}}}
\newcommand{\ccl} {\ensuremath{\overline{c}_{cl}}}
\newcommand{\mcl} {\ensuremath{\overline{M}_{cl}}}
\newcommand{\mucl}{\ensuremath{\overline{\mu}_{cl}}}
\newcommand{\recl}{\ensuremath{\overline{Re}_{cl}}}
\newcommand{\rocl}{\ensuremath{\overline{\rho}_{cl}}}
\newcommand{\retau} {\ensuremath{Re_{\tau}}}
\newcommand{\retaus}{\ensuremath{Re_{\tau}^*}}
\newcommand{\ratio}{\ensuremath{\theta/\theta_\text{rms}}}
\newcommand{\ratiomt}{\ensuremath{\theta/\theta_\text{rms}M_{t_0}^2}}
\newcommand{\bomega}{\ensuremath{\boldsymbol{\omega}}}
\newcommand{\blambda}{\ensuremath{\boldsymbol{\Lambda}}}
\newcommand{\relambda}{\ensuremath{Re_{\lambda}}}

\tikzstyle{longdash}=[dash pattern=on 6pt off 2pt]
\tikzstyle{longdash}=[dash pattern=on 6pt off 2pt]
\tikzstyle{dashdotdot}=[dash pattern=on 3pt off 2pt on \the\pgflinewidth off 2pt on \the\pgflinewidth off 2pt]
%\newcommand{\solid}{\rule[0.1cm]{7mm}{0.2mm}}
%\newcommand{\plm}{\rule[0.1cm]{2mm}{0.2mm}}
%\newcommand{\pli}{\rule[0.1cm]{0.5mm}{0.2mm}}
%\newcommand{\dashed}{\plm\hspace{2.0mm}\plm\hspace{2.0mm}\plm}
%\newcommand{\dasheddot}{\plm\hspace{0.5mm}\pli\hspace{0.5mm}\plm\hspace{0.5mm}\pli}
%\newcommand{\dasheddotdot}{\plm\hspace{0.5mm}\pli\hspace{0.5mm}\pli\hspace{0.5mm}\plm\hspace{0.5mm}
%\pli\hspace
%{0.5mm}\pli}
%\newcommand{\ssolid}{\rule[0.1cm]{4mm}{0.2mm}}
%\newcommand{\sssolid}{\rule[0.1cm]{1mm}{0.2mm}}
%\newcommand{\solidsquare}{\plm\hspace{-1.0mm}$\blacksquare$\hspace{-1.0mm}\plm}
%\newcommand{\solidcircle}{\plm\hspace{-1.0mm}$\bullet$\hspace{-1.0mm}\plm}
%\newcommand{\solidgradient}{\plm\hspace{-1.0mm}$\blacktriangledown$\hspace{-1.0mm}\plm}
%%%%%%%%%%%%%%% Definitions de couleurs %%%%%%%%%%%%%%%%%%%
\definecolor{brick}{rgb}{0.75,0,0}
\definecolor{blight}{rgb}{0.7,0.65,1}
\definecolor{yellow1}{rgb}{1,0.85,0.1}
\definecolor{yellow2}{rgb}{1,0.95,0.5}
\definecolor{lila}{rgb}{0.7,0.3,0.5}
\definecolor{brick}{rgb}{0.75,0,0}
\definecolor{rltgreen}{rgb}{0,0.5,0}
\definecolor{oneblue}{rgb}{0,0,0.75}
\definecolor{marron}{rgb}{0.64,0.16,0.16}
\definecolor{forestgreen}{rgb}{0.13,0.54,0.13}
\definecolor{purple}{rgb}{0.62,0.12,0.94}
\definecolor{dockerblue}{rgb}{0.11,0.56,0.98}
\definecolor{freeblue}{rgb}{0.25,0.41,0.88}
\definecolor{myblue}{rgb}{0,0.2,0.4}
\definecolor{rosy}{rgb}{.971,.951,.92} % A rosy manila
\definecolor{gris}{rgb}{0.6,0.6,0.6}
\definecolor{grisclair}{rgb}{0.9,0.9,0.9}
\definecolor{rouge}{rgb}{0.6,0.,0.}
\definecolor{darkgreen}{rgb}{0.12, 0.3, 0.17}

\newcommand{\vA}{{\bf A}}
\newcommand{\va}{{\bf a}}
\newcommand{\vB}{{\bf B}}
\newcommand{\vb}{{\bf b}}
\newcommand{\vC}{{\bf C}}
\newcommand{\vc}{{\bf c}}
\newcommand{\vD}{{\bf D}}
\newcommand{\vd}{{\bf d}}
\newcommand{\vE}{{\bf E}}
\newcommand{\ve}{{\bf e}}
\newcommand{\vF}{{\bf F}}
\newcommand{\vf}{{\bf f}}
\newcommand{\vG}{{\bf G}}
\newcommand{\vg}{{\bf g}}
\newcommand{\vH}{{\bf H}}
\newcommand{\vh}{{\bf h}}
\newcommand{\vI}{{\bf I}}
\newcommand{\vi}{{\bf i}}
\newcommand{\vJ}{{\bf J}}
\newcommand{\vj}{{\bf j}}
\newcommand{\vk}{{\bf k}}
\newcommand{\vK}{{\bf K}}
\newcommand{\nk}{{k}}
\newcommand{\vL}{{\bf L}}
\newcommand{\vl}{{\bf l}}
\newcommand{\vM}{{\bf M}}
\newcommand{\vn}{{\bf n}}
\newcommand{\vP}{{\bf P}}
\newcommand{\vp}{{\bf p}}
\newcommand{\Pb}{\bar P}
\newcommand{\vq}{{\bf q}}
\newcommand{\vR}{{\bf R}}
\newcommand{\vr}{{\bf r}}
\newcommand{\vS}{{\bf S}}
\newcommand{\vs}{{\bf s}}
\newcommand{\Tb}{{\bar T}}
\newcommand{\vT}{{\bf T}}
\newcommand{\vt}{{\bf t}}
\newcommand{\vU}{{\bf U}}
\newcommand{\vUb}{\bar {\bf U}}
\newcommand{\Ub}{\bar U}
\newcommand{\vu}{{\bf u}}
\newcommand{\vuo}{{\bf u}_o}
\newcommand{\vV}{{\bf V}}
\newcommand{\Vb}{\bar V}
\newcommand{\vv}{{\bf v}}
\newcommand{\vW}{{\bf W}}
\newcommand{\vw}{{\bf w}}
\newcommand{\vX}{{\bf X}}
\newcommand{\vx}{{\bf x}}
\newcommand{\vY}{{\bf Y}}
\newcommand{\vy}{{\bf y}}
\newcommand{\vz}{{\bf z}}
\begin{frontmatter}

\title{Assessment of a high-order shock-capturing central-difference scheme for hypersonic turbulent flow simulations}

\author[ensam]{Luca Sciacovelli\corref{mycorrespondingauthor}}
\cortext[mycorrespondingauthor]{Corresponding author}
\ead{luca.sciacovelli@ensam.eu}
\author[ensam,poliba]{Donatella Passiatore}
\author[ensam]{Paola Cinnella}
\author[poliba]{Giuseppe Pascazio}

\address[ensam]{Arts et Métiers ParisTech, DynFluid Laboratory, 75013 Paris, France}
\address[poliba]{Polytechnic University of Bari, DMMM, 70120 Bari, Italy}

% \phantomsection

\begin{abstract}
High-speed turbulent flows are encountered in most space-related applications (including exploration, tourism and defense fields) and represent a subject of growing interest in the last decades. A major challenge in performing high-fidelity simulations of such flows resides in the stringent requirements for the numerical schemes to be used. These must be robust enough to handle strong, unsteady discontinuities, while ensuring low amounts of intrinsic dissipation in smooth flow regions. Furthermore, the wide range of temporal and spatial active scales leads to concurrent needs for numerical stabilization and accurate representation of the smallest resolved flow scales in cases of under-resolved configurations.
In this paper, we present a finite-difference high-order shock-capturing technique based on Jameson's artificial diffusivity methodology. The resulting scheme is ninth-order-accurate far from discontinuities and relies on the addition of artificial dissipation close to large gradients. The shock detector is slightly revised to enhance its selectivity and avoid spurious activations of the shock-capturing term.
A suite of test cases ranging from 1D to 3D configurations (namely, shock tubes, Shu-Osher problem, isentropic vortex advection, under-expanded jet, compressible Taylor-Green Vortex, supersonic and hypersonic turbulent boundary layers) is analysed in order to test the capability of the proposed numerical strategy to handle a large variety of problems, ranging from calorically-perfect air to multi-species reactive flows. Results obtained on under-resolved grids are also considered to test the applicability of the proposed strategy in the context of implicit Large-Eddy Simulations.
\end{abstract}


\end{frontmatter}

\linenumbers

%===============================================================================
\section{Introduction}
Hypersonic flight has gained renewed attention in recent years, due to its importance for multiple breakthrough applications in the defense and military fields, as well as in the areas of spatial tourism and trans-atmospheric flight (see \cite{leyva2017relentless}).
The accurate prediction of hypersonic flow fields is a challenging task, due to the
massive conversion of kinetic energy from the hypersonic free stream into internal energy as the fluid approaches the body.
The shock waves generated in such flight conditions produce a sharp increase in the fluid temperature, possibly causing vibrational excitation and gas dissociation, and resulting in a nonequilibrium thermochemical state. These processes have major effects on aerodynamic performance, heat transfer rates, fluid-surface interaction (e.g., ablation), and hydrodynamic instabilities leading to boundary layer transition and breakdown to turbulence \citep{candler2019rate}. Their accurate prediction is of crucial importance for the design of the thermal protection system and the prediction of the overall force and heat transfer coefficients, and requires advanced numerical solvers and models.
In this paper, the focus is put on so-called high-fidelity numerical models for Direct Numerical Simulation (DNS) and Large Eddy Simulation (LES), two major enablers for deeper understanding of out-of-equilibrium flow regions dominated by laminar-to-turbulent transition and turbulent regimes.

A major difficulty in DNS and LES of high-speed flows is the extreme sensitivity of small flow scales to numerical approximation errors. The occurrence of shockwaves, with physical thicknesses of the order of a few mean free paths, leads to unfeasible resolution requirements for numerical simulations, at least in the strict DNS sense. On the other hand, velocity fluctuations of the order of the sound speed \cite{lee1991eddy} may lead to the formation of eddy shocklets, embedded in the turbulent flow.
Usually, both kinds of structures are dealt with by using shock-capturing techniques. These consist in locally injecting controlled amounts of numerical dissipation to spread shocks over a few mesh cells wherever the mesh is not sufficiently fine to resolve the shock thickness.
This technique, corresponding to a regularization method, contrasts with the need of using low-dissipation schemes not to alter the fine-scale turbulent motions. These opposite requirements become even more critical as the Mach and Reynolds numbers increase, because of the growing difficulty to distinguish strong gradients due to shocks from those related to turbulent fluctuations.
In such a challenging framework, much effort has been done in the literature to devise numerical methods able handle strong shock waves, while ensuring minimal amounts of dissipation elsewhere.

Two great families of discretization method for the non-linear terms in the Euler and Navier--Stokes equations can be distinguished \cite{pirozzoli2011numerical}: the first one, originally designed to handle inviscid flows with strong discontinuities, relies on some form of upwinding along with flux or slope limiters ensuring non-linear stability; the second one --in principle more suited for smooth flows-- uses central schemes supplemented with some form of filtering or artificial dissipation, or alternatively ensuring discrete conservation of solution invariants such as the overall kinetic energy.
Despite the large number of comparative studies \cite{johnsen2010assessment,lo2010high,lusher2019assessment}, a global consensus on the ``best'' numerical strategy for high-speed turbulent flows has not been reached, since each method proposes a different compromise among concurrent needs: namely, high accuracy, robustness, low computational cost, few tuning parameters and suitability for different configurations.

A well-known drawback of the first class of schemes is the excess of numerical viscosity introduced in the solution, leading to spurious entropy generation and kinetic energy losses in the low Mach number limit \cite{thornber2008entropy}.
Weighted essentially non-oscillatory (WENO) schemes are probably the most popular upwind schemes in the context of LES and DNS of compressible flows. First introduced by Liu \emph{et al.} \cite{liu1994weighted} and later improved by Jiang \& Shu \cite{jiang1996efficient}, they rely on the assembly of high-order numerical fluxes from linear combination of lower-order polynomial reconstructions using suitable weighting coefficients. Many variants have been proposed, e.g., to improve their dispersion and dissipation properties \cite{martin2006bandwidth,hu2010adaptive} and to reduce the nonlinear dissipation \cite{borges2008improved,acker2016improved}.
Coupling with purely central schemes has led to hybrid methods \cite{pirozzoli2002conservative,ren2003characteristic} and enhanced weighting strategies and smoothness sensors \cite{van2017embedded,fu2017targeted,fu2019very,zhao2020shock}.
Contrary-wise, the family of central schemes generally introduces very low dissipation, provided that a selective enough numerical filter or artificial viscosity term is used, but is generally limited to compressible flows at moderately supersonic Mach numbers, i.e. with weak shocks.
These schemes must be supplemented with selective nonlinear filtering \cite{kim2001adaptive,visbal2005shock,hixon2006shock,bogey2009shock}, artificial diffusive fluxes \cite{jameson1981numerical,tadmor1989convergence,kim2001adaptive}, or localized artificial diffusivity (LAD) under the form of modified transport coefficients \cite{cook2004high,cook2007artificial,kawai2010assessment} for damping grid-to-grid oscillations in smooth flow regions and to ensure shock capturing.
The amount of numerical dissipation introduced at a point of the computational domain is adjusted by means of properly-devised sensors, allowing to switch on shock-capturing capabilities where needed. Shock-capturing high-order central-difference schemes have been successfully applied, e.g., to overexpanded jet flows with shock cells \cite{decacqueray2014noise} and high-speed boundary layers of perfect gases up to Mach 6 \cite{franko2013breakdown}, as well as to the direct and large eddy simulation of high-speed flows of single-species, molecularly complex, heavy gases at thermodynamic conditions of the order of magnitude of the liquid/vapor critical point up to Mach 6 \cite{sciacovelli2016dense,sciacovelli2017direct,sciacovelli2017small,sciacovelli2020numerical}.
However, their suitability for the numerical simulation of severe hypersonic, chemically reacting flows with strong shocks and stiff chemical reactions has not yet been assessed.

For LES, where only the dynamics of the large scales is computed while the effects of sub-grid scales (SGS) are modeled, the choice of the numerical scheme is possibly even more critical than in DNS. Scale separation is indeed difficult to establish since the cut-off between resolved and modeled scales is not sharp and arises from a complex combination of implicit filtering by the grid and the discretization schemes.
The intricate interactions between numerical errors and SGS modeling errors has been investigated by numerous authors
(a recent discussion can be found in \cite{gloerfelt2019large}), leading once again to two separate modeling strategies: one relying on the explicit introduction of a SGS model and the other one using the dissipative part of the discretization scheme for ensuring regularization of the unresolved SGS scales \cite{rizzetta88numerical,boris92insights,mathew03explicit,mathew06new,bogey2009turbulence}. The latter has become increasingly spread in the scientific community, due to the good tradeoff between computational cost and accuracy offered for a wide range of applications, provided that a high-resolution scheme is used along with a sufficiently resolved computational grids \cite{aubard2013comparison,gloerfelt2019large}.

The goal of the present study is twofold: i) to assess the capability of a high-order shock-capturing central scheme, used in our previous works \cite{sciacovelli2016dense,sciacovelli2020numerical}, to robustly predict compressible flows with shock waves and chemical non-equilibrium effects while accurately resolving fine-scale turbulent structures; and ii) to demonstrate the suitability of the non-linear numerical dissipation of the scheme to act as a SGS regularization in under-resolved turbulent flow simulations. The scheme uses tenth-order accurate finite-difference approximations of the non-linear fluxes, supplemented with a higher-order extension of Jameson's adaptive artificial dissipation \cite{jameson1981numerical}. The order of accuracy of the artificial viscosity term is chosen to obtain an overall dissipative-dominant truncation error, which reduces the appearance and amplification of spurious oscillations \cite{thomee1965stability} and limits the activation of lower-order nonlinear viscosity. The latter is triggered by a highly-selective shock sensor, built on a combination of the original Ducros' extension \cite{ducros1999large} of Jameson's pressure-based sensor with the Bhagatwala \& Lele \cite{bhagatwala2009modified} modification proposed in the context of LAD methods (more details are given in section~\ref{sec:numeth}).
The scheme is applied to a suite of well-documented test cases of increasing complexity, ranging from 1D and 2D inviscid flow problems to the 3D simulation of a fully turbulent boundary layer at Mach 10 in chemical nonequilibrium conditions and the results are systematically assessed against exact or numerical reference solutions.
%
%===============================================================================
\section{Governing Equations}
\label{sec:goveq}
Our goal is to simulate flows governed by the compressible Navier-Stokes equations for multicomponent chemically-reacting gases, written in differential form:
\begin{align}
\frac{\partial \rho}{\partial t} + \frac{\partial \rho u_j }{\partial x_j} & =0 \\
\label{eq:momentum}
\frac{\partial \rho u_i}{\partial t} + \frac{\partial \left( \rho u_i u_j + p \delta_{ij} \right)}{\partial x_j} &=\frac{\partial \tau_{ij}}{\partial x_j} \\
\label{eq:energy}
\frac{\partial \rho E}{\partial t} + \frac{\partial \left[\left(\rho E + p \right)u_j\right]}{\partial x_j} & =\frac{\partial (u_i \tau_{ij} - q_j)}{\partial x_j} -\frac{\partial}{\partial x_j}\left(\sum_{n=1}^\text{NS} \rho_n u_{nj}^D h_n \right) \\
\label{eq:species}
\frac{\partial \rho_n}{\partial t} + \frac{\partial \left( \rho_n u_j \right)}{\partial x_j} &= -\frac{\partial \rho_n u_{nj}^D }{\partial x_j} + \dot{\omega}_n \qquad (n = 1,..,\text{NS}-1)
%
% \label{eq:vibr}
% \frac{\partial \rho e_V}{\partial t} + \frac{\partial \rho e_V u_j}{\partial x_j} &= \frac{\partial}{\partial x_j} \left( -q_{Vj} - \sum_{m=1}^\text{NM} \rho_m u_{mj}^D e_{Vm} \right) + \sum_m \left( Q_\text{TV} + \dot{\omega}_m e_{Vm} \right)
\end{align}
In the preceding equations, $\rho_n$ is the density of the $n$-th species, $\rho{=}\sum_{n=1}^\text{NS} \rho_n$ the mixture density, NS the total number of species, $u_i$ the velocity vector components in a Cartesian coordinate system, $p$ the pressure, $\delta_{ij}$ the Kronecker symbol and $\tau_{ij}$ the viscous stress tensor; in equation~\eqref{eq:energy}, $E = e + \frac{1}{2}u_i u_i$ represents the specific total energy (with $e$ the mixture specific internal energy) and $q_j$ the heat flux; moreover, $u_{nj}^D$, $h_n$, and $\dot{\omega}_n$ denote the $n$-th species diffusion velocity, specific enthalpy and rate of production, respectively. For temperature values lower than \SI{9000}{K}, ionization and electronic processes can be usually neglected; air is then modeled as a five-species mixture of N$_2$, O$_2$, NO, O and N ($\text{NS}=5$ \cite{park1989nonequilibrium}). To ensure total mass conservation, we solve for the mixture density and NS$-1$ species conservation equations, the NS-th species being computed as $\rho_\text{NS} = \rho - \sum_{n=1}^{\text{NS}-1} \rho_n$. This species is chosen to be Nitrogen since it has the largest mass fraction throughout the computational domain.
The viscous stress tensor is modeled as:
\begin{equation}
 \tau_{ij} = \mu\left(\frac{\partial u_i}{\partial x_j} + \frac{\partial u_j}{\partial x_i} \right) -\frac{2}{3}\mu\frac{\partial u_k}{\partial x_k}\delta_{ij},
% \tau_{ij} = \mu\left(\frac{\partial u_i}{\partial x_j} + \frac{\partial u_j}{\partial x_i} \right) + (\beta-\frac{2}{3}\mu)\frac{\partial u_k}{\partial x_k}\delta_{ij},
 \end{equation}
with $\mu$ the mixture dynamic viscosity. % and $\beta$ the bulk viscosity (supposed to be zero according to Stokes' hypothesis); t
The heat flux is modeled by means of Fourier's law, $q_j = -\lambda \frac{\partial T}{\partial x_j}$, $\lambda$ being the mixture thermal conductivity and $T$ the temperature. Each species is assumed to behave as a perfect gas; Dalton's pressure mixing law leads then to the thermal equation of state:
\begin{equation}
   p = \sum_{n=1}^\text{NS} p_n = \rho T \sum_{n=1}^\text{NS} \frac{\mathcal{R} Y_n}{\mathcal{M}_n} = T \sum_{n=1}^\text{NS} \rho_n R_n,
\end{equation}
$Y_n=\rho_n/\rho$, $R_n$ and ${\cal M}_n$ being the mass fraction, gas constant and molecular weight of the $n$-th species, respectively, and $\mathcal{R} = 8.314$ J/mol~K the universal gas constant. The thermodynamic properties of high-$T$ air species are computed considering the contributions of translational, rotational and vibrational modes \cite{gnoffo1989conservation}; specifically, the internal energy reads:
\begin{equation}
e = \sum_{n=1}^\text{NS} Y_n h_n - \frac{p}{\rho},
\quad \text{with} \quad
h_n = h^0_{f,n} + \int_{T_\text{ref}}^T (c^\text{tr}_{p,n}+c^\text{rot}_{p,n}) \text{ d}T' + e^\text{vib}_n.
\end{equation}
Here, $h^0_{f,n}$ is the $n$-th species enthalpy of formation at the reference temperature ($T_\text{ref} = \SI{298.15}{K}$), $c^\text{tr}_{p,n}$ and $c^\text{rot}_{p,n}$ the translational and rotational contributions to the isobaric heat capacity of the $n$-th species, computed as
\begin{equation}
   c^\text{tr}_{p,n} = \frac{5}{2} R_n \qquad \text{and} \qquad
   c^\text{rot}_{p,n} = \begin{cases}
    R_n & \text{for diatomic species} \\
    0 & \text{for monoatomic species}
   \end{cases},
\end{equation}
and $e^\text{vib}_n$ the vibrational energy of species $n$, given by
\begin{equation}
 e^\text{vib}_n =  \frac{\theta_n R_n}{\exp{(\theta_n/T_V)} - 1},
\end{equation}
with $\theta_n$ the characteristic vibrational temperature of each molecule (\SI{3393}{K}, \SI{2273}{K} and \SI{2739}{K} for N$_2$, O$_2$ and NO, respectively \cite{park1989nonequilibrium}).
After the numerical integration of the conservation equations, temperature is computed from the specific internal energy by means of an iterative Newton--Raphson method.

Pure species' viscosity and thermal conductivities are computed using curve-fits by Blottner \emph{et al.} \cite{blottner1971chemically} and Eucken's relations \cite{vincenti1965introduction}:
\begin{equation}
   \mu_n = 0.1 \exp[ (A_n \ln T + B_n)\ln T + C_n ], \qquad \lambda_n = \mu_n \left( \frac{5}{2} c^\text{tr}_{v,n} + c^\text{rot}_{v,n} + c^\text{vib}_{v,n} \right)
\end{equation}
where $A_n$, $B_n$ and $C_n$ are fitted parameters. The corresponding mixture properties are evaluated by means of Wilke's mixing rules \cite{wilke1950viscosity}:
\begin{equation}
   \mu = \sum_{n=1}^\text{NS} \frac{X_n \mu_n}{\sum_{m=1}^\text{NS} X_m \phi_{nm}}, \qquad \lambda = \sum_{n=1}^\text{NS} \frac{X_n \lambda_n}{\sum_{m=1}^\text{NS} X_m \phi_{nm}}
\end{equation}
where $\displaystyle X_n = \frac{Y_n R_n}{\sum_{m=1}^\text{NS} Y_m R_m}$ denotes the molar fraction of species $n$ and
\begin{equation}
   \phi_{nm} = \frac{1}{\sqrt{8}} \left( 1 + \frac{{\cal M}_n}{{\cal M}_m} \right)^{-\frac{1}{2}} \left[ 1 + \left(  \frac{\mu_n}{\mu_m}\right)^{-\frac{1}{2}} \left( \frac{{\cal M}_m}{{\cal M}_n} \right)^{\frac{1}{4}} \right]^2.
\end{equation}
Mass diffusion is modeled by means of Fick's law:
\begin{equation}
 \rho_n u^D_{nj}= -\rho D_{n} \frac{\partial Y_n}{\partial x_j} + \rho_n \sum_{n=1}^\text{NS} D_n \frac{\partial Y_n}{\partial x_j},
\end{equation}
where the first term on the r.h.s. represents the effective diffusion velocity and the second one is a mass corrector term that should be taken into account in order to satisfy the continuity equation when dealing with non-constant species diffusion coefficients \cite{poinsot2005theoretical,giovangigli1999multicomponent}. Specifically, $D_n$ is an equivalent diffusion coefficient of species $n$ into the mixture, computed following Hirschfelder's approximation \cite{hirschfelder1954molecular} as
\begin{equation}
 D_n= \frac{1-Y_n}{ \sum_{\substack{m = 1 \\ m \neq n}}^\text{NS} \frac{X_n}{D_{mn}} }
 \quad \text{with} \quad
 D_{mn} = \frac{1}{p}\exp{(A_{4,mn})} T^{\left[ A_{1,mn} (\ln T)^2 + A_{2,mn} \ln T + A_{3,mn} \right]}
\end{equation}
where $D_{mn}$ is the binary diffusion coefficient of species $m$ into species $n$, and $A_{1,mn}, ..., A_{4,mn}$ are curve-fitted coefficients computed as in Gupta \textit{et al.} \cite{gupta1990review}.

The five species interact with each other through a reaction mechanism consisting of five reversible chemical steps \cite{park1989nonequilibrium,park1989review}:
\begin{alignat}{3}
 \notag \text{R1}: & \qquad  \text{N}_2 + \text{M} && \Longleftrightarrow 2\text{N} + \text{M} \\
 \notag \text{R2}: & \qquad  \text{O}_2 + \text{M} && \Longleftrightarrow 2\text{O} + \text{M} \\
        \text{R3}: & \qquad  \text{NO}  + \text{M} && \Longleftrightarrow  \text{N} + \text{O} + \text{M} \\
 \notag \text{R4}: & \qquad \text{N}_2 + \text{O}  && \Longleftrightarrow  \text{NO}+ \text{N} \\
 \notag \text{R5}: & \qquad \text{NO}  + \text{O}  && \Longleftrightarrow  \text{N} + \text{O}_2
\end{alignat}
being M the third body (any of the five species considered). Dissociation and recombination processes are described by reactions R1, R2 and R3; whereas the shuffle reactions R4 and R5 represent rearrangement processes. The mass rate of production of the \textit{n}-th species is governed by the law of mass action:
\begin{equation}
 \dot{\omega }_n =  \mathcal{M}_n \sum_{r=1}^\text{NR} \left( \nu_{nr}'' - \nu_{nr}' \right) \times \left[ k_{f,r} \prod_{n=1}^\text{NS} \left(\frac{\rho Y_n}{\mathcal{M}_n}\right)^{\nu_{nr}'} - k_{b,r} \prod_{n=1}^\text{NS} \left(\frac{\rho Y_n}{\mathcal{M}_n}\right)^{\nu_{nr}''} \right],
 \label{mass_action}
\end{equation}
where $\nu_{nr}'$ and  $\nu_{nr}''$ are the stoichiometric coefficients for reactants and products in the $r$-th reaction for the $n$-th species, respectively, and NR is the total number of reactions. Lastly, $k_{f,r}$ and $k_{b,r}$ denote the forward and backward reaction rates of reaction $r$, modeled by means of Arrhenius' law.

For configurations in which air is modeled as a single-species calorically-perfect gas, a constant specific heat ratio is considered ($\gamma = 1.4$), such that $c_p = \frac{\gamma R}{\gamma-1}$. Since $\text{NS} = 1$ and there is no chemical activity, equation~\eqref{eq:species} is not solved and the diffusion velocity is zero. Viscosity is computed by means of Sutherland's Law, $\displaystyle \mu(T) = \frac{C T^{3/2}}{T+S}$, with $C = \SI{1.457933e-6}{kg/m\,s\,K^{1/2}}$ and $S = \SI{110.4}{K}$, whereas the thermal conductivity follows a constant Prandtl law, for which $\lambda = \text{Pr}/(\mu c_p)$.

\section{Numerical method}
\label{sec:numeth}
\vspace{-0.15cm}
In this section we describe the high-order shock-capturing central-difference numerical scheme under investigation.
We first present the most important ingredient, i.e. the spatial discretization scheme for the nonlinear terms, for a
1D system of hyperbolic conservation laws:
\begin{equation}
\frac{\partial w}{\partial t}+\frac{\partial f(w)}{\partial x}=0
\end{equation}
where $w$ is the vector of conservative variables and $f(w)$ the flux function, such that $A=\partial f/\partial w$ is a diagonalizable matrix with real eigenvalues. Extension to multidimensional cases is straightforwardly carried out by applying the scheme in each direction.
Introducing the classical difference and cell-average operators over one cell:
\begin{equation}
\left(\delta\bullet\right)_{j}:=(\bullet)_{j+\frac{1}{2}}-(\bullet)_{j-\frac{1}{2}}, \qquad
\left(\mu\bullet\right)_{j+\frac{1}{2}}:=\frac{1}{2}\left[(\bullet)_{j+1}+(\bullet)_j\right]
\end{equation}
and considering a regular Cartesian grid with constant mesh spacing $\delta x$ (so that $x_j{=}j\,\delta x$), a conservative semi-discrete approximation of the spatial derivative writes:
\begin{equation}\label{conservative}
\left(\frac{\partial w}{\partial t}\right)_j+\frac{(\delta {\cal F})_j}{\delta x}=0
\end{equation}
The numerical flux at cell interface $j{+}\frac{1}{2}$,  ${\cal F}_{j+\frac{1}{2}}$, can be calculated using a simple upwind scheme,  written hereafter as the sum of a central approximation ${\cal H}$ and a dissipative term ${\cal D}$ \cite{lerat1996schemas}:
\begin{align}\label{foflux}
{\cal F}_{j+\frac{1}{2}}=& {\cal H}_{j+\frac{1}{2}} -{\cal D}_{j+\frac{1}{2}} \\
{\cal H}_{j+\frac{1}{2}} =&\frac{f_{j+1}+f_j}{2}=(\mu f)_{j+\frac{1}{2}}\\
{\cal D}_{j+\frac{1}{2}} =&\frac{1}{2}|Q_{j+\frac{1}{2}}|(w_{j+1}-w_{j})=\frac{1}{2}(|Q|\delta w)_{j+\frac{1}{2}}
\end{align}
with $Q$ a dissipation matrix.
For instance, $Q=\frac{1}{2}I$ (with $I$ the identity matrix) gives Lax-Friedrichs' scheme \cite{leveque1992numerical}, $Q=\rho(A)I$  Rusanov's scheme \cite{rusanov1961calculation}, and $Q=A_R$ (with $A_R$ the Roe matrix \cite{roe1981approximate}) Roe's scheme. All such schemes are monotonicity preserving according to Godunov's theorem \cite{godunov1959finite}, but only first order accurate, which makes them unsuitable for the purpose of turbulence resolving simulations.

A standard way for increasing accuracy consists in using MUSCL extrapolations \cite{van1979towards}. Here we adopt an alternative approach, first introduced in
\cite{lerat1996schemas} to construct a third-order scheme and generalized to any order of accuracy in~\cite{lerat2003approximations}, which consists in recursively correcting the truncation error of Eqs.~\eqref{conservative}, \eqref{foflux}. For a scheme of order $2P+3$ with a stencil of $2(P+2)+1$ points, the numerical flux is of the form:
\begin{equation}\label{HOupwind}
{\cal F}_{j+\frac{1}{2}} = \left[\left(I-\sum_{p=0}^P a_p\delta^{2+2p}\right)\mu f
     \ -\ \frac{a_P}{2}|Q|\delta^{2P+3} \tilde w \right]_{j+\frac{1}{2}}
\end{equation}
The coefficients $a_p$ have alternate negative and positive signs as $p$ increases. The interested reader may refer to \cite{lerat2003approximations} for details about their calculation.

The preceding high-order, constant coefficient schemes are not total variation diminishing (TVD) nor monotonicity preserving, and slope or flux limiters should be introduced to avoid the appearance of spurious oscillations.
Note that only schemes of odd order accuracy are included in the family, with a leading truncation error term given by:
\begin{equation}
\varepsilon=\frac{a_P}{2}\delta x^{2P+3} \frac{\partial^{2(P+2)}w}{\partial x^{2(P+2)}}
\end{equation}
proportional to an even-order derivative, i.e. of dissipative nature. Explicit odd-order schemes with constant coefficients have been shown to remain stable in the maximum norm $L_\infty$ under a CFL conditions for problems with non-smooth initial conditions \cite{thomee1965stability,despres2009uniform}. This implies that, although the considered schemes do generate spurious oscillations around discontinuities, such oscillations remain bounded in space and time.
For the sake of clarity, we give hereafter the expressions of the schemes of order 3 to 9 (i.e. $P=0,1,2$ and 3) of the preceding family:
\begin{align}
{\cal F}_{j+\frac{1}{2}} &= \left[\left(I-\frac{1}{6}\delta^2\right)\mu f
     \ +\ \frac{1}{12}|Q|\delta^3 \tilde w \right]_{j+\frac{1}{2}}& \text{(order 3)} \\
{\cal F}_{j+\frac{1}{2}} &= \left[\left(I-\frac{1}{6}\delta^2+\frac{1}{30}\delta^4\right)\mu f
    \ -\ \frac{1}{60}|Q|\delta^5 \tilde w \right]_{j+\frac{1}{2}}
    & \text{(order 5)} \\
{\cal F}_{j+\frac{1}{2}} &= \left[\left(I-\frac{1}{6}\delta^2+\frac{1}{30}\delta^4
  -\frac{1}{140}\delta^6\right)\mu f
 \ +\ \frac{1}{280}|Q|\delta^7 \tilde w\right]_{j+\frac{1}{2}}
  & \text{(order 7)} \\
  {\cal F}_{j+\frac{1}{2}}&=\left[\left(I-\frac{1}{6}\delta^2+\frac{1}{30}\delta^4
-\frac{1}{140}\delta^6+\frac{1}{630}\delta^8 \right)\mu f
 \ -\ \frac{1}{1260}|Q|\delta^9 \tilde w\right]_{j+\frac{1}{2}}& \text{(order 9)}
\end{align}
The same schemes can be derived in the finite volume framework by applying MUSCL reconstruction to the physical fluxes~\cite{lerat2003approximations}. Specifically, for $Q=A_R$, one recovers a flux-extrapolation higher-order extension of Roe's scheme. This choice has the advantage of introducing the minimal amount of numerical damping along each characteristic field. However, the extension of Roe's average to real-gas flows is not unique and can introduce significant overcost (see, e.g. \cite{glaister1988linearised,cinnella2006roe}).

With the purpose of simplifying the application to real-gas flows and reducing computational cost, a different approach is adopted. We first observe that the preceding schemes can be considered as standard central difference approximations of order $2(P+2)$ on $2(P+2)+1$ points of the flux derivative, plus a high-order artificial viscosity of order $2P+3$ on the same stencil depending on matrix $Q$.
Afterwards, we choose $Q=\rho(A)$, as in Rusanov's scheme, which is less optimal than Roe's matrix but avoids complexities associated with the extension of the approximate Riemann solver to real gases. This leads to a scalar numerical dissipation term of the form:
\begin{equation}
{\cal D}_{j+\frac{1}{2}}=\left[\frac{a_P}{2}|Q|\delta^{2P+3} \tilde w \right]_{j+\frac{1}{2}}
\end{equation}
Finally, we nonlinearly combine the preceding high-order dissipation with a lower-order term activated in the vicinity of flow discontinuities by means of a highly selective shock sensor. The dissipation ${\cal D}$ then becomes:
\begin{equation}
{\cal D}_{j+\frac{1}{2}}={\rho(A)}_{j+\frac{1}{2}}
\left[\varepsilon_2\delta w + (-1)^{(P+1)}\varepsilon_{2(P+2)}\delta^{(2P+3)} w\right]_{j+\frac{1}{2}}
\end{equation}
with $\rho(A)$ the spectral radius of the flux Jacobian matrix $A=\partial f/\partial w$, and
\begin{equation}
{\varepsilon_2}_{j+\frac{1}{2}}=k_2\max(\varphi_j,\varphi_{j+1}),\qquad
{\varepsilon_{2(P+2)}}_{j+\frac{1}{2}}=\max(0,k_{2(P+2)}-k_\varepsilon {\varepsilon_2}_{j+\frac{1}{2}}),
\end{equation}
where $k_2$ and $k_{2(P+2)}$ are adjustable dissipation coefficients and $k_\varepsilon$ is a constant equal to $a_0/a_P$, determining the threshold below which the higher-order dissipation is switched off.\\
For schemes of order 3 to 9, this gives the following expressions:
\begin{align}
{\cal D}_{j+\frac{1}{2}}&={\rho(A)}_{j+\frac{1}{2}}
\left[\varepsilon_2\delta w - \varepsilon_{4}\delta^3 w\right]_{j+\frac{1}{2}}&\quad&{\varepsilon_{4}}_{j+\frac{1}{2}}=\max(0,k_{4}-{\varepsilon_2}_{j+\frac{1}{2}})\\
{\cal D}_{j+\frac{1}{2}}&={\rho(A)}_{j+\frac{1}{2}}
\left[\varepsilon_2\delta w + \varepsilon_{6}\delta^5 w\right]_{j+\frac{1}{2}}&\quad&{\varepsilon_{6}}_{j+\frac{1}{2}}=\max(0,k_{6}-\frac{1}{5}{\varepsilon_2}_{j+\frac{1}{2}})\\
{\cal D}_{j+\frac{1}{2}}&={\rho(A)}_{j+\frac{1}{2}}
\left[\varepsilon_2\delta w - \varepsilon_{8}\delta^7 w\right]_{j+\frac{1}{2}}&\quad&{\varepsilon_{8}}_{j+\frac{1}{2}}=\max(0,k_{8}-\frac{3}{70}{\varepsilon_2}_{j+\frac{1}{2}})\\\
{\cal D}_{j+\frac{1}{2}}&={\rho(A)}_{j+\frac{1}{2}}
\left[\varepsilon_2\delta w + \varepsilon_{10}\delta^9 w\right]_{j+\frac{1}{2}}&\quad&{\varepsilon_{10}}_{j+\frac{1}{2}}=\max(0,k_{10}-\frac{1}{105}{\varepsilon_2}_{j+\frac{1}{2}})
\end{align}
For $k_2{=}0$ and $\DPS k_{2(P+2)}{=}\frac{a_P}{2}$ one recovers the upwind schemes of Eq. (\ref{HOupwind}).
%, similarly to the original Jameson's formulation, in which $k_4 {=} \frac{1}{12}$ leads to a third-order upwind scheme.
The activation of the low-order dissipation component rests on the value of the shock-capturing sensor $\varphi_j$, which consists in a combination of different terms. Specifically, one has:
\begin{equation}
\label{eq:sensor}
\varphi_j =
\underbrace{\frac{1}{2} \left[1 - \tanh \left(2.5 + 10 \frac{\delta x}{c} \nabla \cdot \mathbf u \right) \right]}_{\text{I}} \times
\underbrace{\frac{(\nabla\cdot \mathbf u)^2}{(\nabla\cdot \mathbf u)^2 + |\nabla\times\mathbf u|^2  + \epsilon}}_{\textrm{II}} \times
\underbrace{\left|\frac{p_{j+1} - 2p_j + p_{j-1}}{p_{j+1}+2p_j + p_{j-1}}\right|}_{\textrm{III}}
\end{equation}
The second and third terms denote the classical Ducros' \cite{ducros1999large} and Jameson's pressure-based \cite{jameson1981numerical} shock sensors, respectively, $\epsilon$ being a small positive value ($\epsilon = 10^{-16}$) to avoid division by zero. Their combination palliates to some of the deficiencies related to the stand-alone application of the Ducros' sensor, which takes into account only the relative magnitudes of dilation and vorticity and may result in unwanted activations of the shock-capturing term in regions where both of these two quantities are small (e.g., in the irrotational flow outside boundary layers or mixing layers).
To correct this deficiency, the constant $\epsilon$ can be parametrized by introducing suitable characteristic velocity and length scales depending on the flow under investigation \cite{pirozzoli2011numerical}. The main drawback of this method resides in the loss of generality, $\epsilon$ being transformed in a configuration-dependent parameter. The introduction of the pressure-based sensor allows one to bypass the activation of Ducros' sensor, strongly reducing the amount of low-order dissipation injected and leaving the acoustic perturbations crossing the domain much less affected.
The first term of equation~\eqref{eq:sensor} takes into account the Ducros' sensor modification of Bhagatwala \& Lele \cite{bhagatwala2009modified}, initially proposed to enhance the selectivity of the artificial bulk viscosity in the Localized Artificial Diffusivity (LAD) technique. In regions of positive dilation $\varphi_j$ is switched off, whereas its value increases slowly with the magnitude of the negative dilation. Moreover, the scaling factor $10 \delta x / c$ has the twofold role of i) normalize the grid-dependent numerical dilation and ii) make it invariant with the mesh-size. The sensor is ${\cal O}(1)$ in high-divergence regions and tends to zero in vortex-dominated regions, allowing the capture of flow discontinuities with minimal damping of the vortical structures inside the flow.

\begin{figure}[!tb]
\centering
 \begin{tikzpicture}
   \node[anchor=south west,inner sep=0] (a) at (0,0) {\includegraphics[width=0.475\textwidth, trim={10 10 50 40}, clip]{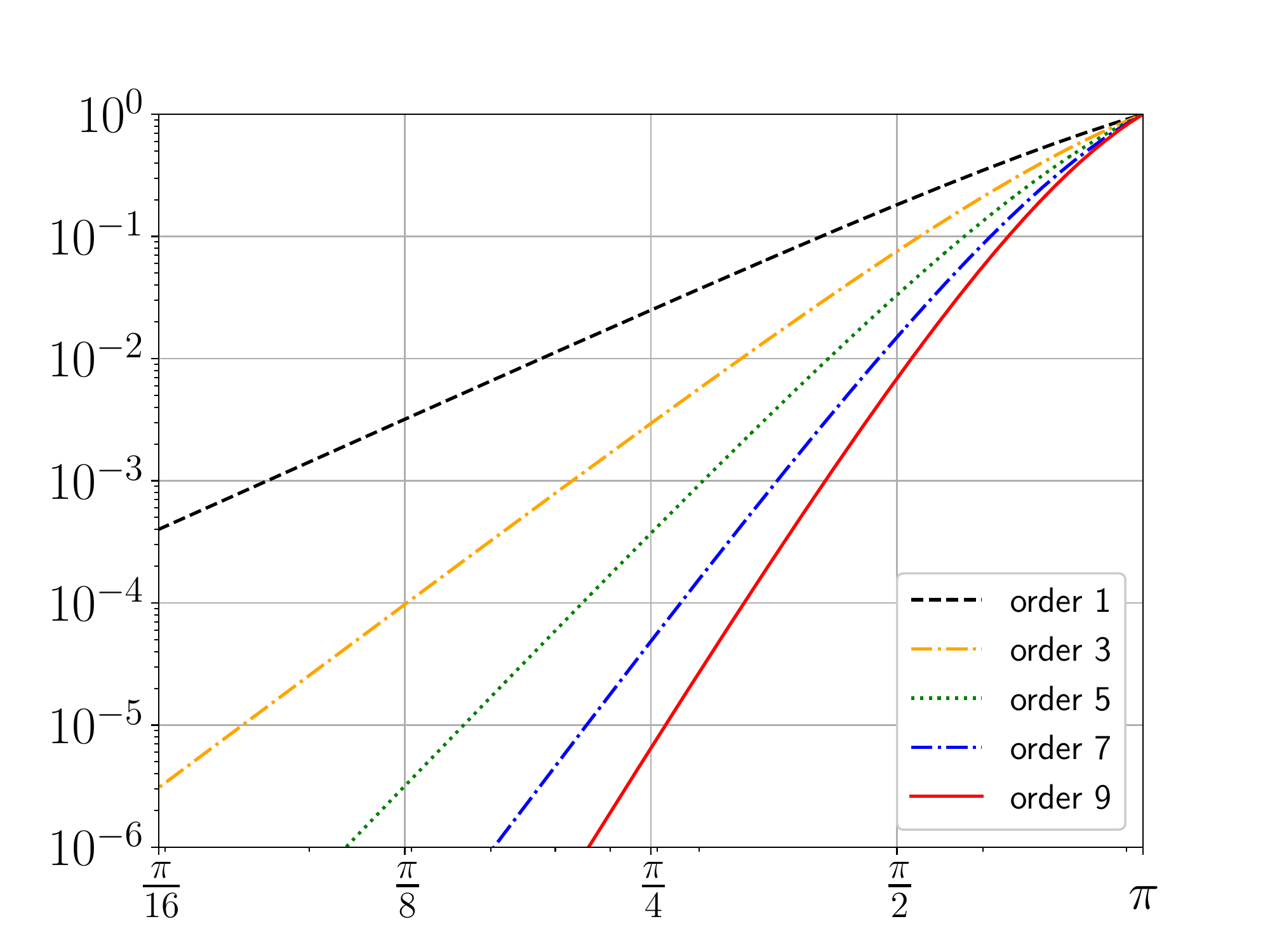}};
   \begin{scope}[x={(a.south east)},y={(a.north west)}]
     \node [align=center] at (-0.01,0.95) {(a)};
     \node [align=center] at (0.55,-0.025) {$k\delta x$};
     \node [align=center,rotate=90] at (-0.015,0.55) {$\Phi(k\delta x)$};
   \end{scope}
 \end{tikzpicture}
 \hspace{-0.36cm}
 \begin{tikzpicture}
   \node[anchor=south west,inner sep=0] (a) at (0,0) {\includegraphics[width=0.475\textwidth, trim={10 10 50 40}, clip]{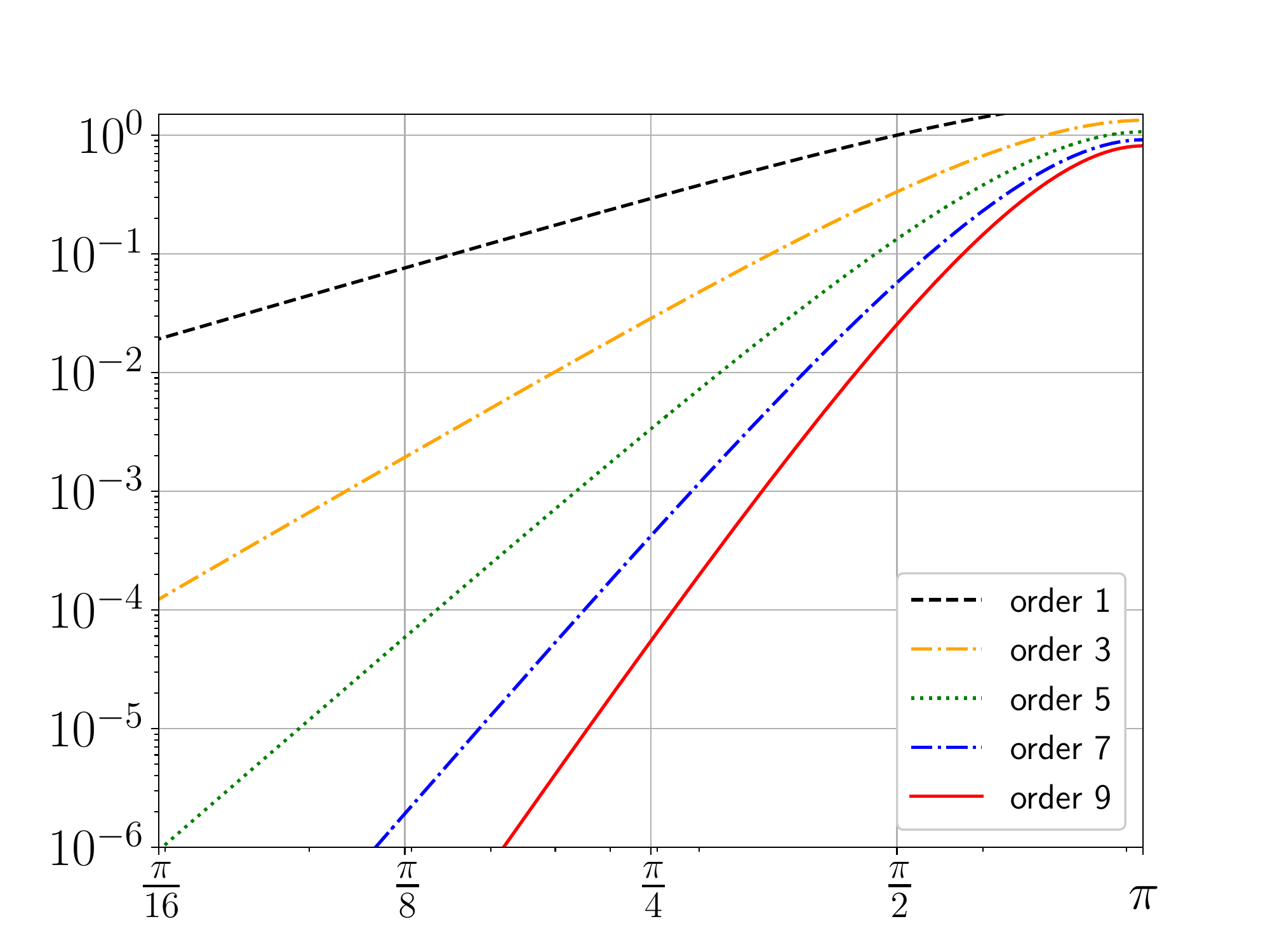}};
   \begin{scope}[x={(a.south east)},y={(a.north west)}]
     \node [align=center] at (-0.01,0.95) {(b)};
     \node [align=center] at (0.55,-0.025) {$k\delta x$};
     \node [align=center,rotate=90] at (-0.015,0.55) {$D(k\delta x)$};
   \end{scope}
 \end{tikzpicture}
 \vspace{-0.75cm}
\caption{Dispersion  and dissipation errors as a function of the reduced wavenumber $k\delta x$ up to of the ninth-order scheme. (a) dispersion error $\Phi$, (b) dissipation error $D$.}\label{fig:errors}
%\caption{Dispersion ($\Phi$) and dissipation ($D$) errors as a function of the reduced wavenumber up to of the ninth-order scheme. (a) $k^*\Delta x$ vs $k\Delta x$, (b) dispersion error, and dissipation function in (c) linear scale and (d) logarithmic scale.}\label{fig:errors}
\end{figure}
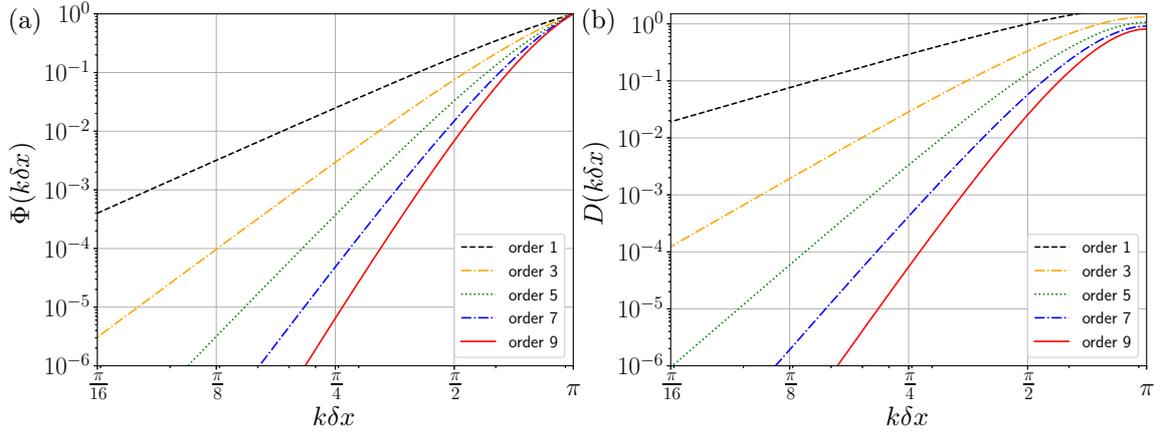

In the following, we mostly focus on the ninth-order accurate scheme of the preceding family. Far from flow discontinuities, such scheme has  low phase and dissipation errors. Its leading truncation error term is of the form $\DPS k_{10}\delta x^9 \frac{\partial^{10} w}{\partial x^{10}}$, i.e. it is consistent with a tenth-order viscosity. Such viscosity term acts differently according to the wavenumber, dissipating scales characterized by reduced wavenumbers of about $0.35 \pi$ or higher (i.e. wavelengths that are discretized with less than 6 mesh points), while leaving larger scales almost unaffected. In figure~\ref{fig:errors} we report the dissipation and phase errors of the ninth-order scheme for the approximation of a linear advection problem, as a function of the reduced wavenumber $k\delta x$. Lower-order schemes of the same family are also reported to illustrate the effect of increasing accuracy. For all schemes, the dispersion errors is exactly the same as for the standard central scheme of order $2(P+2)$. Thanks to its selectiveness in the wavenumber space, the ninth-order dissipation constitutes a suitable implicit subgrid regularization term for LES simulations \cite{gloerfelt2019large}, with the capability of seamlessly converging to DNS in smooth flow regions as the grid is refined. Unless otherwise stated, we set $k_2{=}1$ and $k_{10}{=}\frac{1}{1260}$ for all computations.\\

In Navier--Stokes calculations, the viscous flux derivatives are approximated by fourth-order-accurate central formulae, if not specified differently.
Finally, in all of the following calculations time advancement is carried out by means of an explicit third-order TVD Runge--Kutta scheme~\cite{gottlieb1998total}.
The non-uniformity of the wall-normal mesh spacing is taken into account by a suitable 1-D coordinate transformation. Near the non-periodic boundaries, the finite-difference stencil for the convective terms is progressively reduced down to the fourth order (and, correspondingly, the numerical dissipation term); then, both the convective and viscous fluxes are evaluated from the interior points by using fourth-order backward differences.

\section{Preliminary Validations}
\label{sec:validations}
The ninth-order shock-capturing central scheme under investigation is first applied to selected inviscid test cases, in order to verify its convergence order in smooth flow regions and to assess its shock-capturing capabilities. %===============================================================================
\subsection{Isentropic vortex advection problem}
The accuracy of the discretization scheme in smooth inviscid flow is quantified for the well-known two-dimensional isentropic vortex advection problem \cite{shu1998essentially,davoudzadeh1995accuracy,yee2000entropy}, in which an inviscid vortex is superimposed to an uniform, perfect-gas ($\gamma = 1.4$) air flow. The perturbations in velocity and temperature are given by
\begin{equation}
   \begin{cases}
   \displaystyle \delta u = - \frac{y}{R} \Omega \\
   \displaystyle \delta v = + \frac{x}{R} \Omega \\
   \displaystyle \delta T = - \frac{\gamma -1}{2} \Omega^2
   \end{cases}
   \text{with} \quad \Omega = \beta \exp\left( - \frac{1}{2\sigma} \left[ \left(\frac{x}{R}\right)^2 + \left(\frac{y}{R}\right)^2 \right] \right).
\end{equation}
These allow to define, along with the isentropic relations, the initial flow conditions of the primitive variables as
\begin{equation}
   \begin{cases}
   \tilde{\rho}_0 = \left(1 + \delta T\right)^{\frac{1}{\gamma-1}} \\
   \tilde{u}_{0} = M_\infty \cos \alpha + \delta u \\
   \tilde{v}_{0} = M_\infty \sin \alpha + \delta v \\
   \tilde{p}_{0} = \frac{1}{\gamma}(1 + \delta T)^{\frac{\gamma}{\gamma-1}}
   \end{cases}
\end{equation}
where the subscript $(\bullet)_0$ denotes a quantity given at $t=0$, and the symbol $\widetilde{(\bullet)}$ indicates a nondimensional quantity ($\rho_\infty$, $T_\infty$ and $a_\infty$ being the characteristic density, temperature and velocity, respectively).
In the classical case \cite{shu1998essentially}, one has $R = \sigma = 1$, $M_\infty = \sqrt{\frac{2}{\gamma}}$ and $\alpha = \SI{45}{\degree}$. Periodic conditions are applied at the boundaries. The length of the computational domain has been increased from $[L_x, L_y] = [-5 ,5]$ to $[-10, 10]$ in order to reduce the influence of the small artificial shear layers generated near the boundaries by the non-zero velocity perturbations, which can pollute the results when considering smaller domains \cite{spiegel2015survey}.

The error with respect to the exact solution (pure advection of the initial vortex) is measured as
\begin{equation}
   \varepsilon_{\tau,h} = L_2(\Psi^h_{(i,j)}) = \sqrt{ \frac{1}{h} \sum_{i,j} \left( \Psi^h_{(i,j)} - \Psi^e_{(i,j)} \right)^2 },
\end{equation}
with $\Psi^{\tau,h}_{(i,j)}$ and $\Psi^e_{(i,j)}$ the computed and exact value of a generic flow variable at grid point $(i,j)$, and $h = L_x/(N-1)$ the spatial grid size, $N=N_x=N_y$ being the number of grid points.
For an unsteady problem, the numerical error may be written as $\varepsilon_{\tau, h} = C_\tau\tau^p + C_h h^q$, where $C_\tau$ and $C_h$ are some constants, $\tau$ the time-step, $h$ the grid size, and $p$ and $q$ the order of the temporal integration and spatial discretization schemes, respectively.
%The numerical error should tend to zero (or, similarly, the numerical solution tends to the exact solution) as $\tau \rightarrow 0$ and $h \rightarrow 0$.
The ratio of error decay between numerical solutions using time steps $\tau$ and $m\tau$ and grid sizes $h$ and $nh$ writes:
\begin{equation}
   r_e = \frac{\varepsilon_{m\tau, nh}}{\varepsilon_{\tau,h}} = \frac{C_\tau (m\tau)^p + C_h (nh)^q}{C_\tau \tau^p + C_h h^q} = \frac{\delta m^p + n^q}{\delta + 1} \quad \text{with} \quad
   \delta = \frac{C_\tau\tau^p}{C_h h^q}.
\end{equation}
Assuming $m, n > 1$, the overall accuracy ranges between $q$ for small $\delta$ (meaning that the temporal error is negligible with respect to spatial error) and $p$ for large $\delta$.

%The present section aims at verifying quantitatively the numerical accuracy of the scheme presented in section~\ref{sec:numeth} for configurations without shocks and/or strong gradients.
%At any point $(i,j,k)$ of a given computational grid, one can define $\Psi^{\tau,h}_{(i,j,k)}$ and $\Psi^e_{(i,j,k)}$ being respectively the discrete value of the flow variable of interest $\Psi$ (e.g., density, velocity components, temperature, pressure, ..) and the corresponding value of the exact solution. A metric of the computational error, denoted $\varepsilon_{\tau,h}$, may be defined by considering the $L_2$-norm of the quantity $\Psi$:
%\begin{equation}
%   \varepsilon_{\tau,h} = L_2(\Psi^h_{(i,j,k)}) = \sqrt{ \frac{1}{h} \sum_{i,j,k} \left( \Psi^h_{(i,j,k)} - \Psi^e_{(i,j,k)} \right)^2 },
%\end{equation}
%where $h = 1/N_\text{DOF}$, $N_\text{DOF}$ denoting the total number of grid points.\\
%
%
%

\begin{figure}[!tb]
\centering
 \begin{tikzpicture}
   \node[anchor=south west,inner sep=0] (a) at (0,0) {\includegraphics[width=0.75\textwidth]{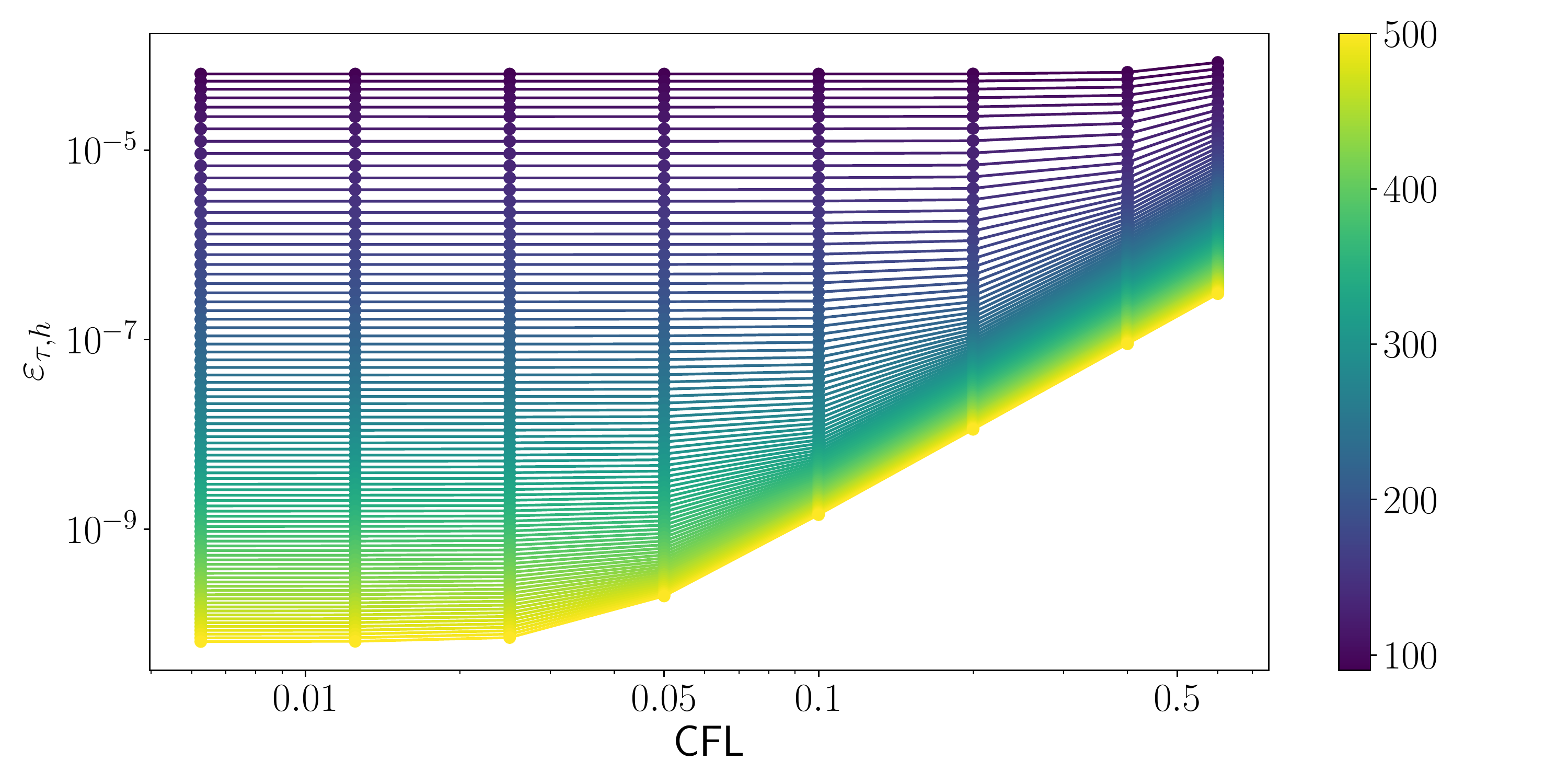}};
   \begin{scope}[x={(a.south east)},y={(a.north west)}]
     \node [align=center] at (0.025,0.95) {(a)};
   \end{scope}
 \end{tikzpicture}

 \begin{tikzpicture}
   \node[anchor=south west,inner sep=0] (a) at (0,0) {\includegraphics[width=0.49\textwidth]{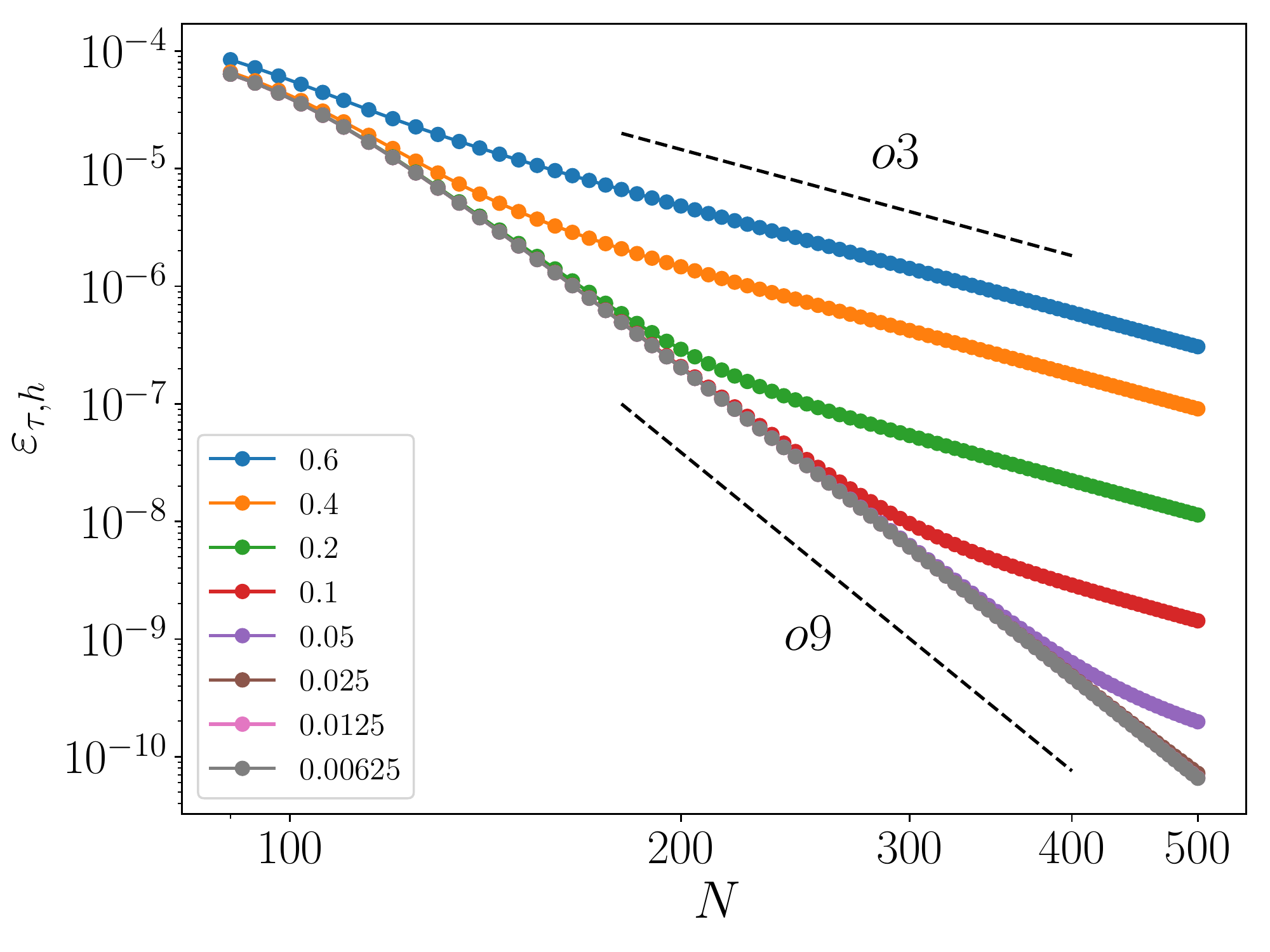}};
   \begin{scope}[x={(a.south east)},y={(a.north west)}]
     \node [align=center] at (0.025,0.95) {(b)};
   \end{scope}
 \end{tikzpicture}
 \hspace{-0.25cm}
 \begin{tikzpicture}
   \node[anchor=south west,inner sep=0] (a) at (0,0) {\includegraphics[width=0.49\textwidth]{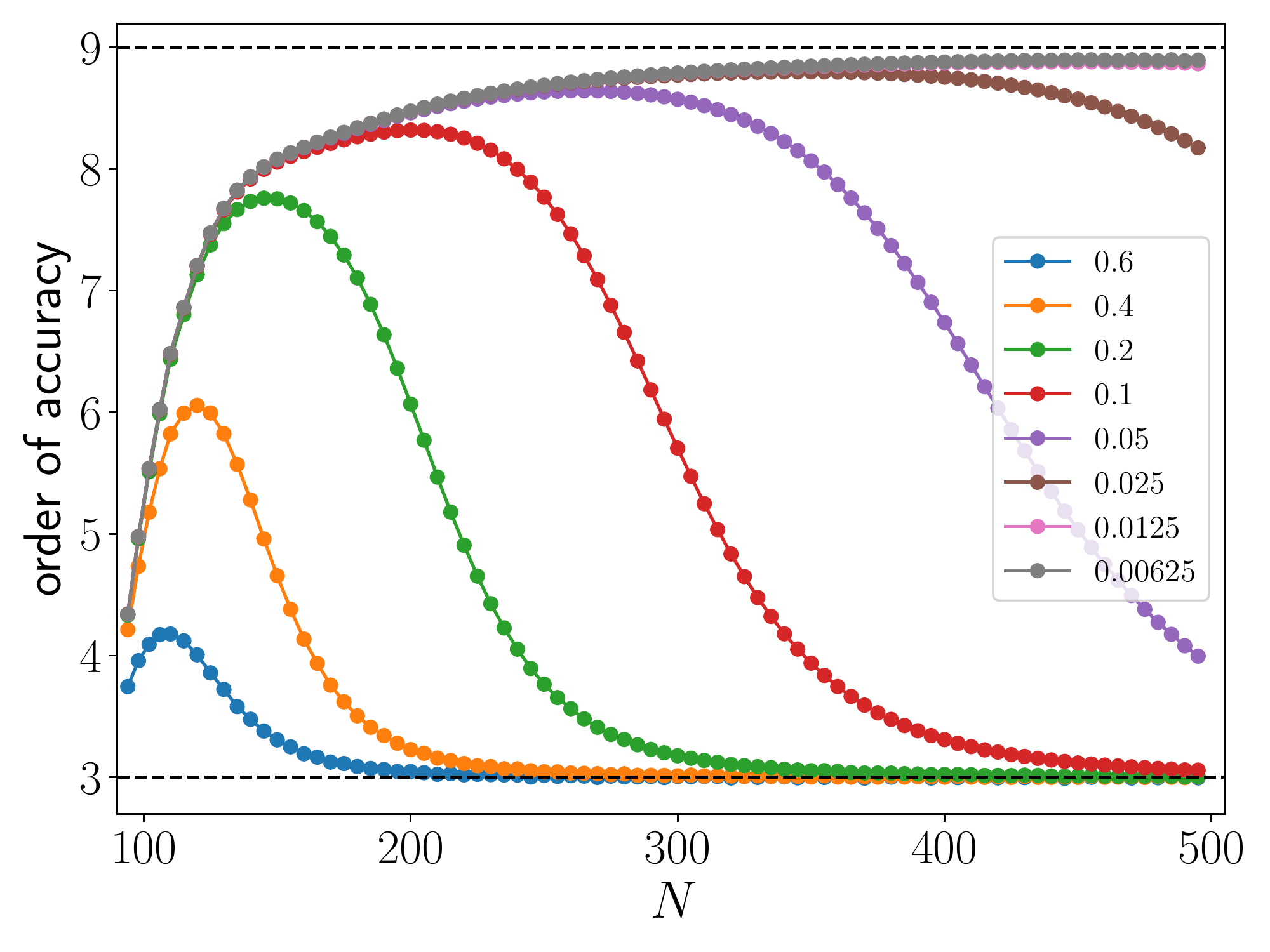}};
   \begin{scope}[x={(a.south east)},y={(a.north west)}]
     \node [align=center] at (0.025,0.95) {(c)};
   \end{scope}
 \end{tikzpicture}
\caption{Analysis of the order of convergence of the error norm for the isentropic vortex convection configuration. (a): error of the $L_2$-norm as a function of the CFL number for computational grids ranging from $100^2$ to $500^2$; (b): error of the $L_2$-norm as a function of the number of grid points for CFL numbers ranging from $0.6$ to $0.00625$; (c): overall spatio-temporal order of accuracy as a function of the number of grid points for CFL numbers ranging from $0.6$ to $0.00625$.}\label{fig:isentropic_vortex_rk3}
\end{figure}
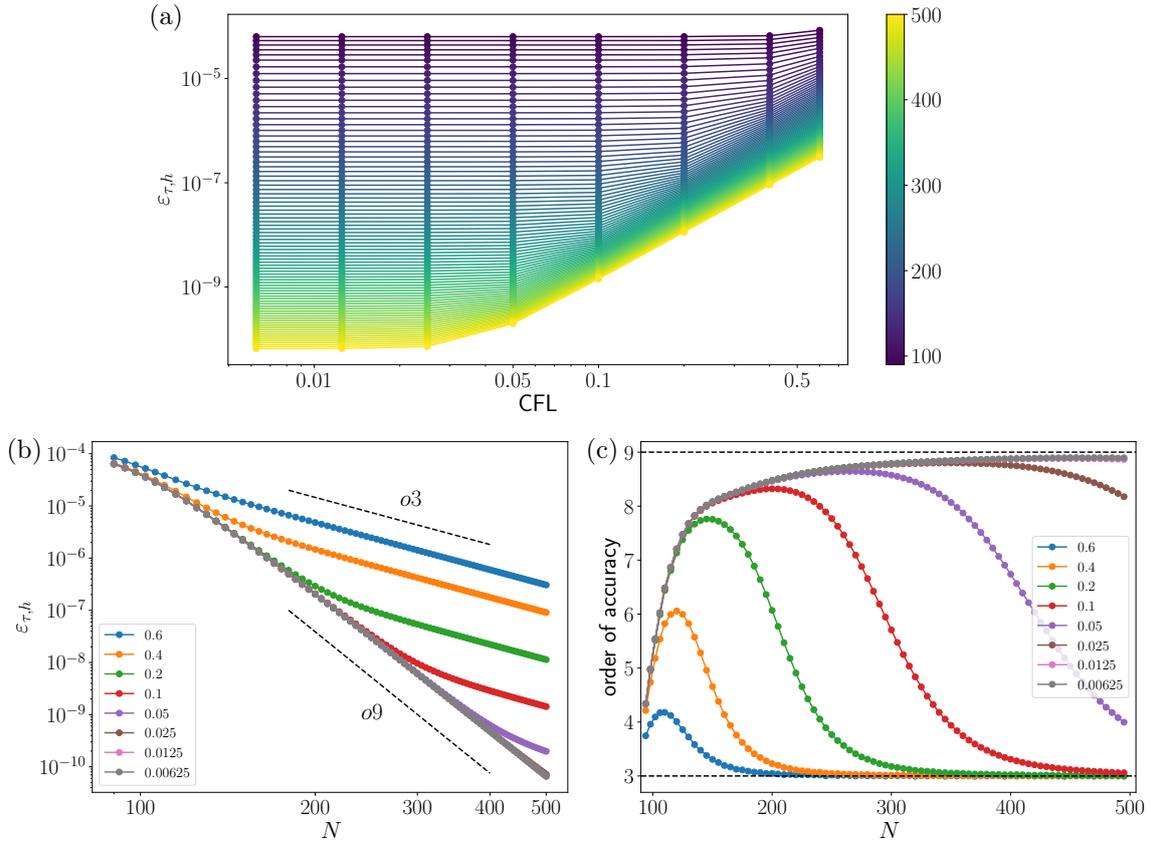

Several runs are carried out by varying the number of grid points (between $100^2$ and $500^2$) and the CFL number (from 0.6 to 0.00625); the errors measured after one cycle (i.e. when the vortex returns to the initial position for the first time) are shown in figure~\ref{fig:isentropic_vortex_rk3}, where each symbol denotes a run.
Figure~\ref{fig:isentropic_vortex_rk3}a displays the value of the numerical error as a function of the CFL number for several grids (colour plot). In order to recover the formal order of accuracy of the spatial discretization scheme for a given grid, the CFL number should be small enough not to affect $\varepsilon_{\tau,h}$; i.e., $\varepsilon_{\tau,h}$ should read a plateau for sufficiently small values of the CFL numbers. This is clearly visible in figure~\ref{fig:isentropic_vortex_rk3}a, which also highlights that even smaller CFL numbers should be considered for grids finer than $500^2$. In Figure~\ref{fig:isentropic_vortex_rk3}b we report the error as a function of $N$ for various CFL. The order of accuracy of the numerical solution varies as expected between 3 to 9 (formal orders of the time integration and spatial discretization numerical schemes, respectively) as the CFL number gets smaller.
For CFL numbers of 0.2 or smaller, a more or less extended region with ninth-order slope is observed; the slope tends to decrease for finer grids, due to the larger relative importance of the temporal error.
The local slopes are shown in figure~\ref{fig:isentropic_vortex_rk3}c versus the number of grid points. The presence of a maximum can be explained by considering the temporal-to-spatial error ratio, $\delta$. In order to keep a constant $\delta$ when refining the grid, the timestep should be decreased as $\tau \propto h^{q/p} = h^{9/3} = h^3$; whereas a constant CFL imposes $\tau \propto h$ (i.e., a varying $\delta$).
At $\text{CFL} = \num{1.25e-2}$ and \num{6.25e-3} the formal order of the spatial discretization scheme is recovered over the whole range of considered grid resolutions; this confirms previous studies suggesting the use of $\text{CFL} \lesssim 0.01$ \cite{direnzo2020htr}.

%===============================================================================
\subsection{Ideal-gas shock tube problems}
The next test case is a classical benchmark for the assessment of shock-capturing capabilities. Specifically, we consider the well-known Sod \cite{sod1978survey} and Lax \cite{lax1954weak} 1D shock tube problems, corresponding to the following Riemann problems:
\begin{equation}
   (\rho, u, p)_\text{SOD} =
   \begin{cases}
   (1,0,1) & x < 0\\
   (0.125,0,0.1) & x \geq 0
   \end{cases} ; \qquad
   (\rho, u, p)_\text{LAX} =
   \begin{cases}
   (0.445,0.698,3.528) & x < 0\\
   (0.5,0,0.571) & x \geq 0
   \end{cases}
\end{equation}
\begin{figure}[!tb]
\centering
\includegraphics[width=0.50\textwidth]{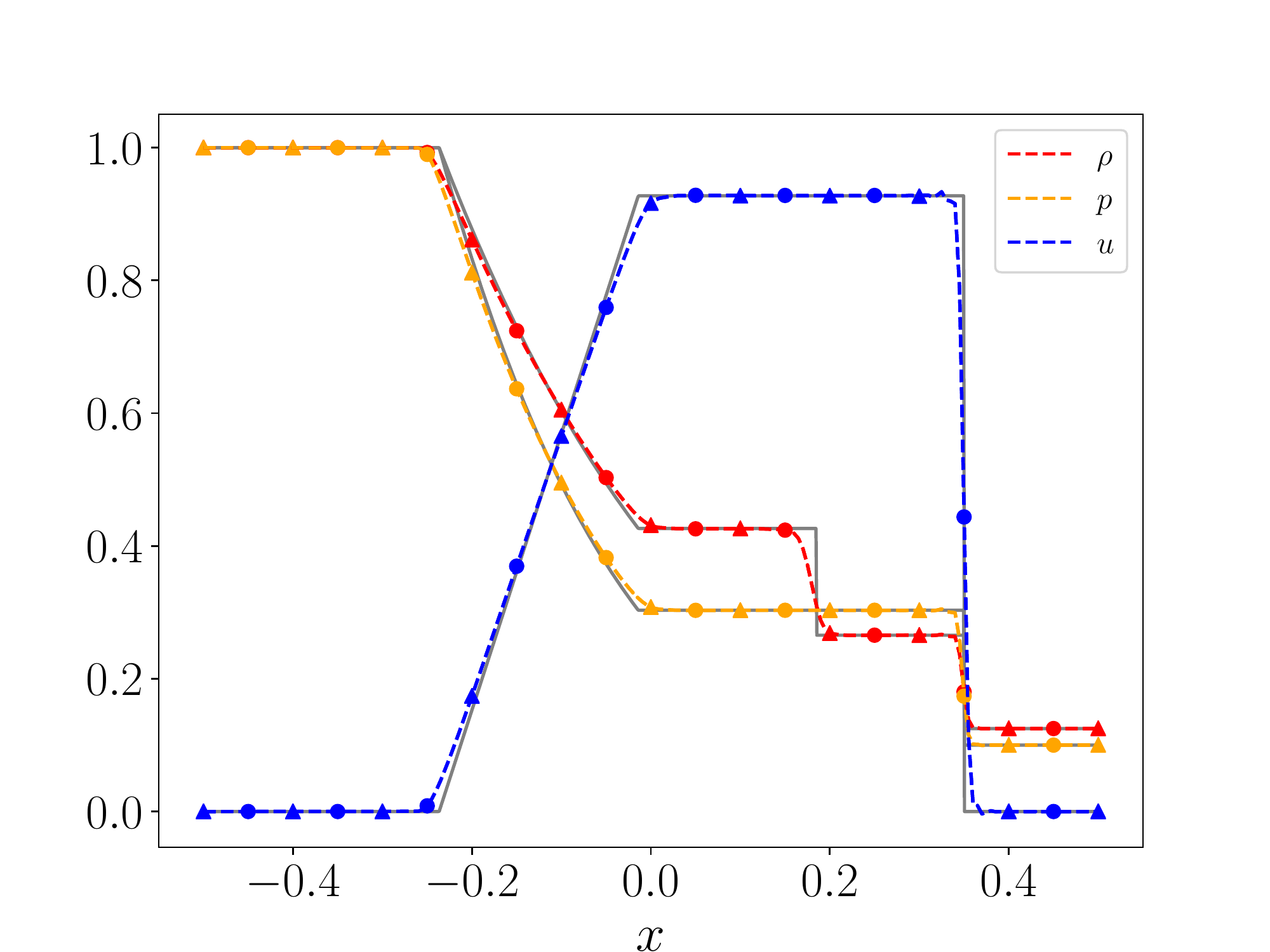}\hspace{-0.5cm}
\includegraphics[width=0.50\textwidth]{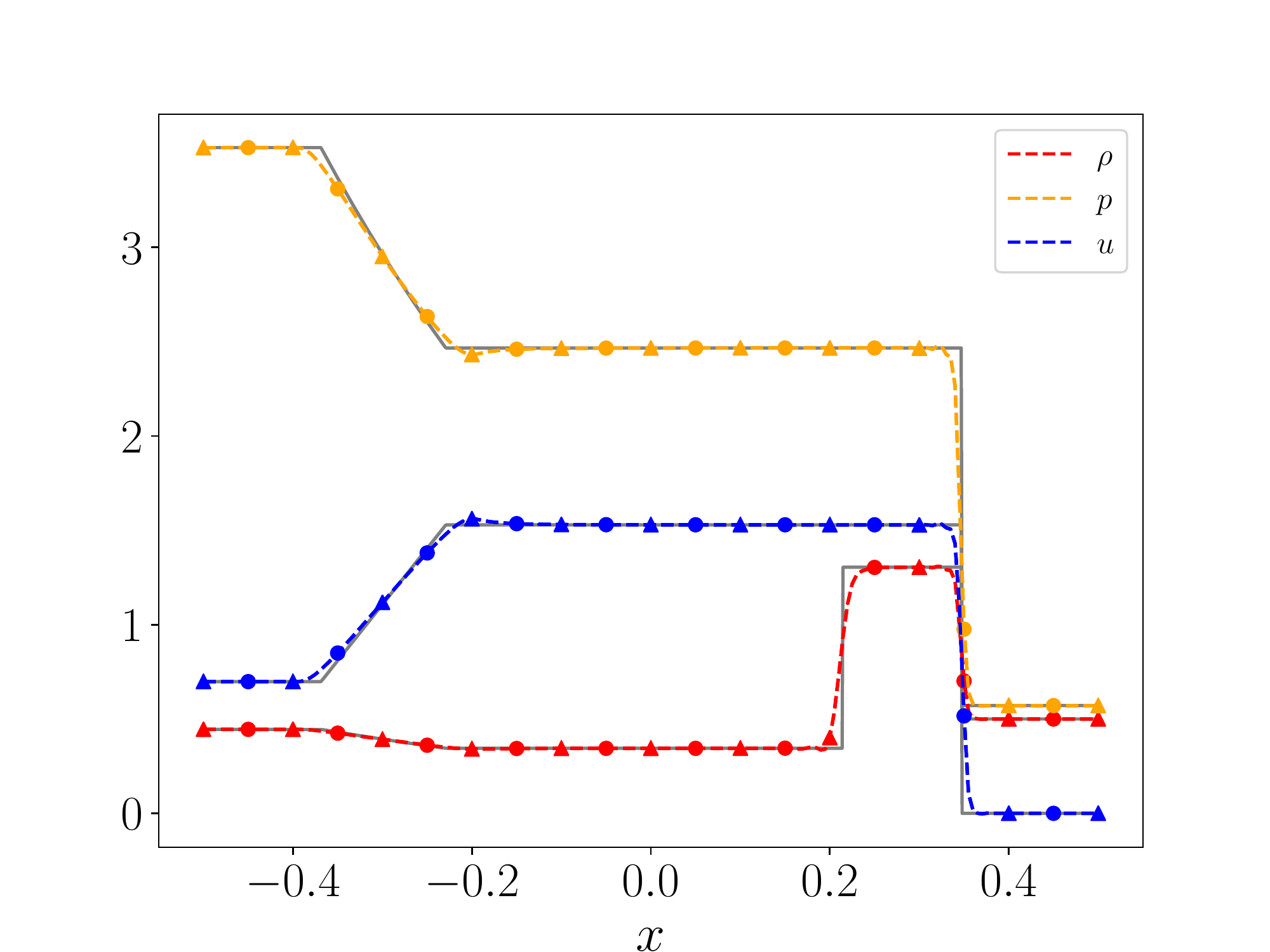}
\caption{Reference and numerical solutions for Sod's (left) and Lax's (right) shock tubes. Triangles: single species, circles: N$_2$-O$_2$ mixture. The gray solid line denotes the reference solution.}\label{fig:shock_tube_sodlax}
\end{figure}
Numerical results are compared to the exact solutions at the nondimensional time $t^*=0.2$ ($\Delta t^* {=} \num{5e-4}$) and $t^*=0.13$ ($\Delta t^* {=} \num{e-3}$), respectively.
For these test cases, the artificial viscosity coefficients are chosen equal to $k_2=2$, $k_{10}=1/630$. The numerical domain is discretized with 200 evenly-spaced grid points; air is considered either as a single-species, calorically-perfect gas ($\gamma=1.4$) or as a two-species mixture of Nitrogen ($Y_{\text{N}_2}=79\%$) and Oxygen ($Y_{\text{O}_2}=21\%$).
Figure~\ref{fig:shock_tube_sodlax} shows the results for density, velocity and pressure, whose profiles are compared to the corresponding exact solutions. A good agreement is shown for both cases and the single-species and multi-species numerical solutions are perfectly superposed. The slight smearing of the numerical solution across the contact discontinuity can be attributed to the high-order dissipation term, the shock-capturing component being switched off in that region due to the constant value of the pressure. Note that a similar behavior is observed for other high-order schemes \cite{fu2017targeted}.

%===============================================================================
\subsection{Shu-Osher problem}
\begin{figure}[!tb]
\centering
\includegraphics[width=\textwidth]{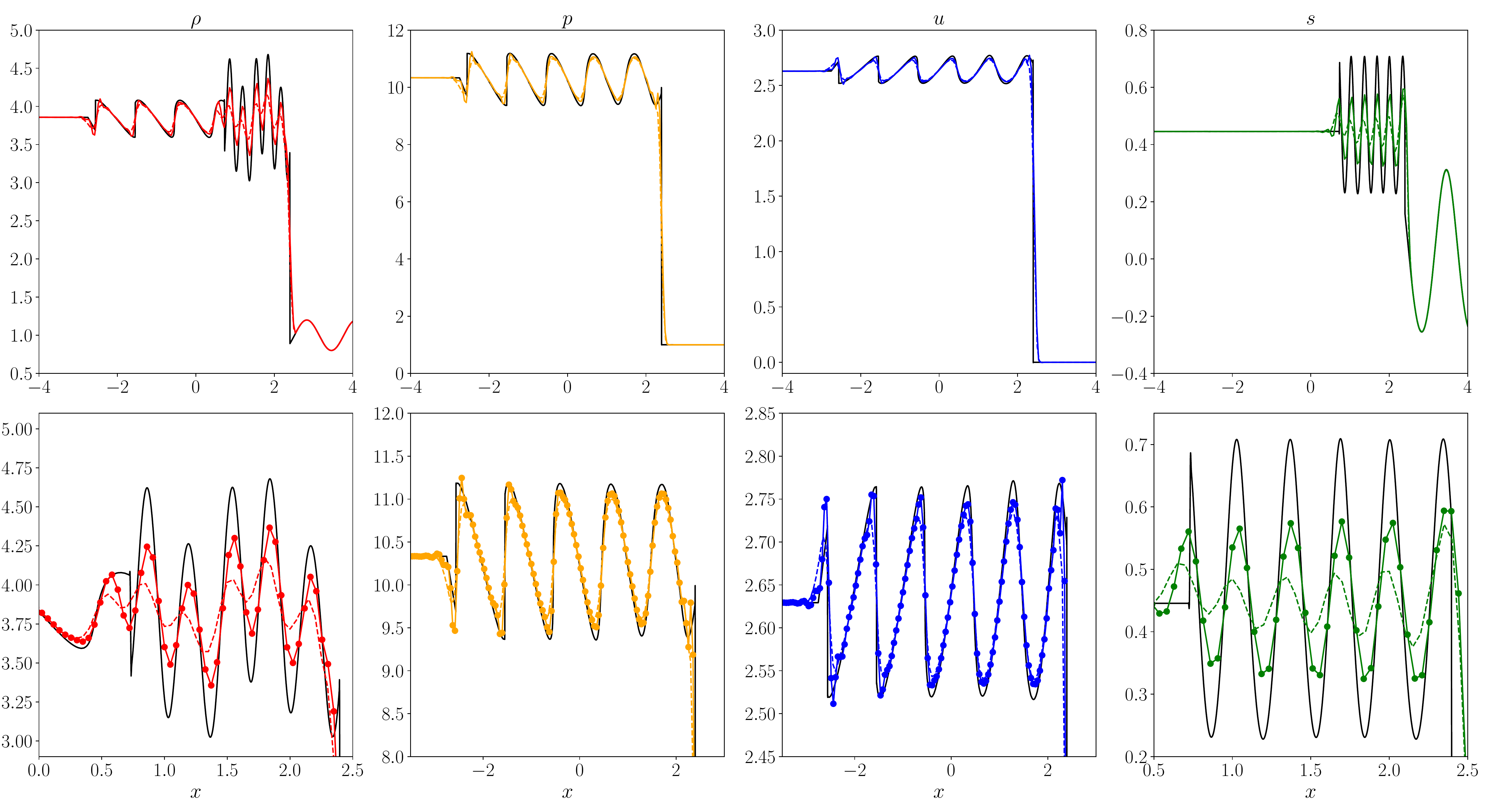}
\caption{Numerical solution for the Shu-Osher problem at $t^*=1.8$. Top: profiles of density, pressure, velocity and entropy; bottom: zoom on the post-shock regions. In each subfigure, the black solid lines denote the reference solution, and the colored solid and dashed lines the present solution, respectively with and without Bhagatwala \& Lele's correction.}
\label{fig:shu_osher}
\end{figure}

The Shu-Osher problem \cite{shu1989efficient} consists of a $M{=}3$ shock propagating in a perturbed density field and allows to evaluate the behavior of the numerical scheme for a simplified shock-turbulence interaction configuration. The extent of the computational domain is $[-5, 5]$ and the initial conditions are
\begin{equation}
   (\rho, u, p)_\text{SHU} =
   \begin{cases}
   (3.857143, 2.629369, 10.33333) & x<-4\\
   (1 + 0.2\sin(5x), 0, 1) & x \geq -4
   \end{cases}
\end{equation}
The 1D Euler equations are solved on a uniform mesh with $N=200$ and a reference solution is computed with the same numerical scheme on a mesh with $N=2000$. Figure~\ref{fig:shu_osher} shows results for density, pressure, velocity and entropy at the final time $t^*=1.8$. The profiles of pressure and velocity are in good agreement with the reference solution, with some oscillations registered near the shock (which is captured reasonably well) that remain bounded to small values throughout the simulation. On the contrary, the density and entropy profiles exhibit a stronger damping after the shock train passage. This is partly due to the use of a scalar artificial viscosity, which introduces the same amount of dissipation for all characteristic fields, whether they be of acoustic or entropic nature. Present results are in agreement with those of other classical numerical schemes \cite{johnsen2010assessment}.
Of note, in this 1D case the Ducros sensor is inactive, and the introduction of the Bhagatwala \& Lele correction to the shock-capturing term strongly enhances the entropy waves resolution (contrary to the shock tube cases, where no appreciable differences were found), mitigating the spurious activation of Jameson's pressure-based sensor.

%===============================================================================
\subsection{Real-gas shock tube problem}
\begin{figure}[!tb]
\centering
    \begin{tikzpicture}
      \node[anchor=south west,inner sep=0] (a) at (0,0) {\includegraphics[width=0.48\textwidth]
      {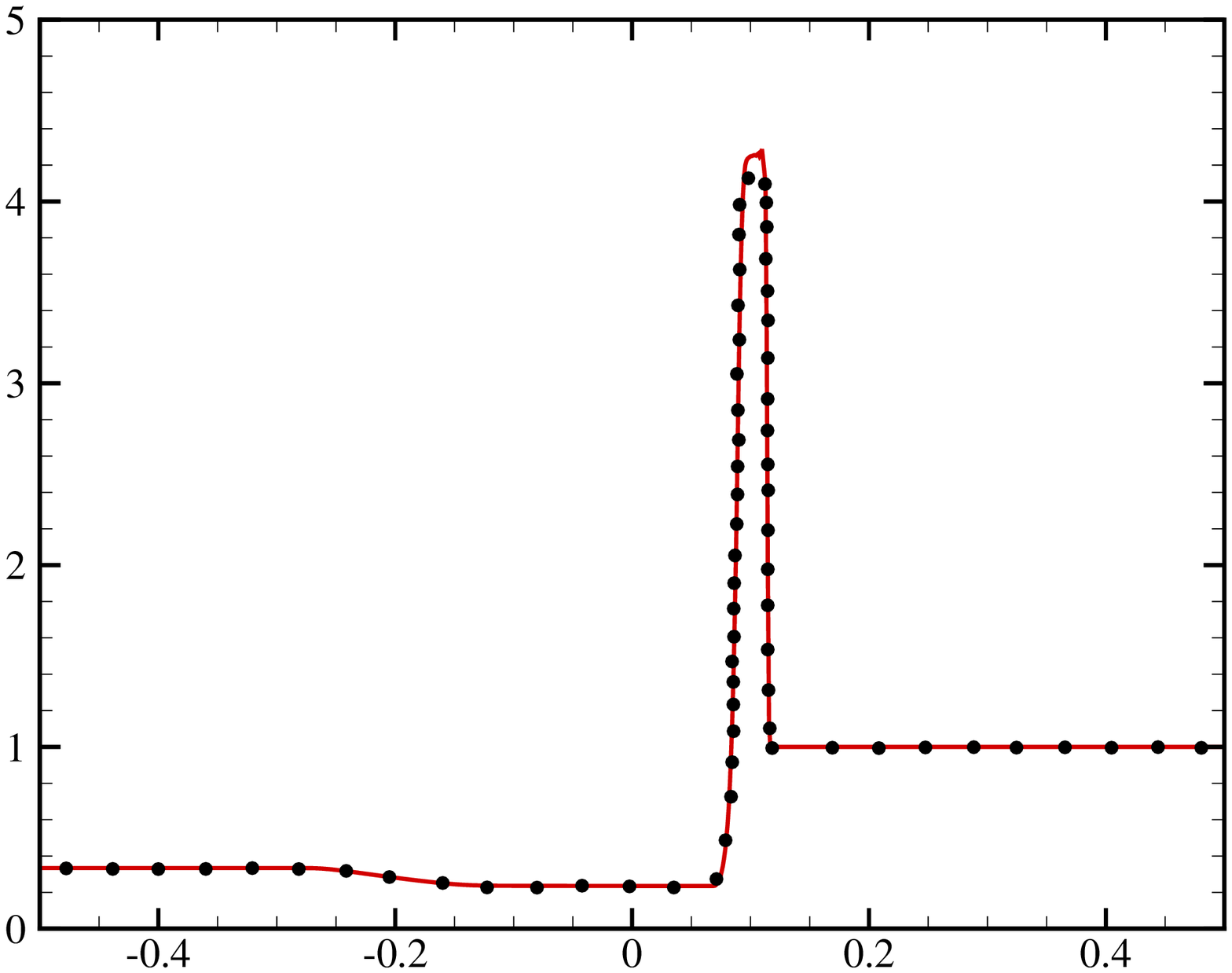}};
      \begin{scope}[x={(a.south east)},y={(a.north west)}]
        \node [align=center] at (0.55,0.025) {$x$};
        \node [align=center,rotate=90] at (0.05,0.55) {$\rho/\rho_R$};
      \end{scope}
    \end{tikzpicture}
  \hspace{-0.3cm}
    \begin{tikzpicture}
      \node[anchor=south west,inner sep=0] (a) at (0,0) {\includegraphics[width=0.48\textwidth]
      {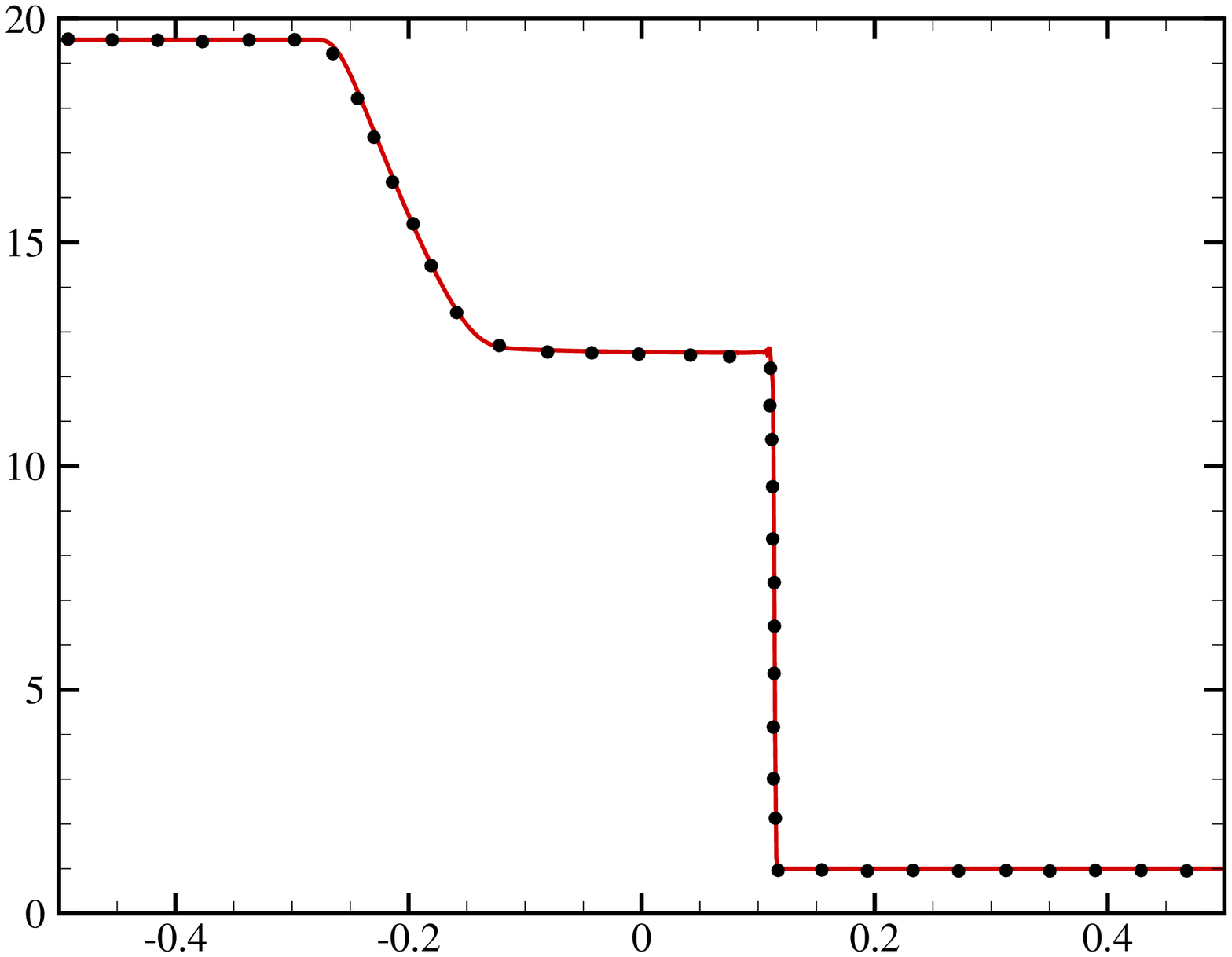}};
      \begin{scope}[x={(a.south east)},y={(a.north west)}]
        \node [align=center] at (0.55,0.025) {$x$};
        \node [align=center,rotate=90] at (0.05,0.55) {$p/p_R$};
      \end{scope}
    \end{tikzpicture}

    \begin{tikzpicture}
      \node[anchor=south west,inner sep=0] (a) at (0,0) {\includegraphics[width=0.48\textwidth]
      {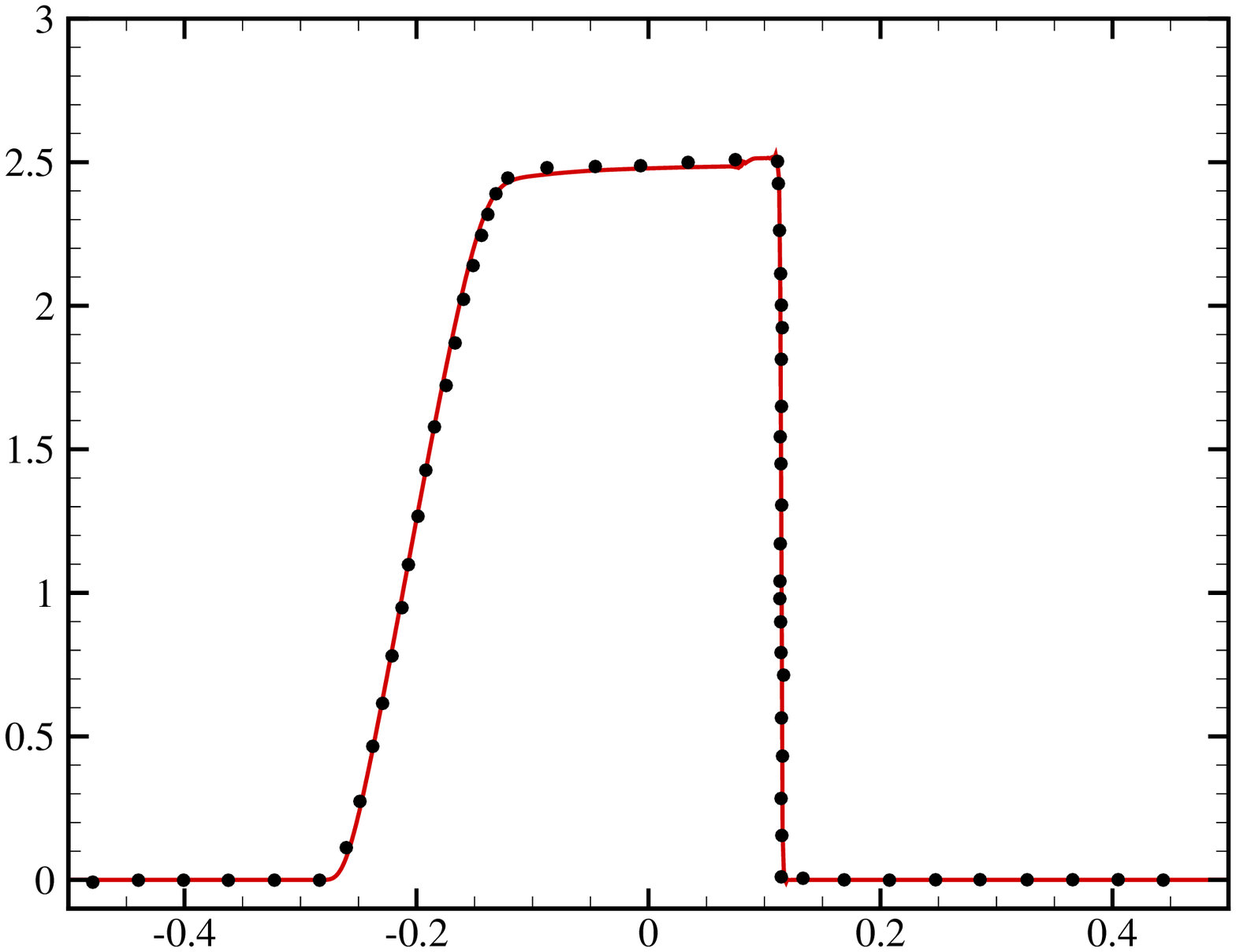}};
      \begin{scope}[x={(a.south east)},y={(a.north west)}]
        \node [align=center] at (0.55,0.025) {$x$};
        \node [align=center,rotate=90] at (0.05,0.55) {$u/c_R$};
      \end{scope}
    \end{tikzpicture}
  \hspace{-0.3cm}
    \begin{tikzpicture}
      \node[anchor=south west,inner sep=0] (a) at (0,0) {\includegraphics[width=0.48\textwidth]
      {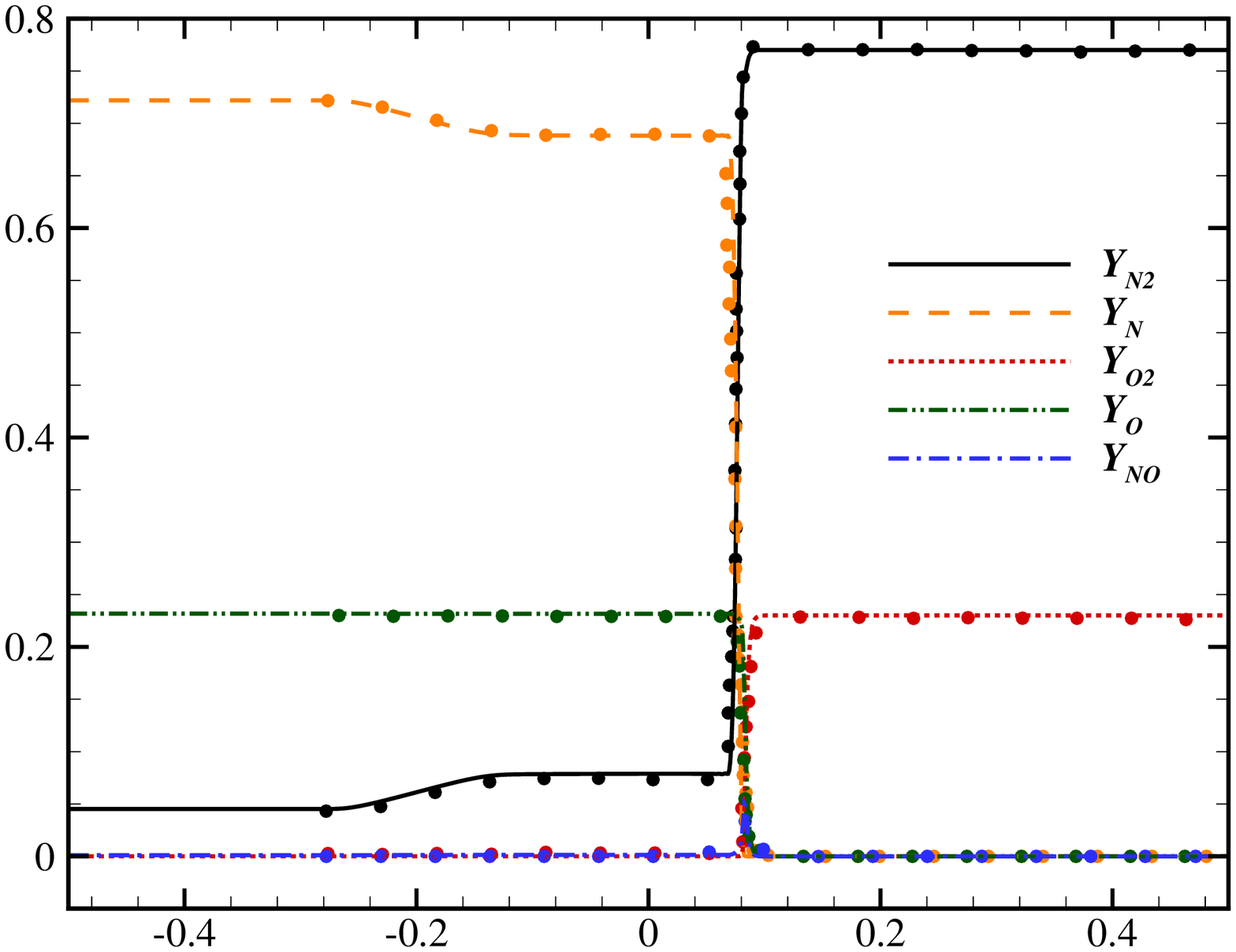}};
      \begin{scope}[x={(a.south east)},y={(a.north west)}]
        \node [align=center] at (0.55,0.025) {$x$};
        \node [align=center,rotate=90] at (0.05,0.55) {$Y_n$};
      \end{scope}
    \end{tikzpicture}
\vspace{-0.15cm}
\caption{Profiles of density, pressure, velocity and mass fractions for the reacting, real gas shock tube case. Lines: current simulation; symbols: reference solution \cite{grossman1990flux}.}\label{fig:grossman}
\end{figure}

The last inviscid test case consists of a multi-species, high-temperature shock tube designed to study thermochemical effects \cite{grossman1990flux}. At $t=0$, the left (L) and right (R) initial conditions are the following:
\begin{alignat}{3}
   &P_L = \SI{1.95256e5}{Pa}, \qquad && u_L=\SI{0}{m/s}, \qquad && T_L=\SI{9000}{K} \\
   &P_R = \SI{e4}{Pa},        \qquad && u_R=\SI{0}{m/s}, \qquad && T_R=\SI{300}{K}
\end{alignat}
Chemical nonequilibrium is modelled by means of Park's 5-species model (N$_2$, O$_2$, N, O, NO); the initial values for the species mass fractions correspond to the mixture equilibrium composition at the given pressure and temperature for the right and left states, respectively.
Due to the stiffness of the chemical source terms, the solution is advanced in time using a CFL number of 0.02 as suggested in \cite{grossman1990flux} for explicit Runge-Kutta time integrations. The simulation is stopped when the shock-wave reaches the location $x=\SI{0.110}{m}$. Given the severe conditions developed by flow in the present problem, the Bhagatwala \& Lele's modification of the shock sensor was turned off for better numerical robustness. \\
Results at the final time are shown in figure~\ref{fig:grossman}. The numerical scheme is able to capture correctly the rarefaction wave, the contact discontinuity and the shock; moreover, the distributions of the species mass fractions agree very well with data from \cite{grossman1990flux}.

%===============================================================================
\section{Applications to multiscale turbulent flows}
\label{sec:results}
In this section we analyze the performance of the shock-capturing scheme for viscous compressible flows with shocks and fine-detail vortical structures.
The applications range from a two-dimensional under-expanded jet flow to a hypersonic turbulent boundary layer in chemical nonequilibrium.

%===============================================================================
\subsection{Two-dimensional underexpanded jet}
\begin{figure}[tb]
\includegraphics[width=\textwidth, trim={5 5 5 5}, clip]{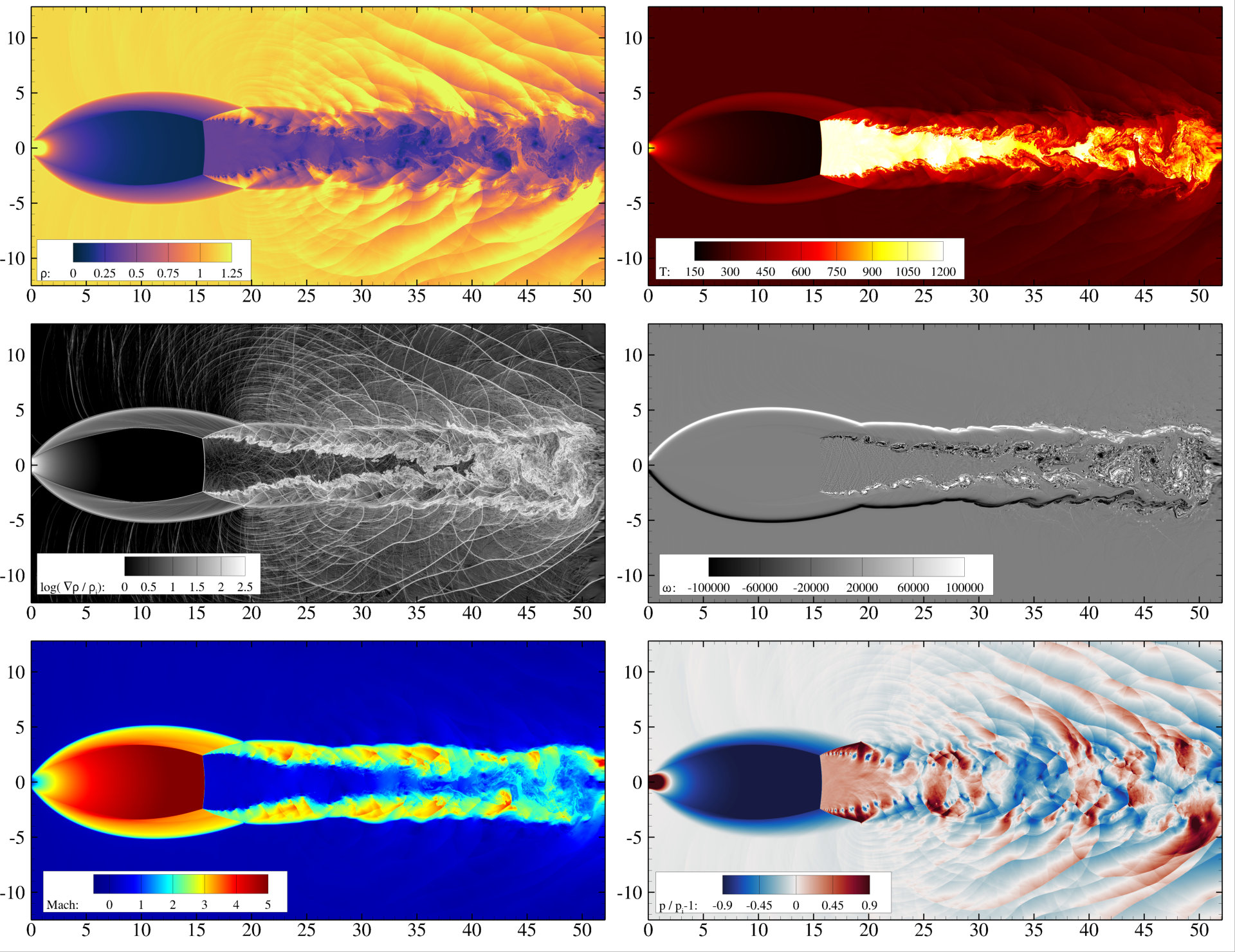}
\caption{Instantaneous snapshot of an N$_2$-O$_2$ inert underexpanded jet. From top to bottom, left to right: isocontours of density $\rho$, temperature $T$, numerical Schlieren $\log(\nabla \rho/\rho_j)$, vorticity $\omega$, Mach number $M$ and pressure fluctuations $p/p_j - 1$.}
\label{fig:jet_insta}
\end{figure}

A N$_2$-O$_2$ inert, highly underexpanded jet has been first considered to test the suitability of the numerical scheme to deal with strong discontinuities and vortical layers in multi-species flows. This configuration represents the starting point for the study of reactive jets and has been widely analyzed in the past years, both experimentally and numerically. In addition, it has been shown that 2D simulations are able to capture reasonably well some detailed features of the problem, such as the location and dimension of the Riemann wave, the scale of the jet and the timewise averages of the thermodynamic variables.\\
The present setup is similar to the one reported in Mart\'inez Ferrer \emph{et al.} \cite{ferrer2014detailed}. Specifically, the pressure $P_0$ and temperature $T_0$ of the mixture in the injection plane are set to 15 atm and 1000 K, whereas the ambient values are 1 atm and 300 K, respectively, resulting in a nozzle pressure ratio (NPR) of 15. The inflow jet Mach number is equal to 1 and a slow coflow at $M=0.05$ is imposed consistently with previous studies \cite{ferrer2014detailed,su2020numerical}. The height of the injector exit is equal to $D = \SI{3}{cm}$ and the extent of the computational domain is $L_x \times L_y = 50 D \times 25 D$.
At the inflow, the jet velocity is prescribed by means of an hyperbolic-tangent profile (see ``Profile 2'' of Michalke \cite{michalke1984survey}):
\begin{equation}
   \frac{u(r)}{U} = \frac{1}{2} \left\{ 1 + \tanh \left[0.25\frac{R}{\theta} \left( \frac{R}{r} - \frac{r}{R} \right)\right]\right\}
\end{equation}
where $U$ is the inflow centerline velocity, $R$ the inflow half-height, $\theta$ the initial momentum thickness of the shear layer and $r$ the local distance from the jet axis. The ratio $R/\theta$ is an important parameter influencing the jet stability, which is mainly related to the introduction of vorticity in the jet shear layer. In our study, we consider $R/\theta = 12.5$. At the remaining outflow boundaries, characteristic conditions are imposed, along with sponge regions and grid stretching to avoid spurious reflections.

\begin{figure}[!tb]
\begin{tikzpicture}
   \node[anchor=south west,inner sep=0] (a) at (0,0) {\includegraphics[width=0.48\columnwidth]{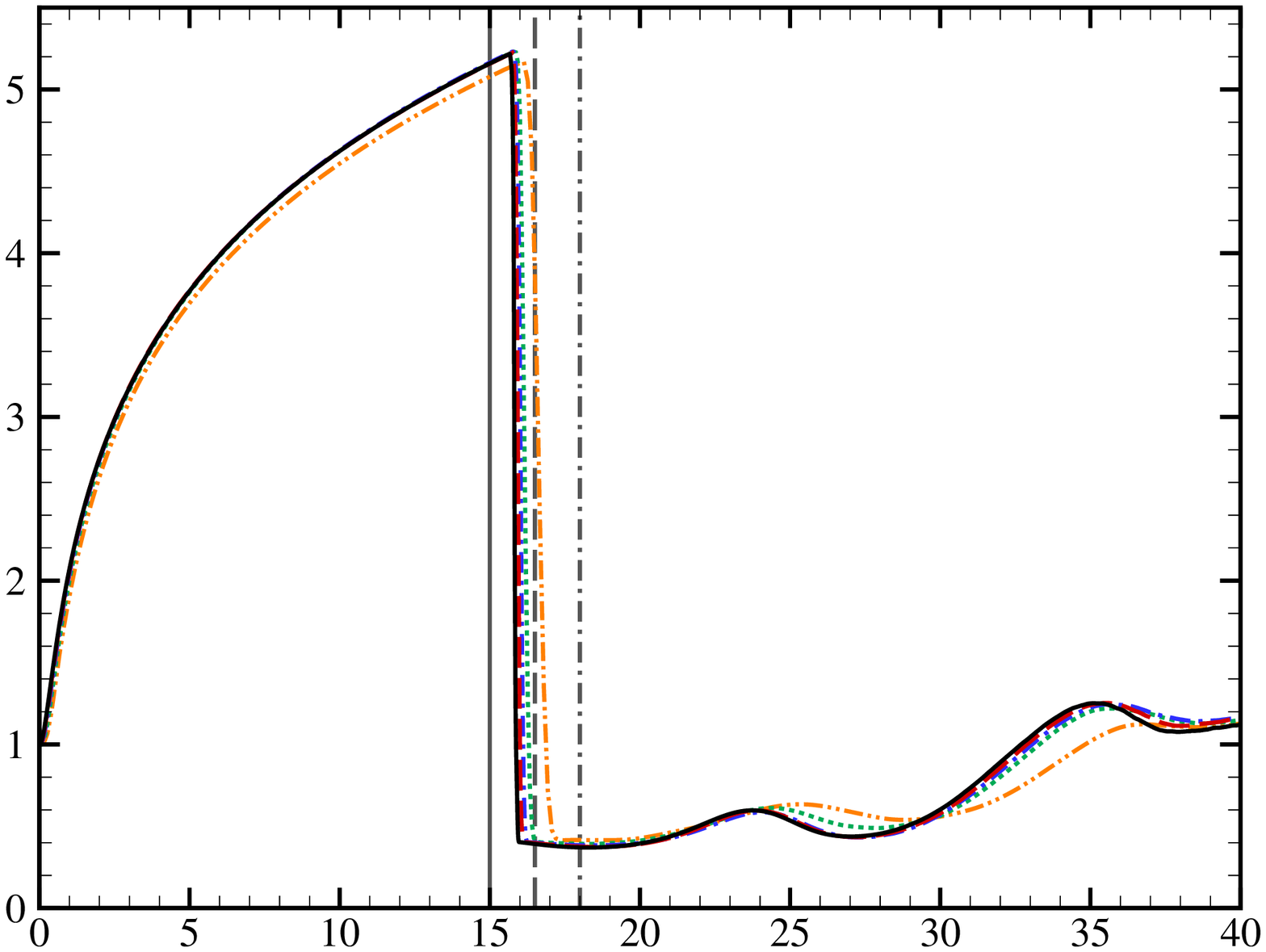}};
   \begin{scope}[x={(a.south east)},y={(a.north west)}]
     \node [align=center] at (0.55,0.00) {$x/D$};
     \node [align=center,rotate=90] at (0.02,0.55)  {$M$};
     \node [align=center] at (0.01,0.95)  {(a)};
   \end{scope}
\end{tikzpicture}
\begin{tikzpicture}
   \node[anchor=south west,inner sep=0] (a) at (0,0) {\includegraphics[width=0.48\columnwidth]{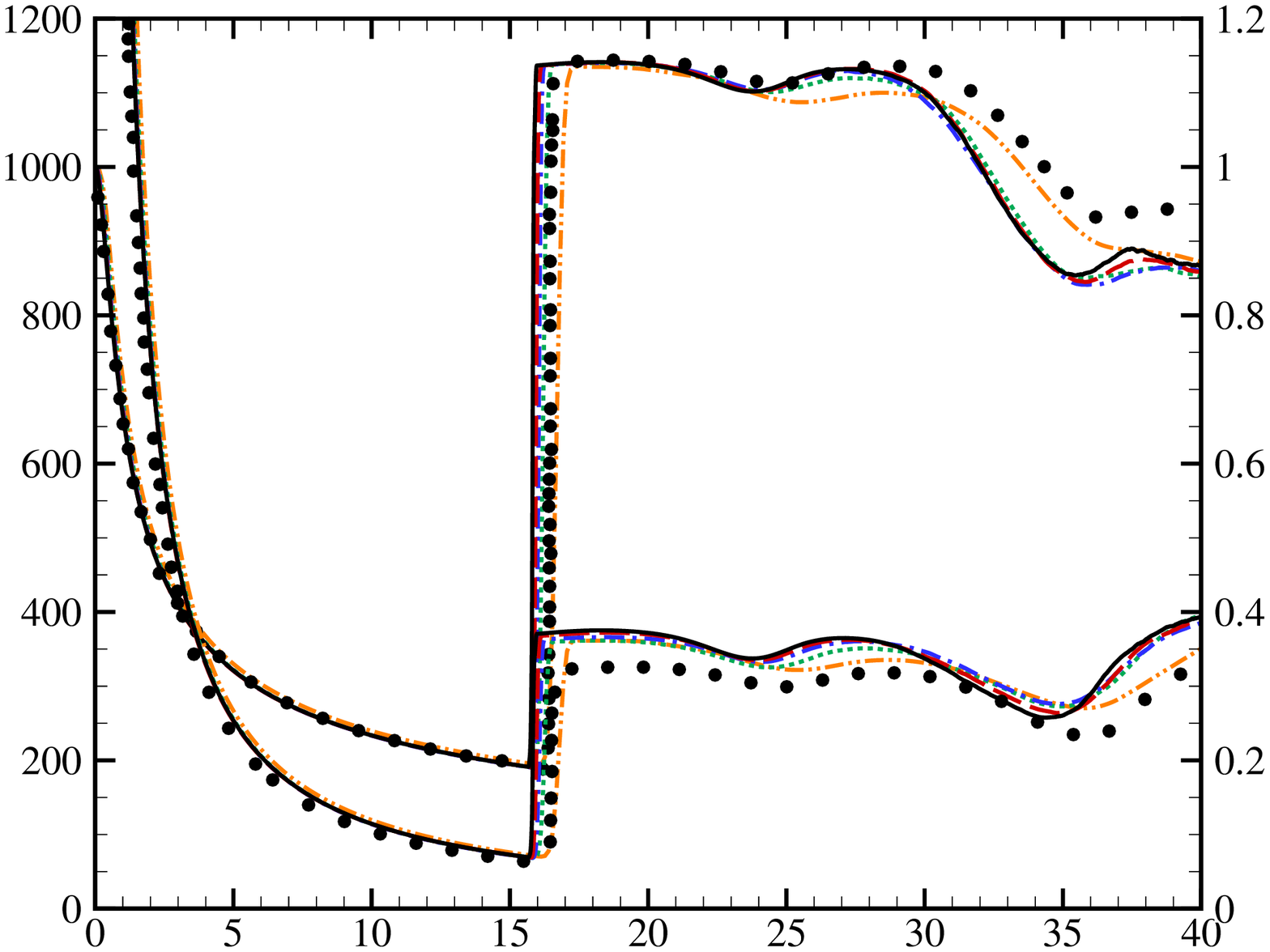}};
   \begin{scope}[x={(a.south east)},y={(a.north west)}]
     \node [align=center] at (0.55,0.00) {$x/D$};
     \node [align=center,rotate=90] at (0.00,0.55)  {$T$ [K]};
     \node [align=center,rotate=270] at (1.,0.55)  {$\rho$ [kg/m$^3$]};
     \node [align=center] at (0.01,0.95) {(b)};
     \draw [->] (0.4,0.8) -- (0.2,0.8);
     \draw [->] (0.45,0.2) -- (0.65,0.2);
   \end{scope}
\end{tikzpicture}

\begin{tikzpicture}
   \node[anchor=south west,inner sep=0] (a) at (0,0) {\includegraphics[width=0.48\columnwidth]{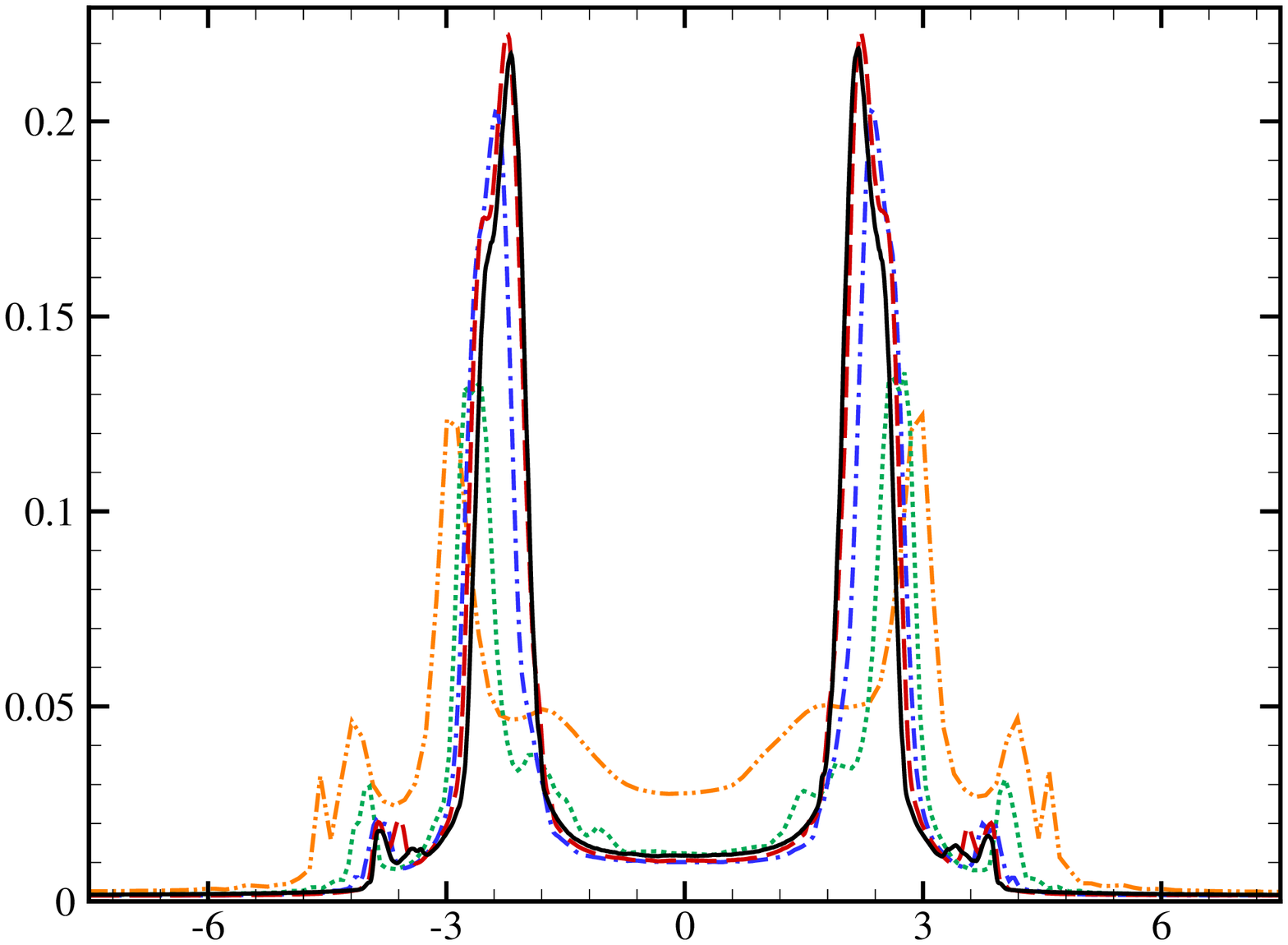}};
   \begin{scope}[x={(a.south east)},y={(a.north west)}]
      \node [align=center] at (0.55,0.00) {$x/D$};
      \node [align=center,rotate=90] at (-0.01,0.55)  {$T_\text{rms}/T_j$};
     \node [align=center] at (0.01,0.95)  {(c)};
      \node [align=center] at (1.50,0.55) {
      \begin{tabular}{cccc}
          \toprule
          $\Delta x$ (mm) &  $N_x$ & $N_y$ & legend\\
          \midrule
          0.25 & 6110 & 3168 & \protect\lsolid{black}      \\
          0.5  & 3100 & 1656 & \protect\ldashlon{red}      \\
          1    & 1590 & 880  & \protect\ldashdot{blue}     \\
          2    & 825  & 500  & \protect\ldott{green}       \\
          4    & 444  & 280  & \protect\ldashdotdot{orange}\\
          \bottomrule[\heavyrulewidth]
      \end{tabular}};
   \end{scope}
\end{tikzpicture}
\caption{Influence of the grid resolution on the underexpanded jet. Axial profiles of Mach number (a); axial profiles of normalized temperature and density (b); rms profiles of temperature along the vertical line $x/D = 20$ (c). In panel (a), the vertical lines denote the streamwise locations of the Riemann wave observed by Mart\'inez Ferrer \emph{et al.} \cite{ferrer2014detailed} (\protect\ldashdot{gray}), Su \emph{et al.} \cite{su2020numerical} (\protect\ldashlon{gray}) and Sheeran \& Dosanjh \cite{sheeran1968observations} (\protect\lsolid{gray}). In panel (b), black circles represent data extracted from Su \emph{et al.} \cite{su2020numerical}.}
\label{fig:jet_ave}
\end{figure}

Five computational grids were considered, with mesh sizes equal to $\Delta x = \Delta y =$ 4, 2, 1, 0.5 and 0.25 mm.
Figure~\ref{fig:jet_insta} shows an instantaneous snapshot of several quantities for the most refined grid used in the current study, revealing the challenging physics of the problem (for a thorough description of the flow physics of free underexpanded jets, the reader may refer to Franquet \emph{et al.} \cite{franquet2015free}).
The influence of the grid resolution is shown in figure~\ref{fig:jet_ave}, where the axial profiles of the Mach number (panel a), temperature and density (panel b) are shown. The location and width of the Mach disk is in agreement with experimental observations \citep{sheeran1968observations}, as well as recent numerical results obtained for similar NPR values \citep{su2020numerical}. Averaged profiles of the thermodynamic quantities are in accordance with results shown by Su \text{et al.} \cite{su2020numerical}; the small post-shock discrepancies in density may be attributed to the slight difference in the inflow N$_2$ and O$_2$ mass fractions.
The r.m.s. temperature values extracted at the vertical line $x/D=20$ reveal nearly-converged profiles for the $\Delta x = \SI{0.5}{mm}$ grid, whereas $\Delta x = \SI{2}{mm}$ grid is already sufficiently fine to capture correctly the mean profiles of first-order quantities. In any case, the numerical scheme is shown to handle satisfactorily under-resolved grids without the insurgence of numerical instabilities nor non-physical results.

\begin{figure}[tb]
\centering
 \begin{tikzpicture}
   \node[anchor=south west,inner sep=0] (a) at (0,0) {\includegraphics[width=0.49\textwidth, trim={5 5 5 5}, clip]{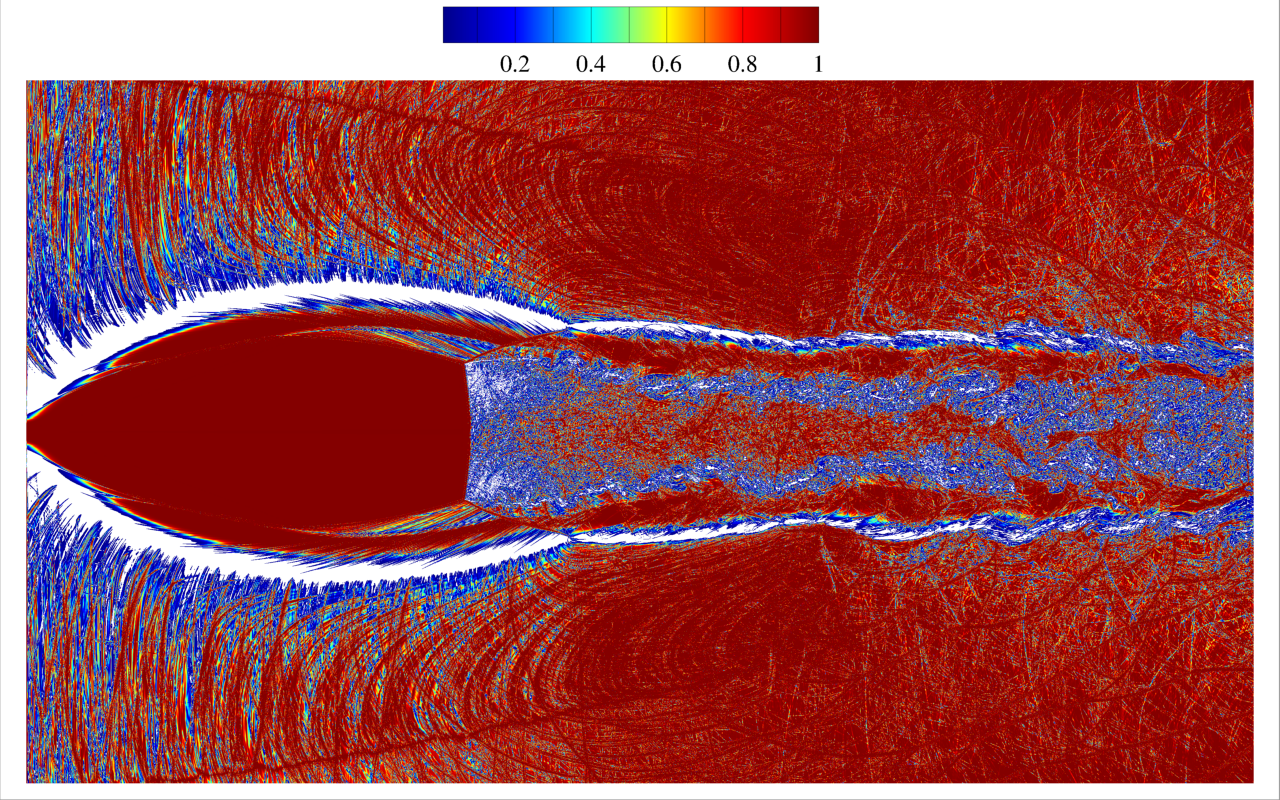}};
   \begin{scope}[x={(a.south east)},y={(a.north west)}]
     \node [align=center] at (0.95,0.95) {(a)};
   \end{scope}
 \end{tikzpicture}
 \hspace{-0.3cm}
 \begin{tikzpicture}
   \node[anchor=south west,inner sep=0] (a) at (0,0) {\includegraphics[width=0.49\textwidth, trim={5 5 5 5}, clip]{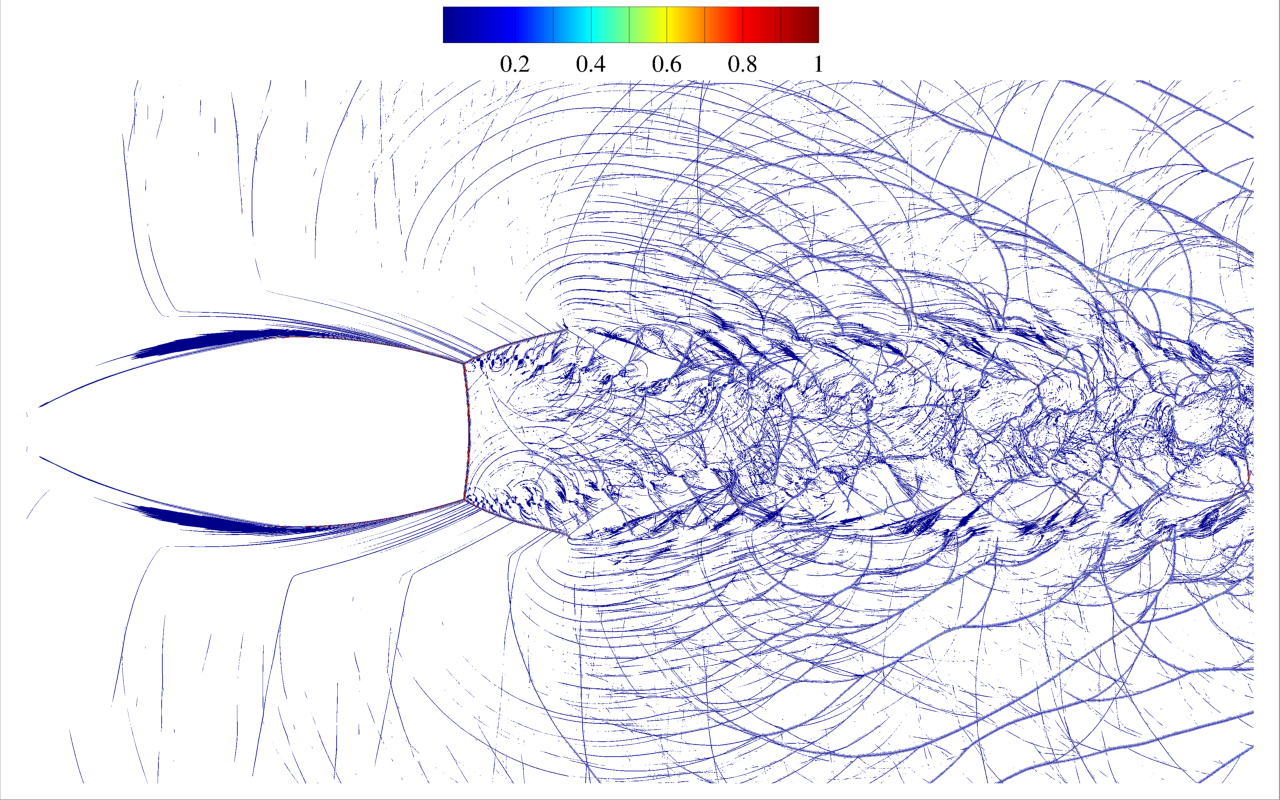}};
   \begin{scope}[x={(a.south east)},y={(a.north west)}]
     \node [align=center] at (0.95,0.95) {(b)};
   \end{scope}
 \end{tikzpicture}

  \begin{tikzpicture}
   \node[anchor=south west,inner sep=0] (a) at (0,0) {\includegraphics[width=0.49\textwidth, trim={5 5 5 5}, clip]{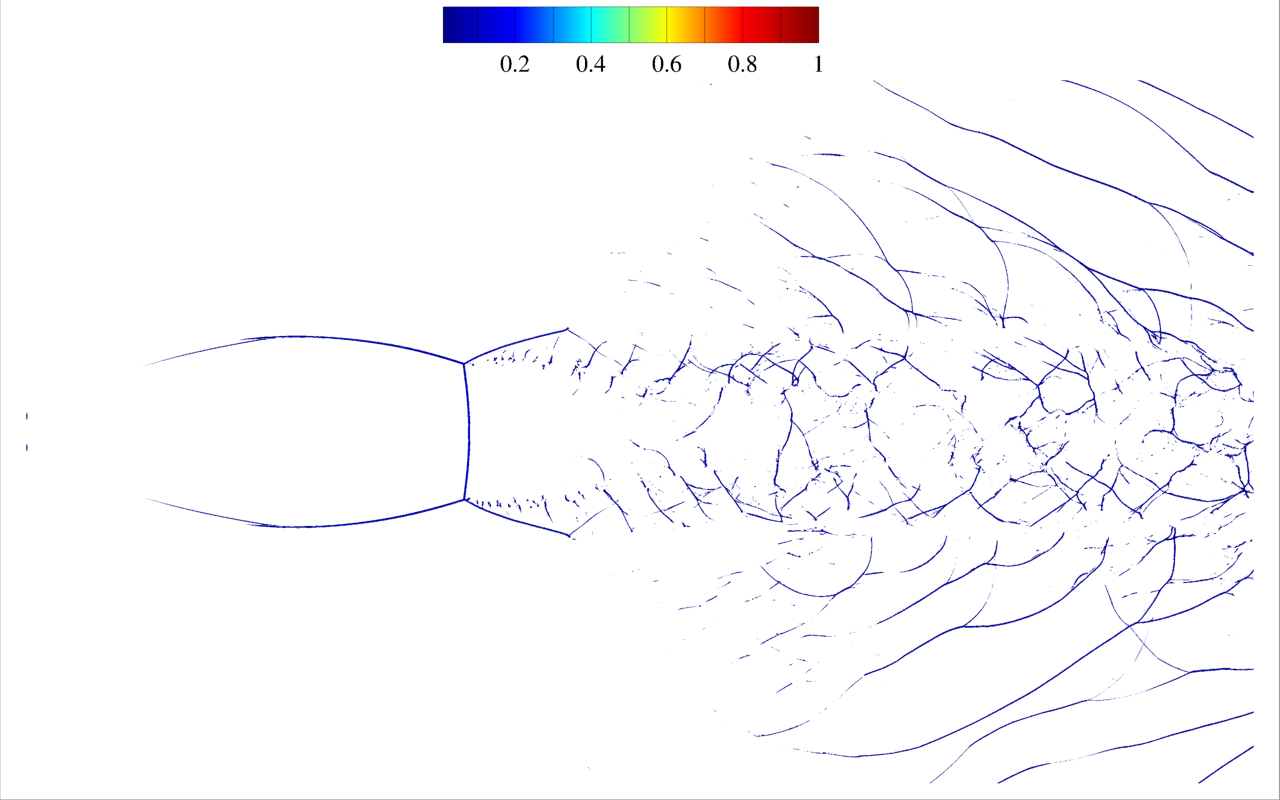}};
   \begin{scope}[x={(a.south east)},y={(a.north west)}]
     \node [align=center] at (0.95,0.95) {(c)};
   \end{scope}
 \end{tikzpicture}
 \hspace{-0.3cm}
 \begin{tikzpicture}
   \node[anchor=south west,inner sep=0] (a) at (0,0) {\includegraphics[width=0.49\textwidth, trim={5 5 5 5}, clip]{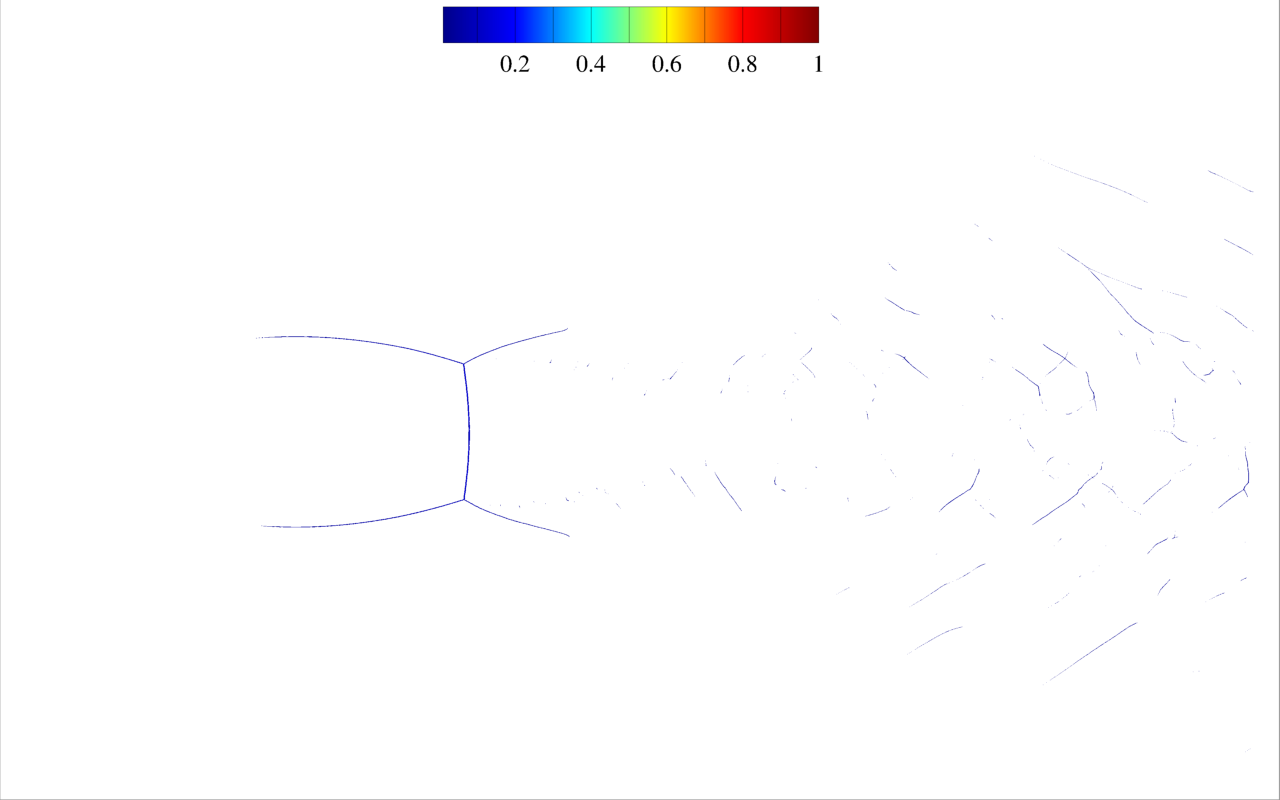}};
   \begin{scope}[x={(a.south east)},y={(a.north west)}]
     \node [align=center] at (0.95,0.95) {(d)};
   \end{scope}
 \end{tikzpicture}
\caption{Isocontours of each component of the shock-capturing sensor. Note that values below \num{e-2} are cropped in order to ease visualization. (a): Ducros' sensor, $\frac{(\nabla\cdot \mathbf u)^2}{(\nabla\cdot \mathbf u)^2 + |\nabla\times\mathbf u|^2  + \epsilon}$; (b): correction of Bhagatwala \& Lele (L2-norm), $\frac{1}{2} \left[1 - \tanh \left(2.5 + 10 \frac{\delta x_i}{c} \nabla \cdot \mathbf u \right) \right]$; (c): Jameson's pressure-based sensor (L2-norm), $\left|\frac{p_{j+1} - 2p_j + p_{j-1}}{p_{j+1}+2p_j + p_{j-1}}\right|$, (d): combination of the previous terms.}
\label{fig:jet_sensor}
\end{figure}

In section~\ref{sec:numeth}, we stressed the importance of using a well-constructed shock-capturing sensor for the numerical simulation of high-speed turbulent flows. The current 2D jet represents a suitable configuration for benchmarking such a sensor, due to the concurrent presence of strong steady shocks, turbulent shear layers and propagation of acoustic waves. The influence of each one of the three components of the sensor presented in equation~\eqref{eq:sensor} is shown in figure~\ref{fig:jet_sensor}, displaying their isocontours on an instantaneous snapshot. Values below \num{e-2} have been cropped in order to highlight the flow regions marked by each term as potential shocks. As expected, the Ducros' sensor (panel a) is not able to properly identify strong gradients if applied alone; its values are shown to be close to unity even in smooth-flow regions, essentially because of the absence of vorticity (shown in figure~\ref{fig:jet_insta}d). The Bhagatwala \& Lele correction and the Jameson's pressure-based sensors (panel b and c, respectively) allow one to obtain a more satisfactory large-gradients tracing, the latter being slightly more selective. One may object the utility of using both of them; however, their combined use palliates each other's deficiency. Specifically, the Jameson's sensor could hardly tell the difference between a shock and a high-vorticity, purely-solenoidal region; on the other hand, the Bhagatwala \& Lele correction can result in excessive damping of all dilatational motions when dealing with slightly under-resolved simulations. The combination of the three sensors (panel d) results in an excellent localization of the regions in which low-order numerical dissipation should be injected; that is, along the lateral intercepting shocks, the Mach disk and the reflected shock waves. A few wave fronts are also marked downstream of the Riemann wave; however, the sensor magnitude is almost negligible there (i.e., lower than 0.02 everywhere), and such is the amount of injected dissipation.

%===============================================================================
\subsection{Compressible Taylor-Green Vortex}
We consider a compressible extension of the Taylor-Green Vortex (TGV) in order to assess the behaviour of the numerical scheme when applied to highly under-resolved cases and the suitability of the non-linear artificial dissipation to act as a regularization term in the context of low- and high-$M$ ILES.
The first authors to extend the TGV problem to strongly compressible configurations were Peng \& Yang \cite{peng2018effects}, with the aim of investigating the evolution of vortex-surface fields at large Mach numbers. They considered Mach numbers ranging from 0.5 to 2, albeit at a lower Reynolds number with respect to the one considered in most of the incompressible and low-$M$ studies (i.e., 400 instead of 1600). Recently, Lusher \& Sandham \cite{lusher2019assessment} carried out simulations of TGVs up to $M{=}1.25$, in order to evaluate the behavior of different shock-capturing schemes for high-$M$ turbulent flows. \\
The initial conditions for velocity and pressure fields are:
\begin{flalign}
   u(x,y,z,t=0) &= \sin\left( \frac{x}{L} \right)\cos\left( \frac{y}{L} \right)\cos\left( \frac{z}{L} \right)\\
   v(x,y,z,t=0) &= - \cos\left( \frac{x}{L} \right)\sin\left( \frac{y}{L} \right)\cos\left( \frac{z}{L} \right)\\
   w(x,y,z,t=0) &= 0 \\
   p(x,y,z,t=0) &= \frac{1}{\gamma M^2} + \frac{1}{16} \left[ \cos\left( \frac{2x}{L} \right) + \cos\left( \frac{2y}{L} \right) \right] \left[ 2 + \cos\left( \frac{2z}{L} \right) \right]
\end{flalign}
When considering the compressible form of the TGV problem, different choices are possible to set the initial conditions for the thermodynamic quantities \citep{peng2018effects}. We selected a constant density initial condition, for which the initial density field is equal to $\rho = \SI{1.292}{kg/m^3}$ everywhere, whereas the initial temperature is computed from density and pressure according to the perfect-gas equation of state.
The Reynolds number, Prandtl number and the specific heat ratio are set to 1600, 0.71 and 1.4, respectively. Comparisons are made by considering the total kinetic energy $K$ and the resolved  enstrophy $\Omega$ integrated over the computational domain:
\begin{equation}
   E_k = \frac{1}{\rho_\text{ref}\mathcal{V}} \int_\mathcal{V} \frac{1}{2} \rho u_i u_i \mdd \mathcal{V}, \qquad \qquad
   \Omega = \frac{1}{\rho_\text{ref} Re} \int_\mathcal{V} \mu \left( \varepsilon_{ijk} \Derp{u_k}{x_j} \right)^2 \mdd \mathcal{V}.
\end{equation}
Note that the enstrophy correspond to the solenoidal part of the total viscous dissipation rate; the dilatational component will not be discussed since its contribution has been shown to be negligible also at Mach numbers higher than unity \citep{lusher2019assessment}.\\
In our test, computational grids ranging from $64^3$ to $1024^3$ and Mach numbers equal to 0.1, 0.5 and 1 have been considered, with $Re=1600$. For each run, a third-order TVD Runge-Kutta method is used, in conjunction with a fixed CFL number of 0.5. Periodic conditions are applied in all directions.

\begin{figure}[!tb]
\hspace{-0.1cm}\includegraphics[width=0.33\textwidth]{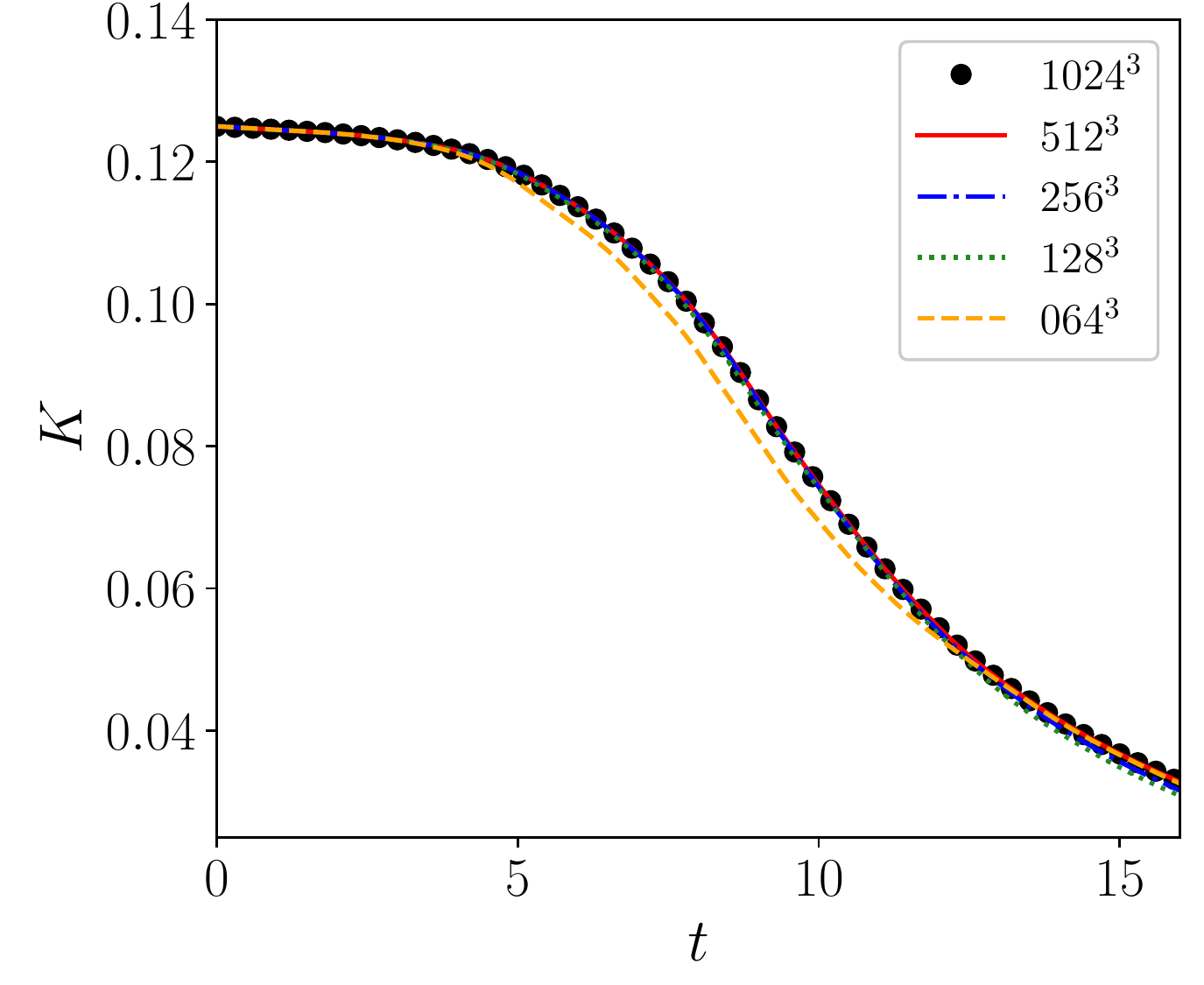}
\hspace{-0.1cm}\includegraphics[width=0.33\textwidth]{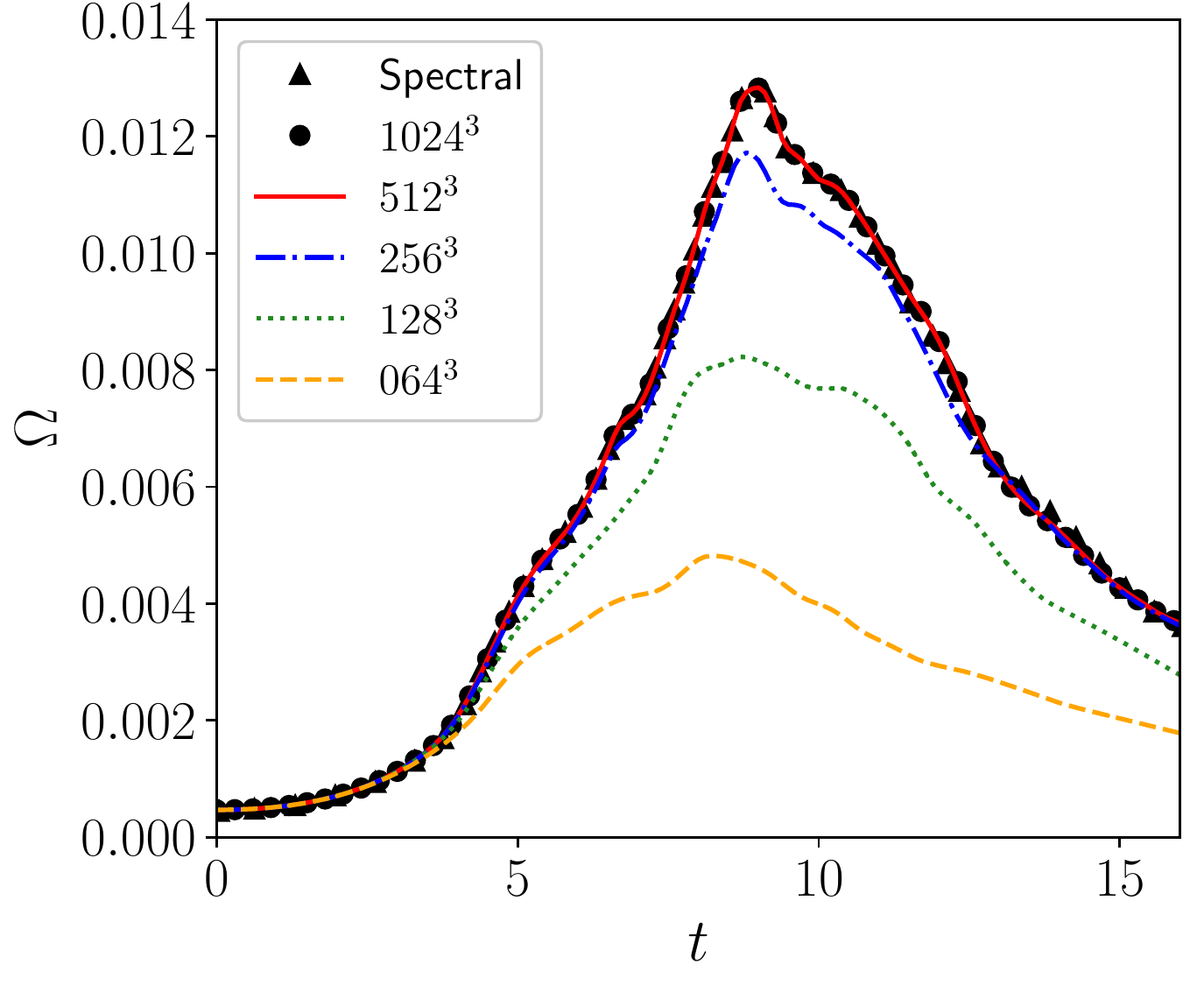}
\hspace{-0.1cm}\includegraphics[width=0.33\textwidth]{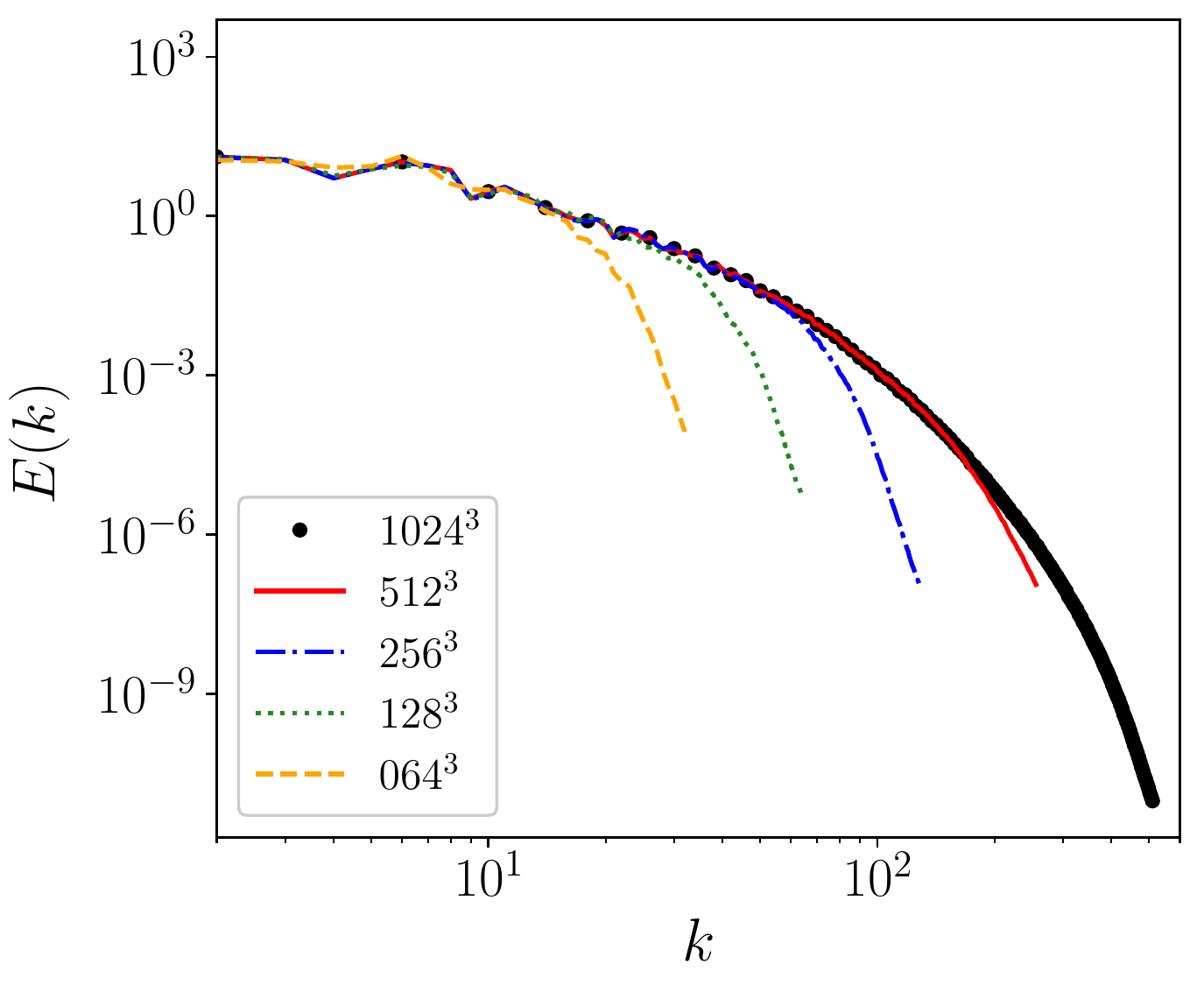}

\hspace{-0.1cm}\includegraphics[width=0.33\textwidth]{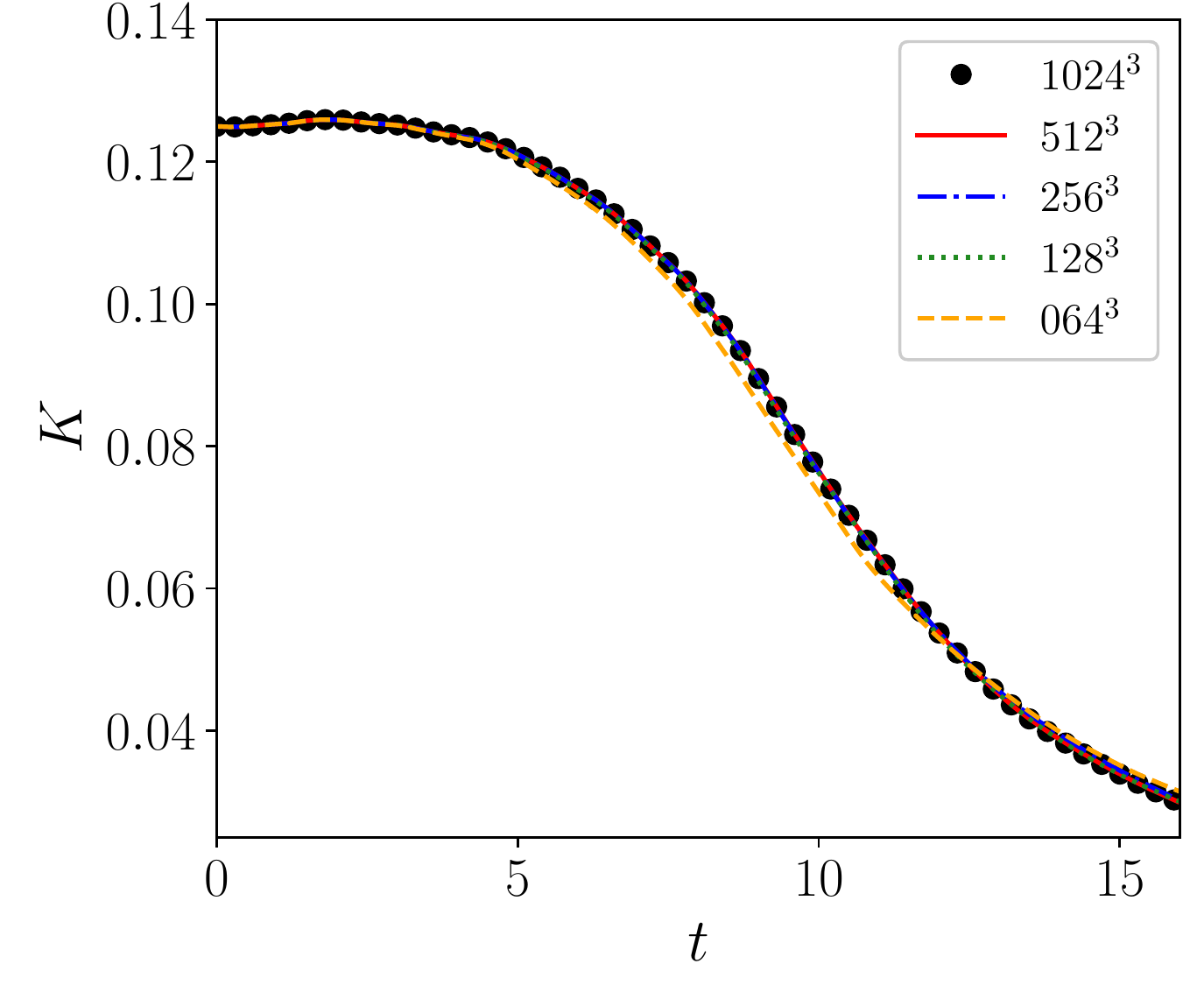}
\hspace{-0.1cm}\includegraphics[width=0.33\textwidth]{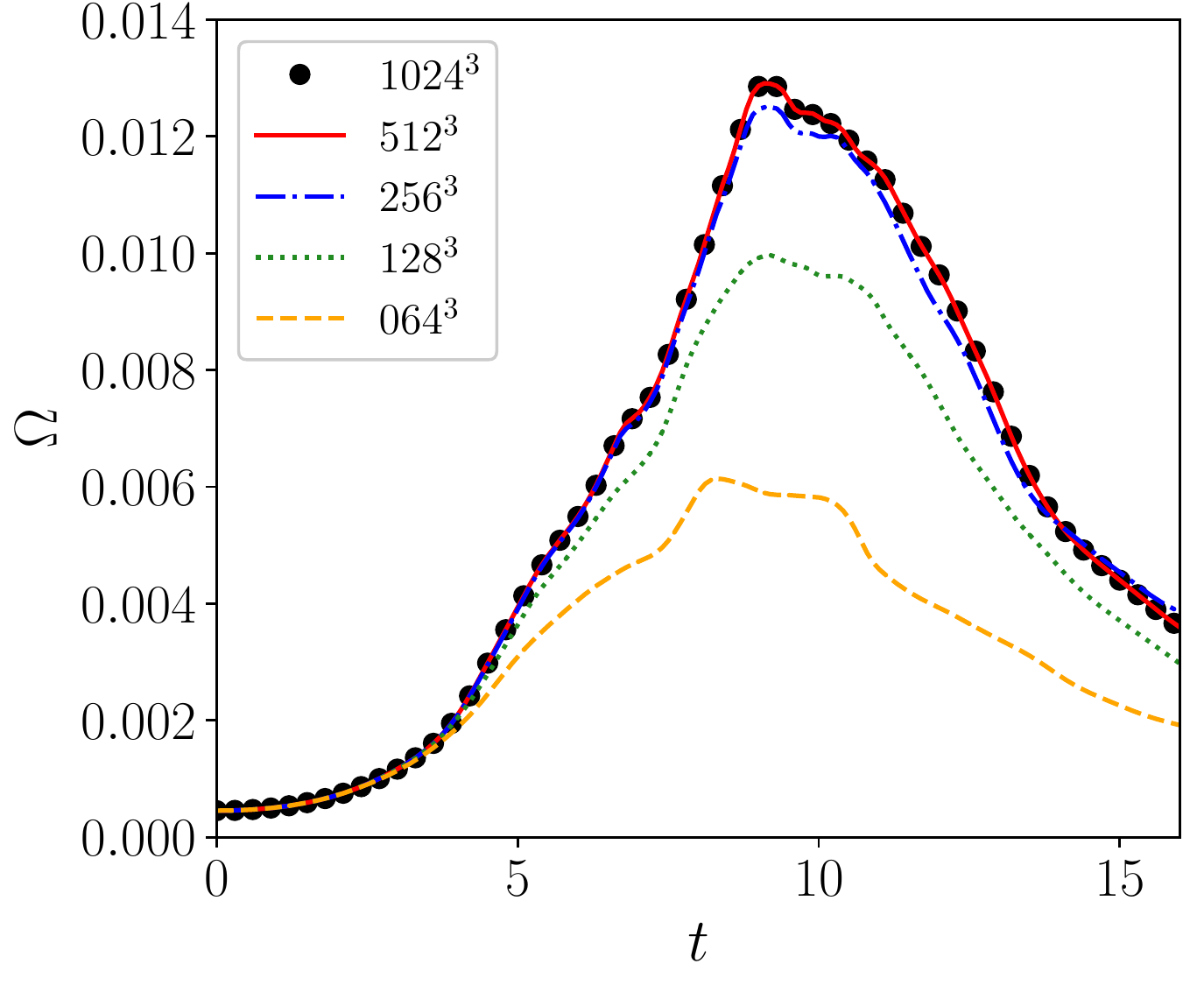}
\hspace{-0.1cm}\includegraphics[width=0.33\textwidth]{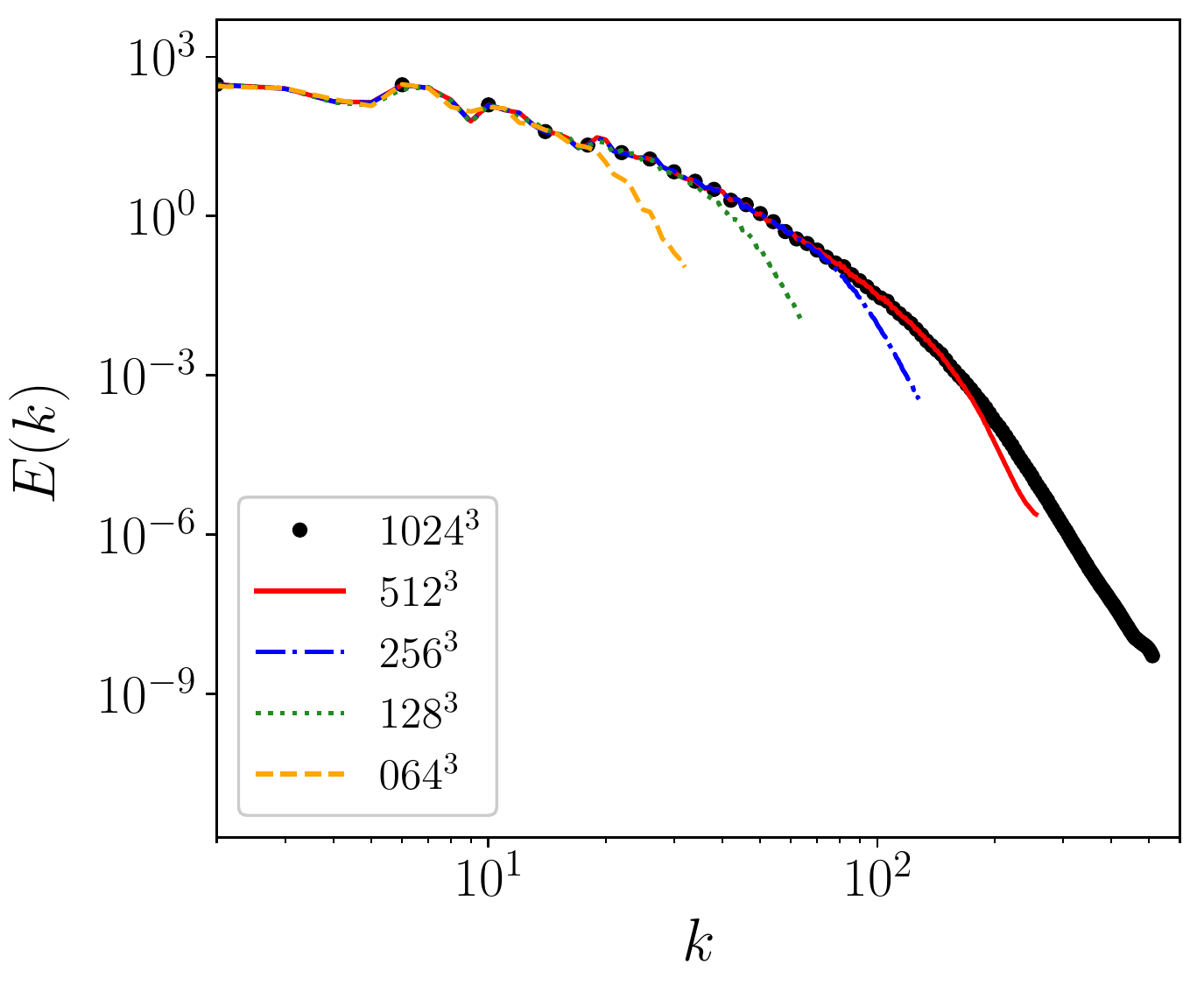}

\hspace{-0.1cm}\includegraphics[width=0.33\textwidth]{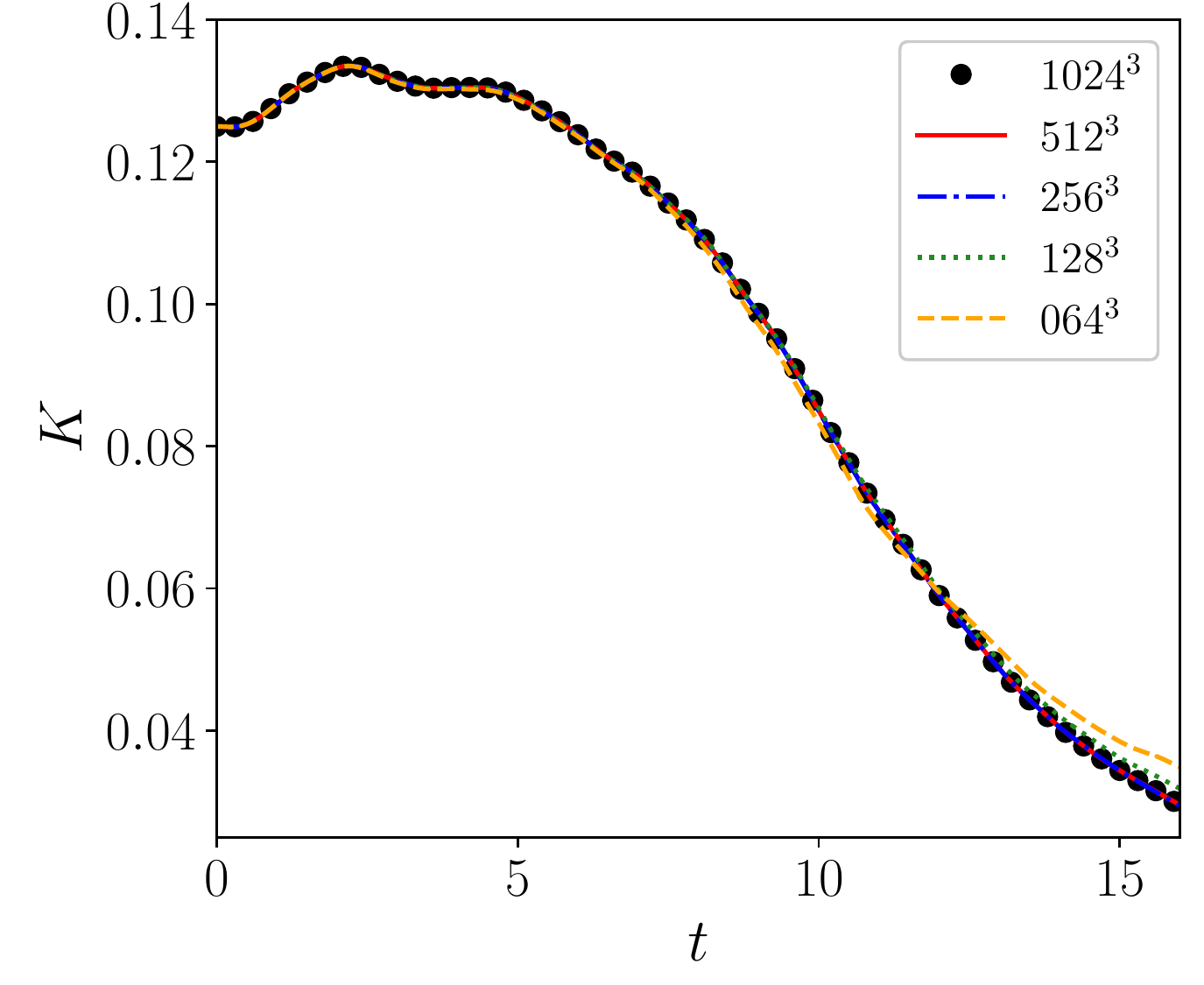}
\hspace{-0.1cm}\includegraphics[width=0.33\textwidth]{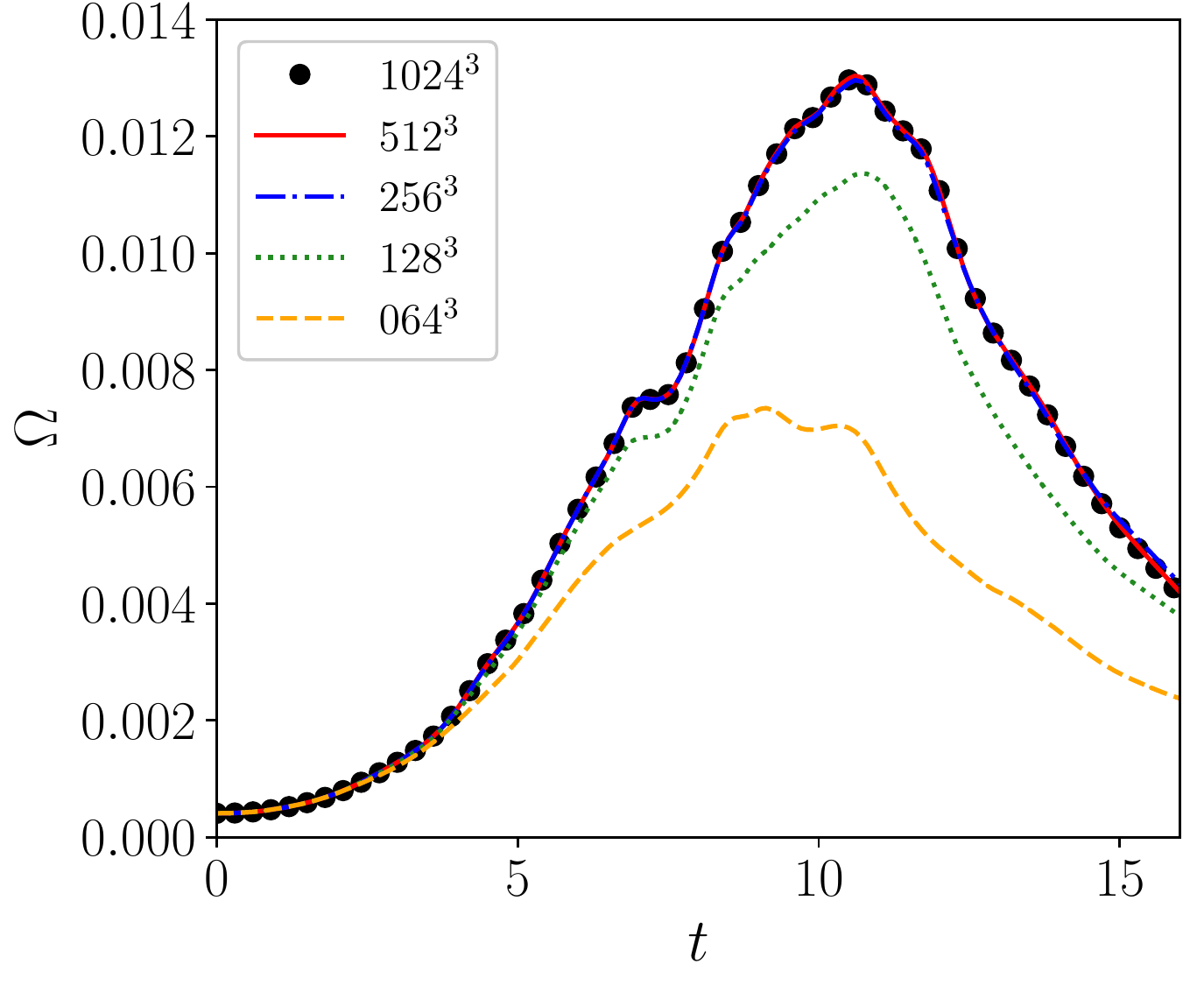}
\hspace{-0.1cm}\includegraphics[width=0.33\textwidth]{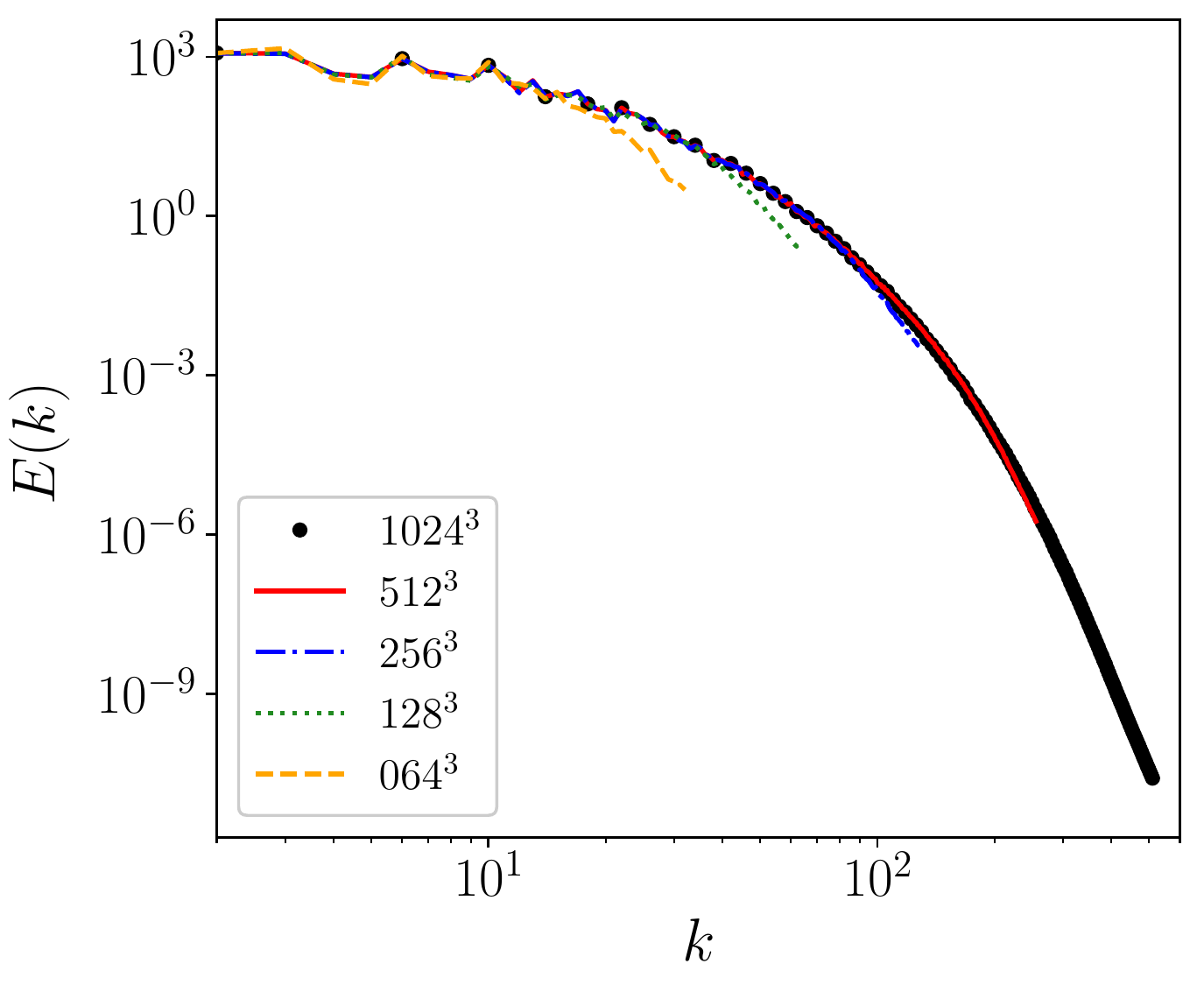}
\caption{Influence of the grid resolution for the Taylor-Green Vortex case. Top: $M{=}0.1$; center: $M{=}0.5$; bottom: $M{=}1$. Left and center columns: timewise evolutions of volume-integrated kinetic energy $K$ and resolved solenoidal enstrophy $\Omega$; right column: turbulent kinetic energy spectra at selected times ($t^* = 8.9$ for $M{=}0.1$, 9.4 for $M{=}0.5$ and 11.1 for $M{=}1$).}
\label{fig:tgv_grids}
\end{figure}

\begin{figure}[!tb]
\centering
\begin{tikzpicture}
      \node[anchor=south west,inner sep=0] (a) at (0,0) {\includegraphics[width=0.45\textwidth, trim={2 5 2 0}, clip]{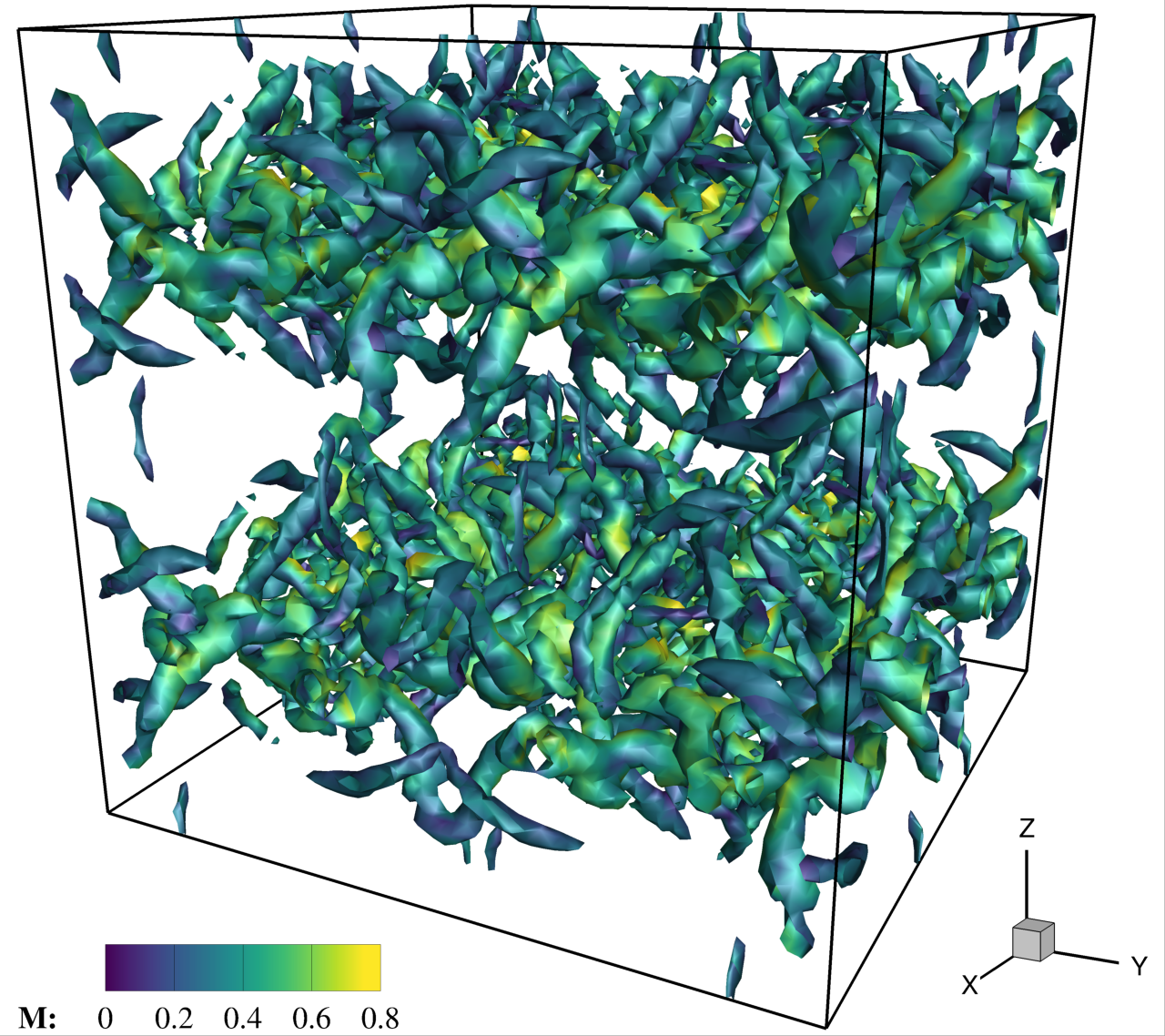}};
      \begin{scope}[x={(a.south east)},y={(a.north west)}]
        \node [align=center] at (0.5,1.04) {\large $64^3$};
      \end{scope}
\end{tikzpicture}
\begin{tikzpicture}
      \node[anchor=south west,inner sep=0] (a) at (0,0) {\includegraphics[width=0.45\textwidth, trim={2 5 2 0}, clip]{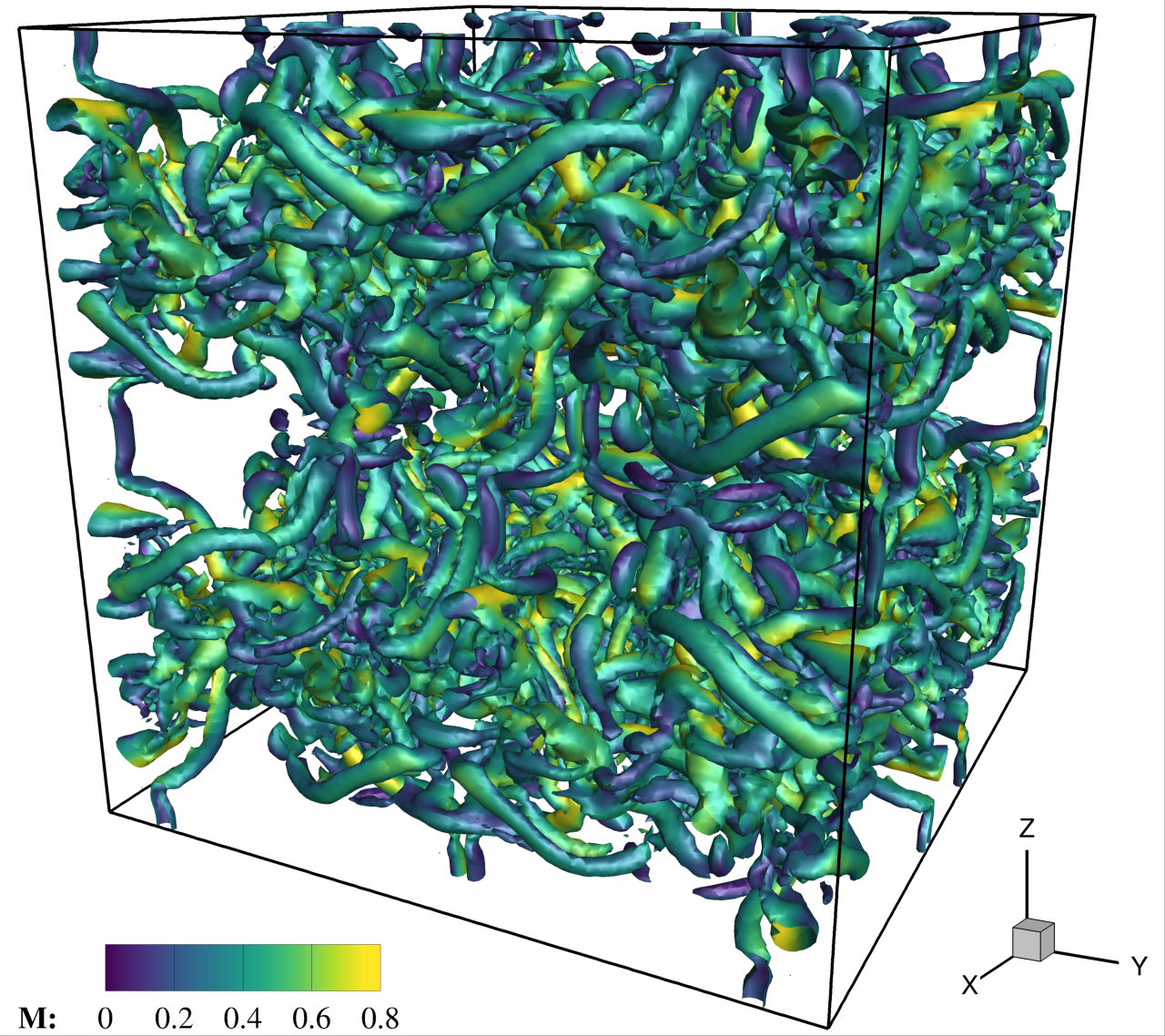}};
      \begin{scope}[x={(a.south east)},y={(a.north west)}]
        \node [align=center] at (0.5,1.04) {\large $128^3$};
      \end{scope}
\end{tikzpicture}

\begin{tikzpicture}
      \node[anchor=south west,inner sep=0] (a) at (0,0) {\includegraphics[width=0.45\textwidth, trim={2 5 2 0}, clip]{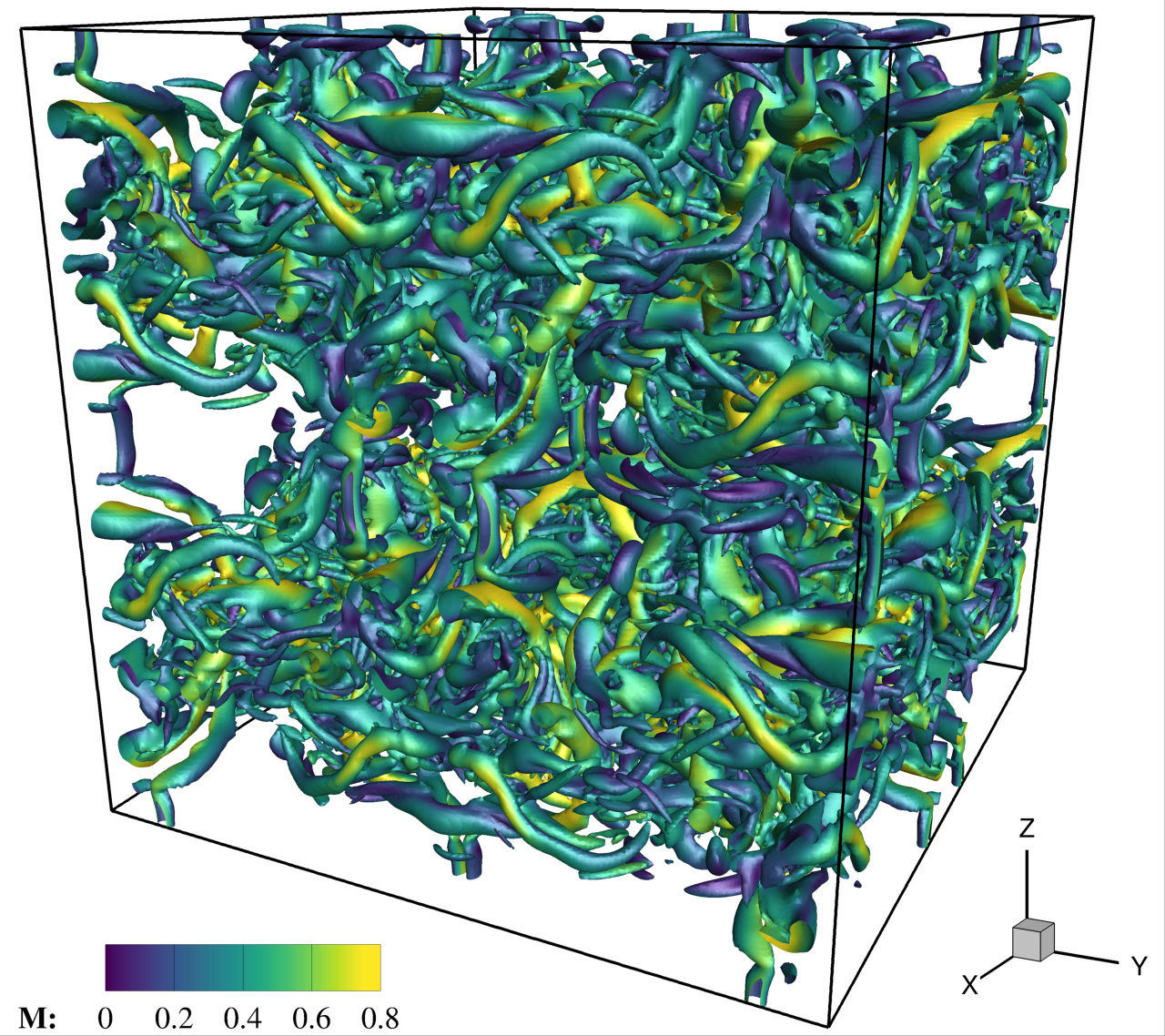}};
      \begin{scope}[x={(a.south east)},y={(a.north west)}]
        \node [align=center] at (0.5,1.04) {\large $256^3$};
      \end{scope}
\end{tikzpicture}
\begin{tikzpicture}
      \node[anchor=south west,inner sep=0] (a) at (0,0) {\includegraphics[width=0.45\textwidth, trim={2 5 2 0}, clip]{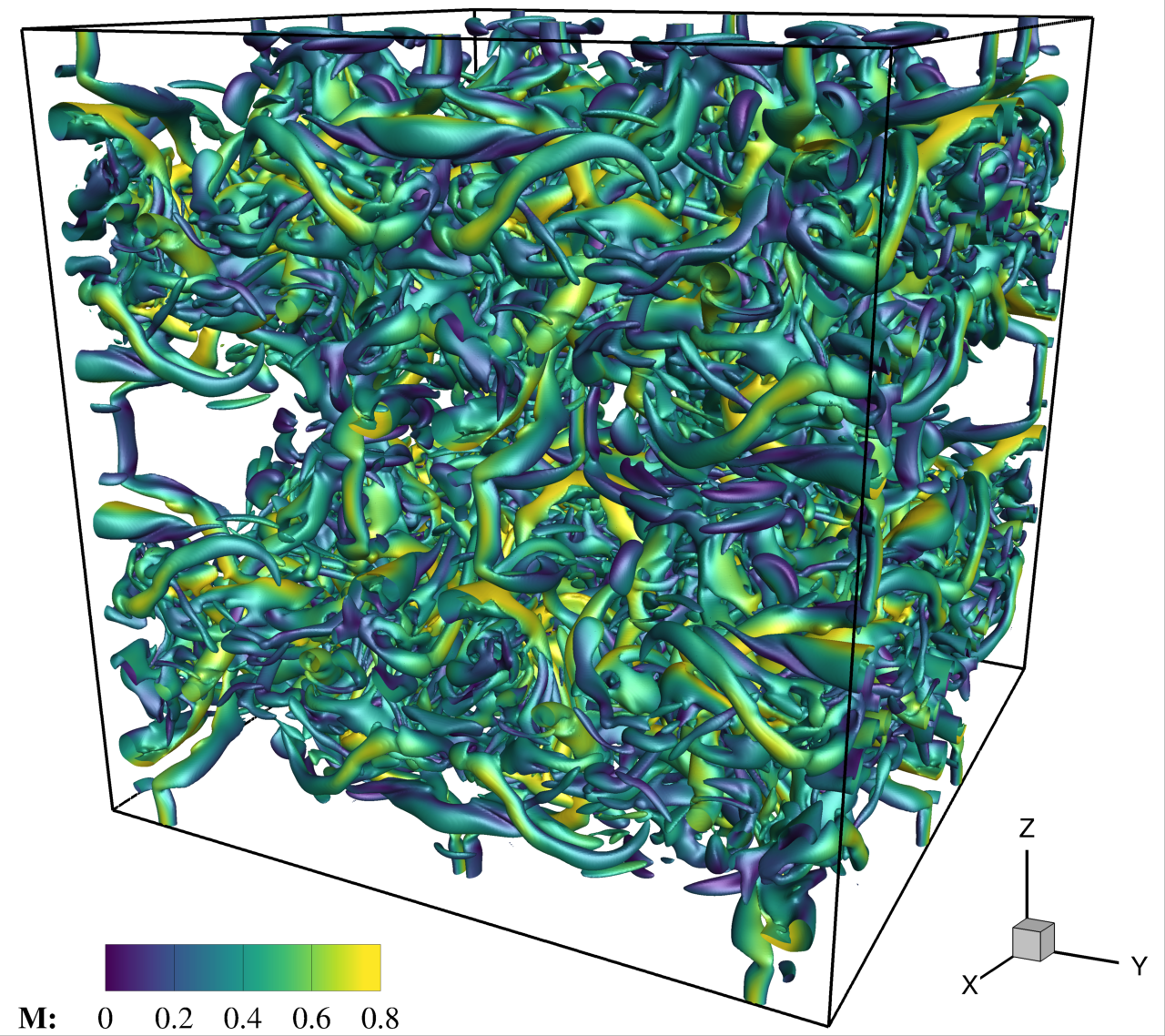}};
      \begin{scope}[x={(a.south east)},y={(a.north west)}]
        \node [align=center] at (0.5,1.04) {\large $512^3$};
      \end{scope}
\end{tikzpicture}
\caption{Isocontours of the $Q-$criterion coloured with local Mach number for computational grids ranging from $64^3$ to $512^3$ ($M=1$ case).}
\label{fig:tgv_Q}
\end{figure}

Figure~\ref{fig:tgv_grids} shows the evolution of the volume-integrated kinetic energy, solenoidal enstrophy and turbulent kinetic energy spectra at selected times (arbitrarily chosen after the time at which enstrophy peaks), as a function of the initial Mach number and of the grid resolution. Several considerations are in order. First, we observe that the low-$M$ case is perfectly superposed to classical spectral results, which confirms the good behaviour of the numerical scheme for low-speed configurations. The low-order artificial nonlinear dissipation is correctly turned off, whereas the higher-order one guarantees an amount of dissipation such that numerical stability is ensured and virtually all active scales are almost untouched. Increasing the Mach number, the flow dynamics are heavily altered: in the early stages of the decay, internal energy is converted into kinetic energy, such that the average pressure-work can exceed the viscous dissipation rate and the total kinetic energy does not exhibit any more the classical monotonous decline of the incompressible case. The enstrophy peak slightly moves towards later times whereas its magnitude is almost unchanged.
%This behavior differ by recent similar results obtained by \cite{lusher2019assessment}, in which a flattening of the curves and slightly lower enstrophy peaks for increasing $M$ were observed. This dissimilarity is most likely due to the different initial conditions chosen by the authors, that considered a constant temperature profile with density fluctuations computed from the equation of state, the opposite of the current study. Indeed, for the $M{=}1$ case we observe a larger internal-to-kinetic energy conversion at the early stages, which translates in higher velocity gradients (and hence, larger dissipation) at later times.\\
The enstrophy is insufficiently resolved on the $64^3$ and $128^3$ grids, albeit the largest scales are correctly captured as displayed by the kinetic energy spectra and the $Q-$criterion visualizations (figure~\ref{fig:tgv_Q}). In any case, no energy pileup is observed at the smaller scales. The numerical scheme is therefore able to handle efficiently highly under-resolved configurations, while ensuring a correct representation of the largest scales. Grid-converged results are observed for the $512^3$ grids; whereas the $256^3$ grids allow one to observe a clear trend; that is, a faster convergence of results for larger initial Mach numbers. Enstrophy values are superposed to the reference solution at $M{=}1$, whereas they are slightly underestimated at $M{=}0.1$. This is coherent with the small-scales representation given by the turbulent kinetic energy spectra: although the cutoff wavenumber $k_c$ (i.e., the wavenumber at which the dissipation term of the numerical scheme starts to have a large influence on the flow field) does not change for a given grid, the spectral energy contents for $k>k_c$ tend to be closer to the DNS distributions for larger $M$. This behavior can be attributed to the time step constraints imposed by acoustic waves in low-$M$ configurations: achieving the same final nondimensional time ($t^*=16$) requires a larger number of time steps, which also implies a larger number of applications of the numerical dissipation operator. This effect (visible only at the smallest, not well resolved scales beyond $k_c$) underlines the satisfactory selectivity properties of the numerical method.

\begin{figure}[!tb]
\centering
\includegraphics[width=0.4\textwidth]{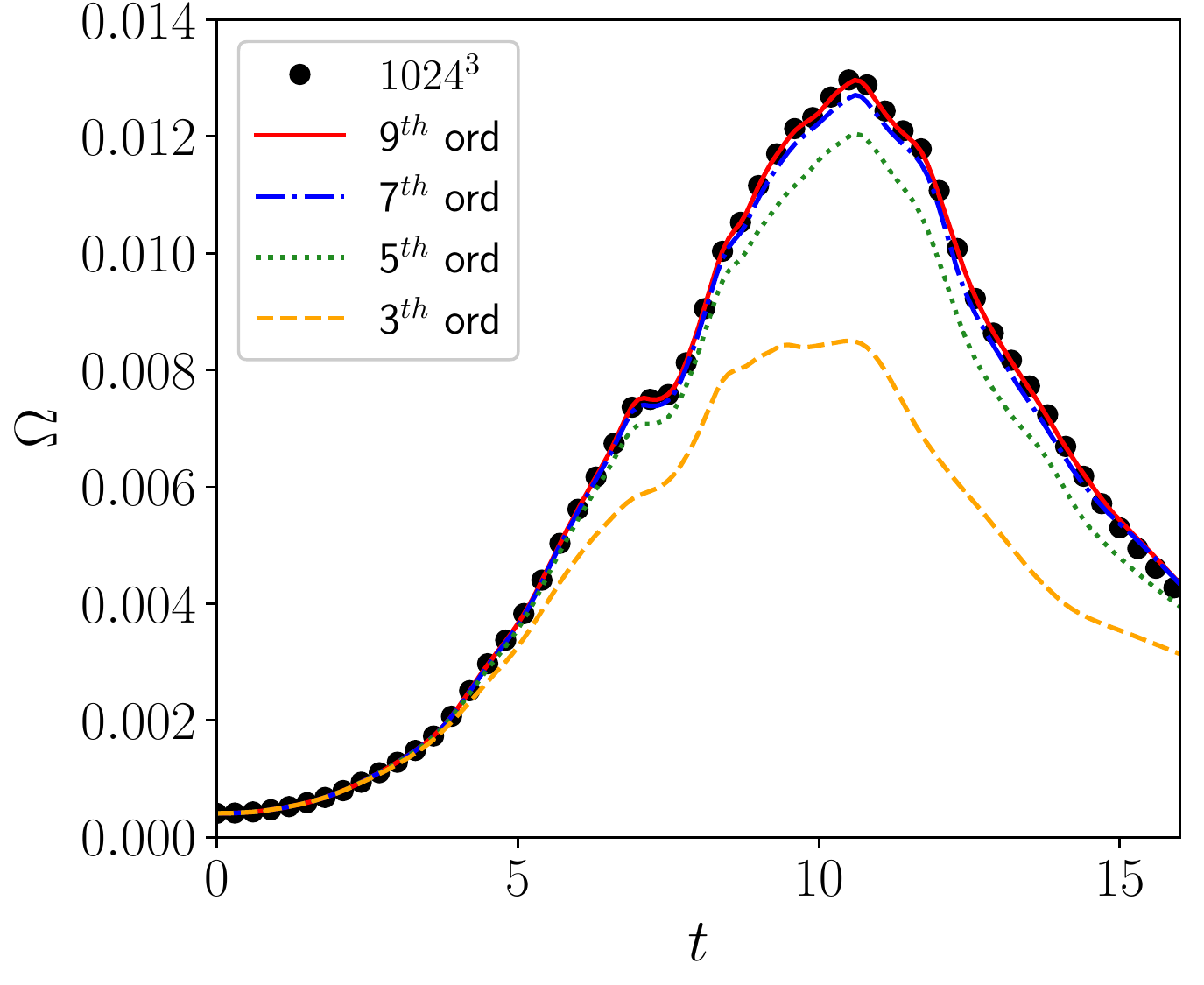}
\includegraphics[width=0.4\textwidth]{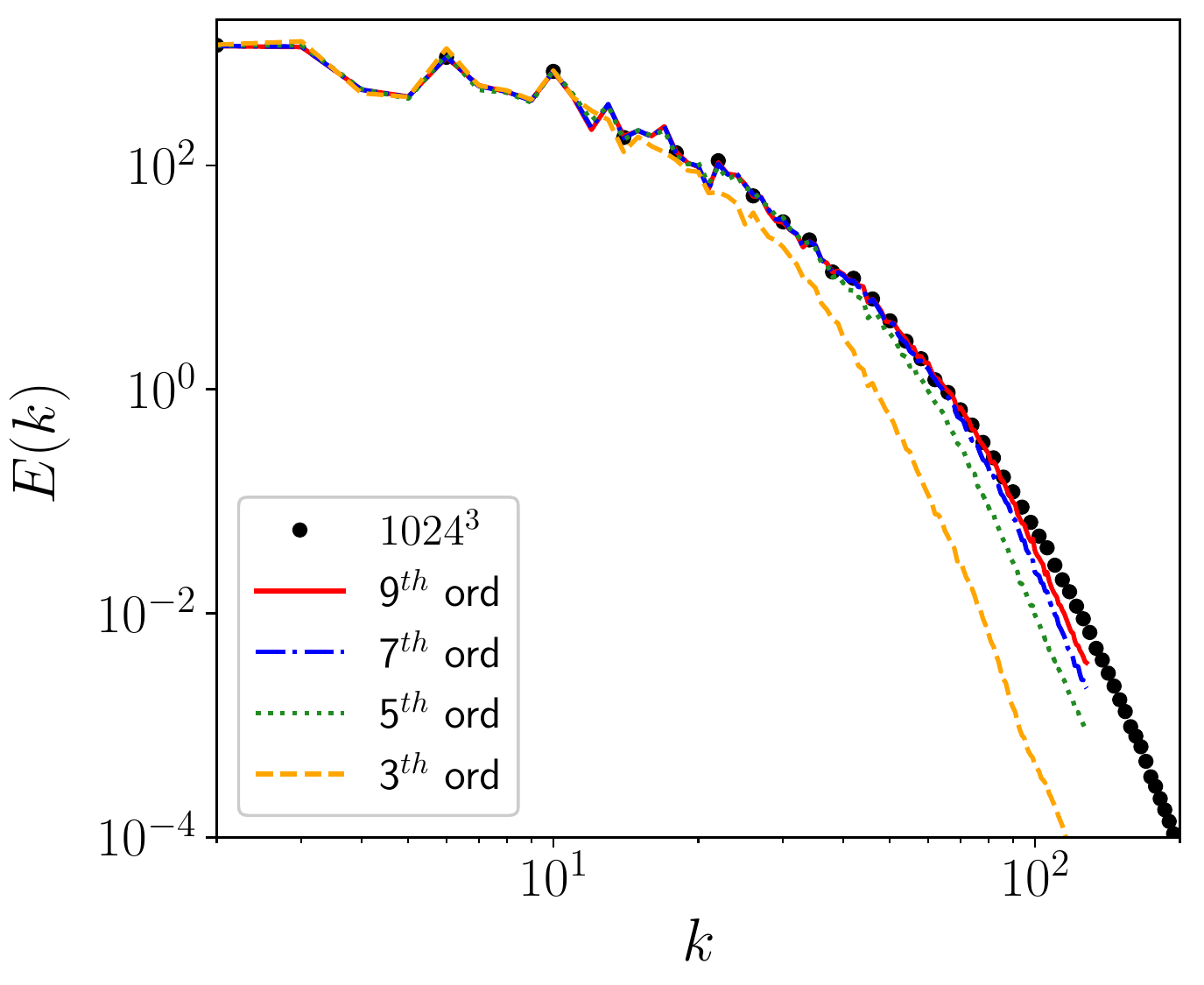}
\vspace{-0.5cm}
\caption{Influence of the order of the nonlinear dissipation term for the TGV case at $M{=}1$ for $256^3$ grid. Left: enstrophy evolution; right: turbulent kinetic energy spectra at $t^*{=}11.1$ (zoom).}
\label{fig:tgv_ordcomp}
\end{figure}

\begin{figure}[!tb]
\centering
\includegraphics[width=0.4\textwidth]{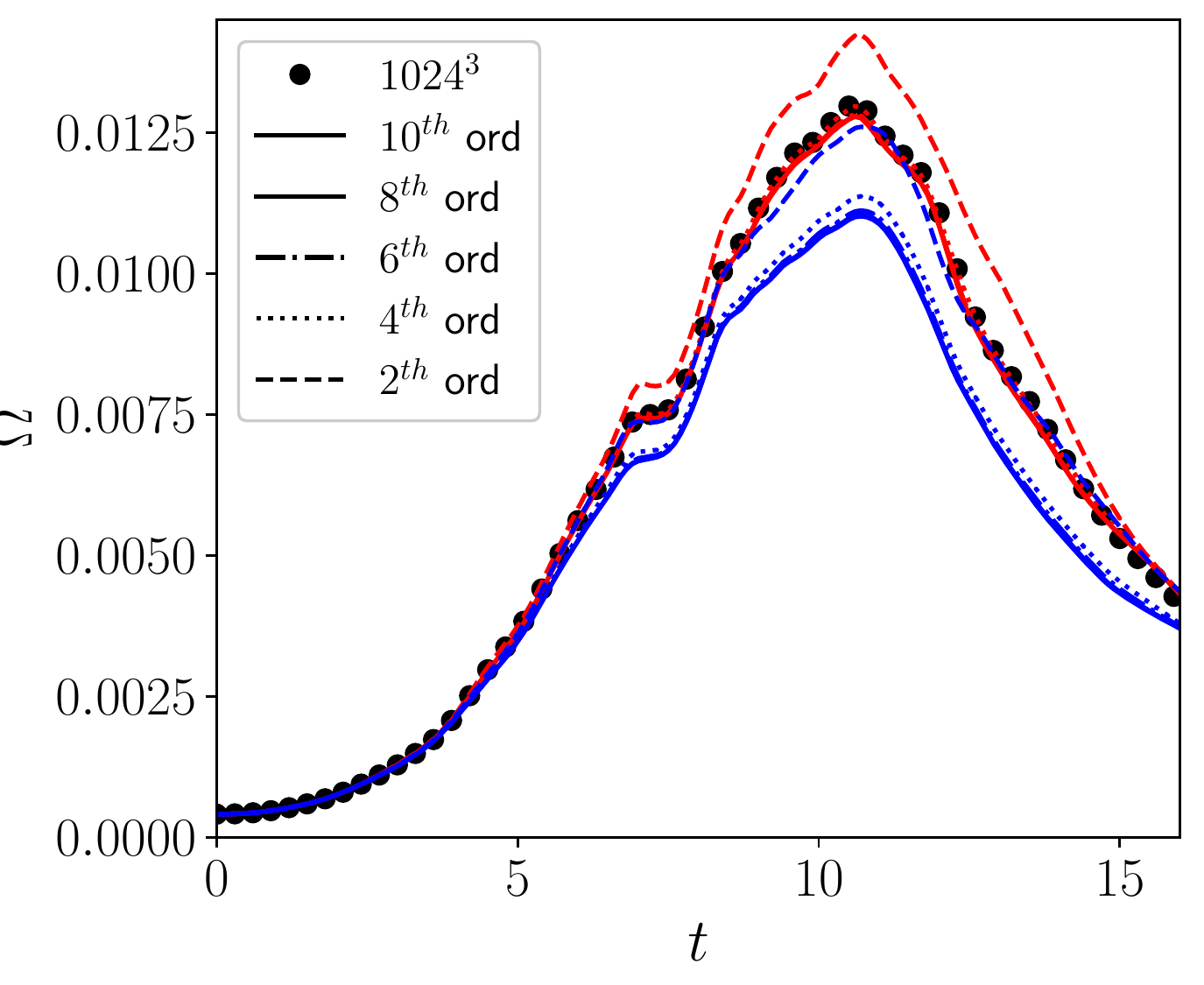}
\includegraphics[width=0.4\textwidth]{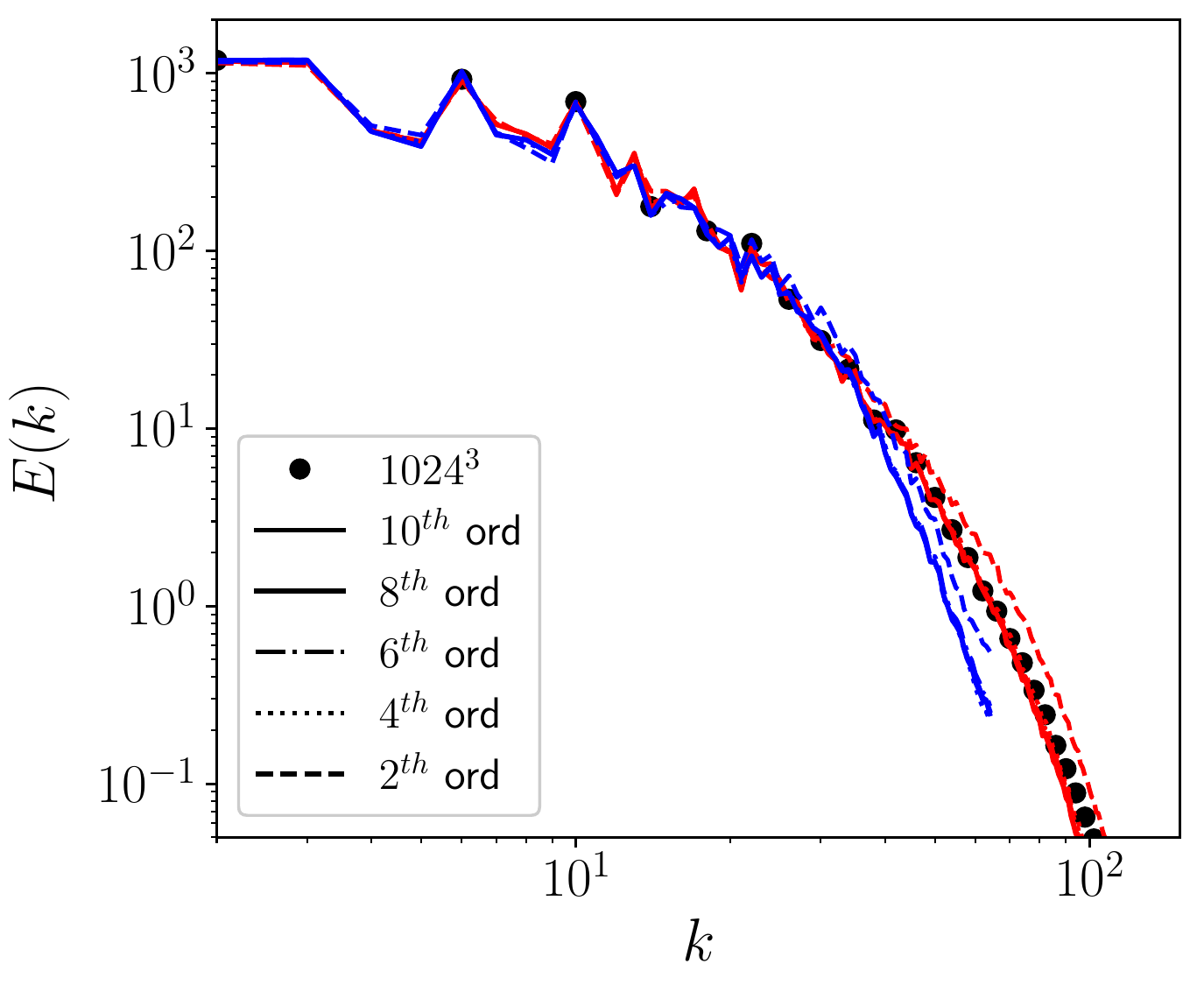}
\vspace{-0.5cm}
\caption{Influence of the order of the numerical scheme for the discretization of the viscous fluxes for the TGV case at $M{=}1$. Left: enstrophy evolution; right: turbulent kinetic energy spectra at $t^*{=}11.1$ (zoom). Red color: $256^3$ grids; blue color: $128^3$ grids.}
\label{fig:tgv_viscfluxcomp}
\end{figure}

To illustrate the importance of using a high-order scheme, in Figure~\ref{fig:tgv_ordcomp} we report enstrophy profiles and kinetic energy spectra for a $256^3$ grid obtained by changing the order of the non-linear dissipation term of the scheme, from $3^\text{rd}$ up to $9^\text{th}$. Results indicate that 1) the $3^\text{th}$-order dissipation on the $256^3$ grid gives results similar to the $9^\text{th}$-order one on a $64^3$ grid, 2) the $9^\text{th}$-order dissipation allows to obtain almost grid-converged results, and 3) the gain of further increasing the dissipation order is almost negligible, even for coarser grids. Of note, results obtained with $3^\text{th}$-order dissipation in conjunction with higher-order discretizations for the convective fluxes (up to $10^\text{th}$ order) are almost superposed, which confirms the fundamental role of the nonlinear dissipation strategy in obtaining high-quality data. Finally, the effect of the order of the numerical scheme for the viscous fluxes discretization was also assessed; results are reported in figure~\ref{fig:tgv_viscfluxcomp} for the $256^3$ and $128^3$ grids. Consistently with the analysis in the incompressible limit of DeBonis \cite{debonis2013solutions}, a negligible influence is observed when considering discretization orders higher than 4. On the contrary, the use of a $2^\text{nd}$-order discretization produces excessive amounts of viscous dissipation, which for the $256^3$ grid is shown to be larger than the one predicted by the DNS. Furthermore, the $128^3$ grid with $2^\text{nd}$-order viscous fluxes seems to closely follow the DNS profile. This is rather fortuitous, since the larger enstrophy values come from an incorrect (overestimated) spectral energy repartition at the smallest scales. These results should warn about possible misleading interpretations of the flow physics in case of unsuitable numeric ingredients and/or strongly under-resolved simulations, as well as the need to resorting to spectral analyses for the study and characterization of turbulent flows, the classical domain-integrated quantities being not necessarily accurate enough.

%===============================================================================
\subsection{Turbulent boundary layer flows}
A crucial aspect for the numerical simulation of wall-bounded turbulent flows is the ability of the numerical scheme to retain good dissipation and dispersion properties near the boundaries. In this section we consider a compressible, calorically-perfect turbulent boundary layer and we carry out a grid-sensitivity analysis to compare results obtained by means of wall-resolved ILES-like grids with respect to reference DNS data. Next, the same numerical strategy is applied to the simulation of an hypersonic turbulent boundary layer of a multi-species, chemically out-of-equilibrium mixture.
%===============================================================================
\subsubsection{Supersonic turbulent boundary layer}
\label{sec:bl_pfg}
The flow conditions for such a configuration are similar to those investigated by several authors \citep{guarini2000direct,pirozzoli2004directa,pirozzoli2008characterization,pirozzoli2011turbulence,wenzel2018dns}; specifically, a nominal Mach number equal to 2.25, a free-stream temperature of $\SI{65}{K}$ and a free-stream density of $\SI{0.13}{kg/m^3}$. The fluid considered is calorically-perfect air; Sutherland's law is used to compute dynamic viscosity, along with a constant Prandtl number equal to 0.72. The rectangular computational domain is discretized with even spacing in the streamwise ($x$) and spanwise ($z$) direction and grid stretching in the wall-normal ($y$) direction, following the profile:
\begin{equation}\label{eq:gridy}
\frac{y(j)}{L_y} = (1-\alpha)\left(\frac{j-1}{N_y-1}\right)^3 + \alpha \frac{j-1}{N_y-1}
\end{equation}
where $L_y$ and $N_y$ denote the domain length and the number of grid points along $y$-direction, respectively; $j\in [1,N_y]$ and $\alpha{=}0.08$. The total extent of the domain is $L_x \times L_y \times L_z = 1600 \delta^*_\text{in} \times 100 \delta^*_\text{in} \times 20\pi\delta_\text{in}^*$, the initial boundary layer thickness $\delta^*_\text{in}$ being used as length scale.
No-slip and isothermal boundary conditions at a temperature close to the laminar adiabatic value ($T_w = \SI{120.18}{K}$) are applied at the wall, whereas characteristic-based conditions are used for the top and right boundaries and periodic conditions in the spanwise direction. A similarity profile is imposed as inlet condition at a distance $x_0$ from the leading edge, corresponding to Re$_{ \delta^*}=1700$. Transition to turbulence is triggered by means of a suction-and-blowing forcing method; this excitation technique consists in applying a time-and-space-dependent wall normal velocity disturbance of the form:
\begin{equation}
\label{eq:forcing}
\frac{v_w}{u_\infty} =A f(x)g(z)\left[\cos(\omega t + \beta z) + \cos (\omega t - \beta z) \right]
\end{equation}
where $A$, $\omega= \tilde{\omega}\delta_\text{in}^*/c_\infty$ and $\beta=\tilde{\beta}\delta^*_\text{in}$ represent the amplitude, non-dimensional pulsation and spanwise wave number, respectively. Here, the symbol $\tilde{(\bullet)}$ denotes dimensional values and $c_\infty$ is the free-stream speed of sound; $f(x)$ and $g(z)$ are two perturbation-modulation functions defined as in Franko \& Lele \cite{franko2013breakdown}. The forcing strip is located near the inlet, at Re$_{\delta^*}=2000$; the parameters prescribed for the current set of simulations are $A=0.025$, $\omega=0.12$ and $\beta=0.2$. Of note, the spanwise extent of the domain has been selected in order to contain exactly two oblique waves.
Statistics are computed by averaging both in time and in the periodic direction; their collection spans approximately two turnover times and starts after that the initial transient has been evacuated and a statistically-steady state is reached. The sampling time interval is constant and equal to $\Delta t=\Delta\tilde{t} c_\infty /\delta^*_\text{in} \approx \num{3.93e-3}$.

\begin{table}[!tb]
 \caption{List of the computational grids (and associated legend) selected for the supersonic boundary layer, along with the total number of grid points and resolutions in inner units. $\Delta y_w^+$ and $\Delta y_\delta^+$ refer to the wall-normal resolutions at the wall and at the boundary layer edge, respectively. \label{tab:grid}}
 \centering
 \begin{tabular}{cccccccccc}
  \toprule
  & Legend & $N_{tot}$ & $N_x$ & $N_y$ & $N_z$ & $\Delta x^+$ & $\Delta y^+_w$ &  $\Delta y^+_\delta$ & $\Delta z^+$ \\
  \midrule
  DNS    & \protect\markerone           & $ 1.23 \times 10^9$ & 8000 & 300 & 512 & 5.78  & 0.77 & 5.23 & 3.55  \\
  ILES-A & \protect\ldash{orange}       & $ 3.07 \times 10^8$ & 4000 & 300 & 256 & 11.69 & 0.78 & 5.38 & 7.17  \\
  ILES-B & \protect\ldashdotdot{blue}   & $ 1.54 \times 10^8$ & 4000 & 300 & 128 & 11.58 & 0.77 & 4.94 & 14.17 \\
  ILES-C & \protect\ldashdot{red}       & $ 7.68 \times 10^7$ & 2000 & 300 & 128 & 22.58 & 0.76 & 4.66 & 13.85 \\
  ILES-D & \protect\ldashlon{darkgreen} & $ 1.92 \times 10^7$ & 1000 & 300 & 64  & 43.00 & 0.71 & 5.44 & 26.36 \\
  \bottomrule[\heavyrulewidth]
 \end{tabular}
\end{table}
\begin{figure}[!tb]
\centering
\begin{tikzpicture}
      \node[anchor=south west,inner sep=0] (a) at (0,0) {\includegraphics[width=0.44\columnwidth]
      {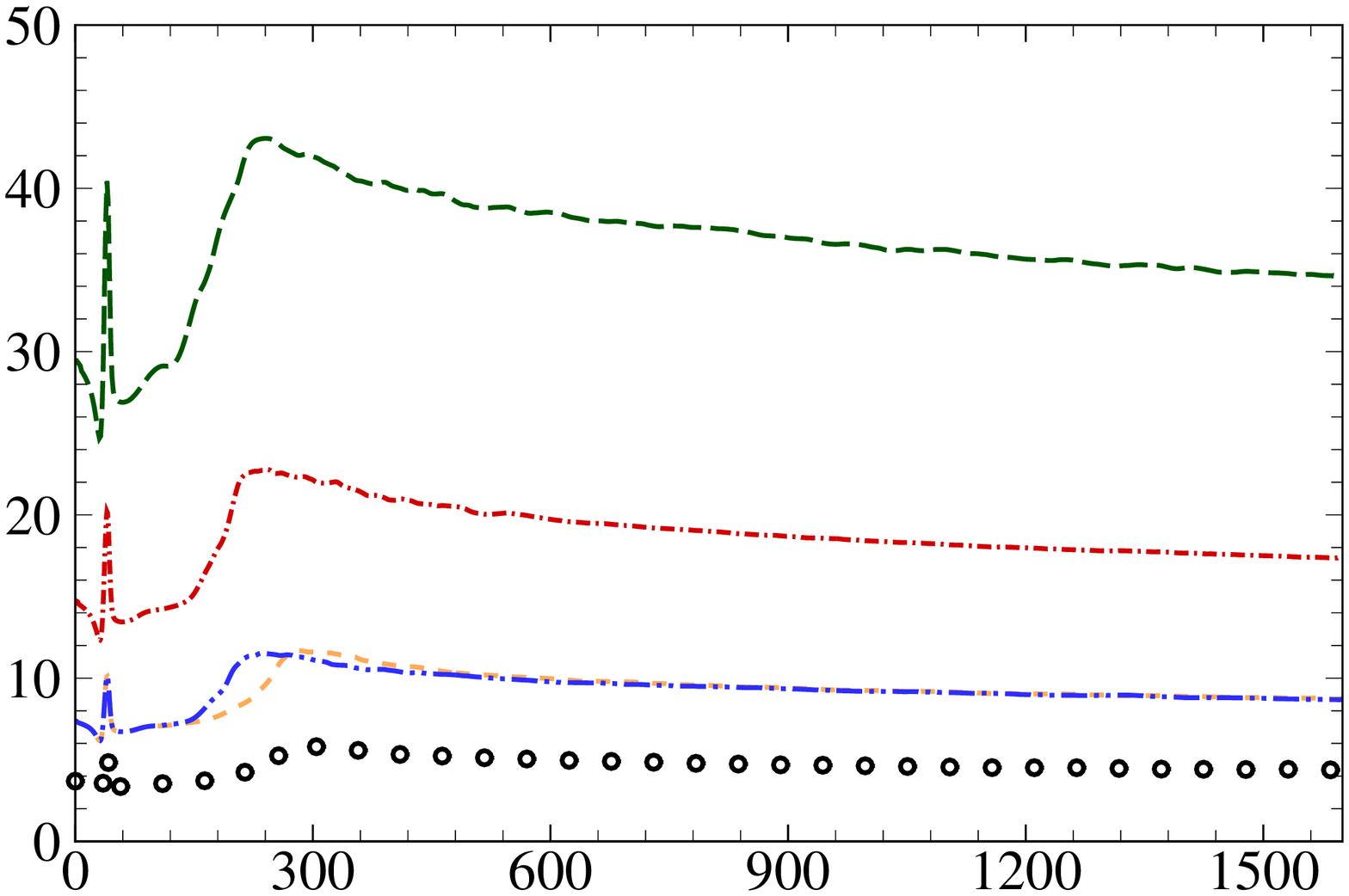}};
      \begin{scope}[x={(a.south east)},y={(a.north west)}]
        \node [align=center] at (0.55,-0.02) {$(x-x_0)/\delta^*_\text{in}$};
        \node [align=center,rotate=90] at (-0.01,0.55)  {\large $\Delta x^+$};
      \end{scope}
\end{tikzpicture}
\begin{tikzpicture}
      \node[anchor=south west,inner sep=0] (a) at (0,0) {\includegraphics[width=0.44\columnwidth]
      {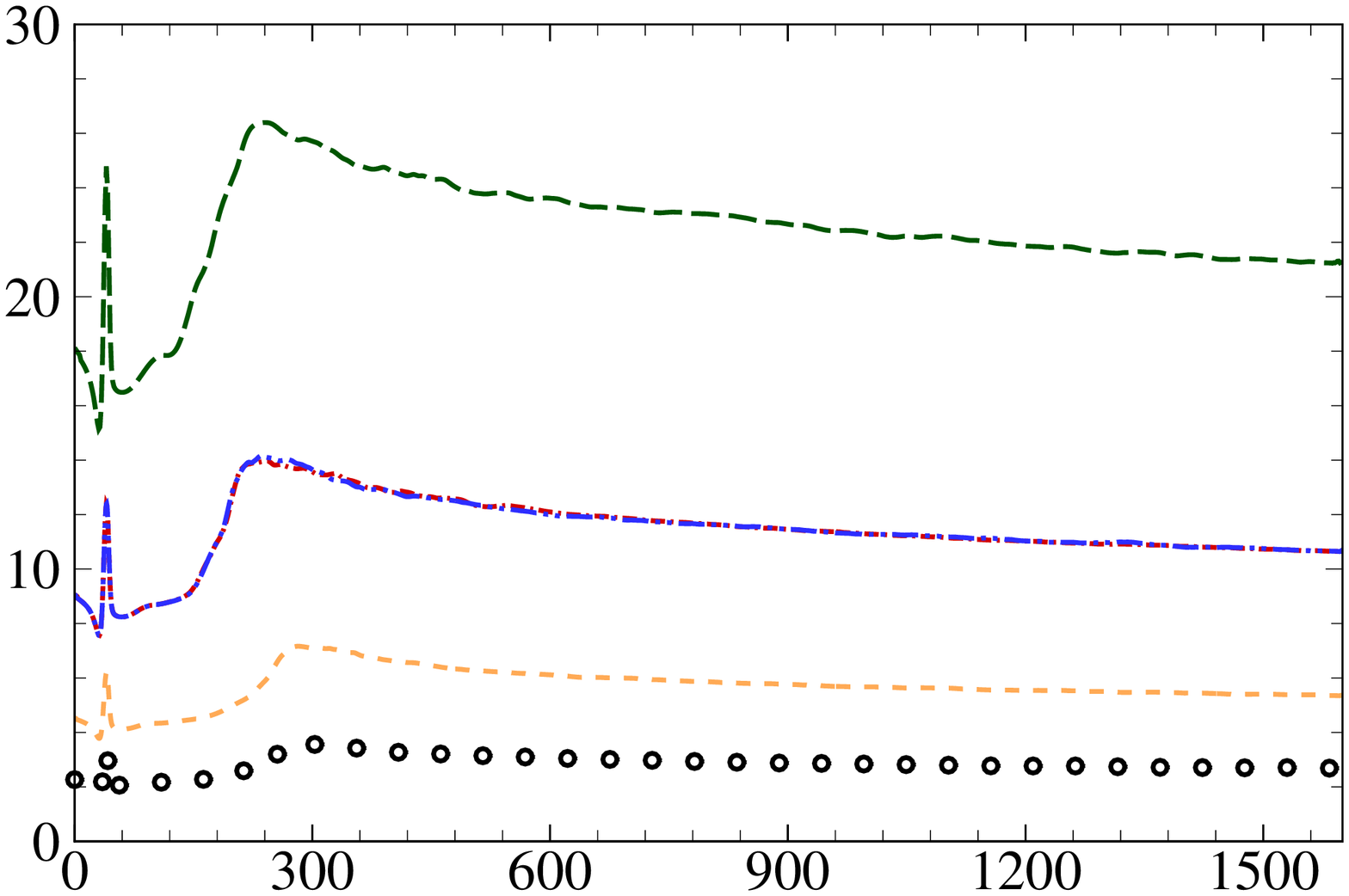}};
      \begin{scope}[x={(a.south east)},y={(a.north west)}]
        \node [align=center] at (0.55,-0.02) { $(x-x_0)/\delta^*_\text{in}$};
        \node [align=center,rotate=90] at (-0.01,0.55)  {\large $\Delta z^+$};
      \end{scope}
\end{tikzpicture}
\vspace{-0.5cm}
\caption{Resolutions of the computational grids along the streamwise direction in wall units. Line legend as in table~\ref{tab:grid}. }
\label{fig:pfg0}
\end{figure}

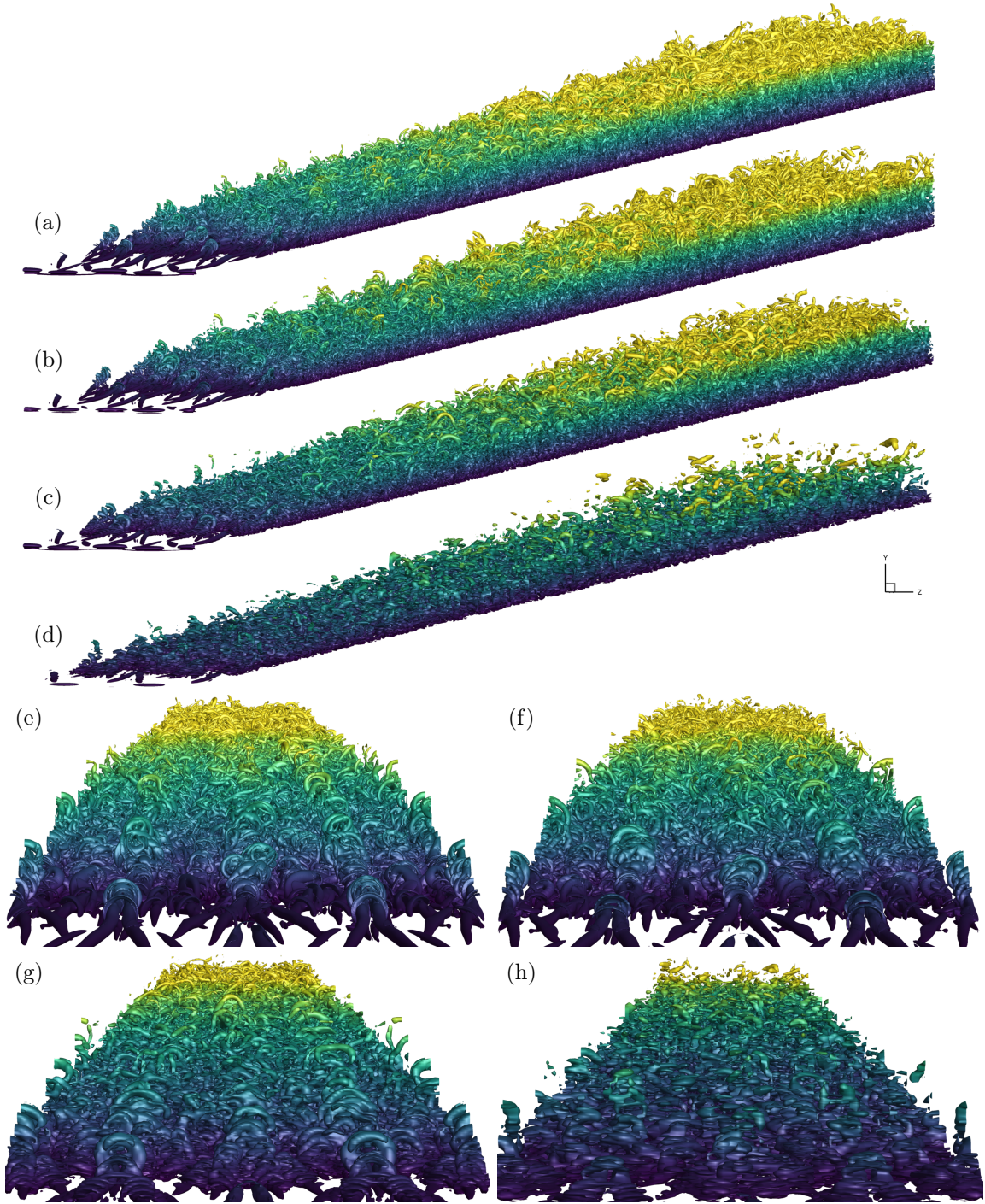
\begin{figure}[!tb]
\vspace{-1cm}
\centering
\begin{tikzpicture}
    \node[anchor=south west,inner sep=0] (a) at (0,0) {\includegraphics[width=\textwidth, trim={5 5 150 5}, clip]{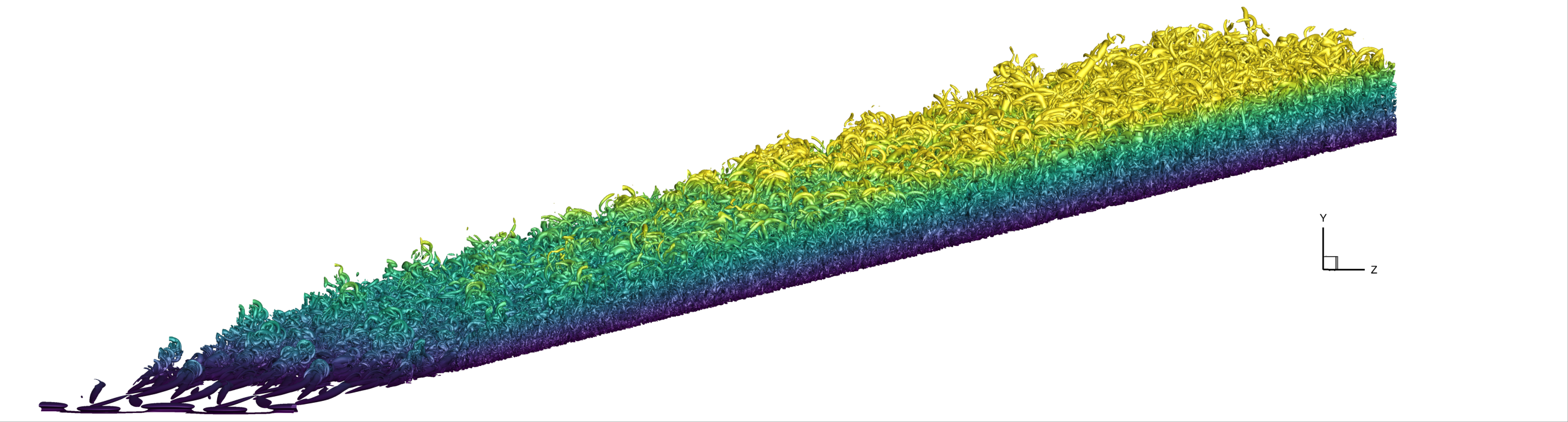}};
    \begin{scope}[x={(a.south east)},y={(a.north west)}]
        \node [align=center] at (0.05,0.2) {(a)};
    \end{scope}
\end{tikzpicture}

\vspace{-2.2cm}
\begin{tikzpicture}
    \savebox\mysavebox{\includegraphics[width=\textwidth, trim={5 5 150 5}, clip]{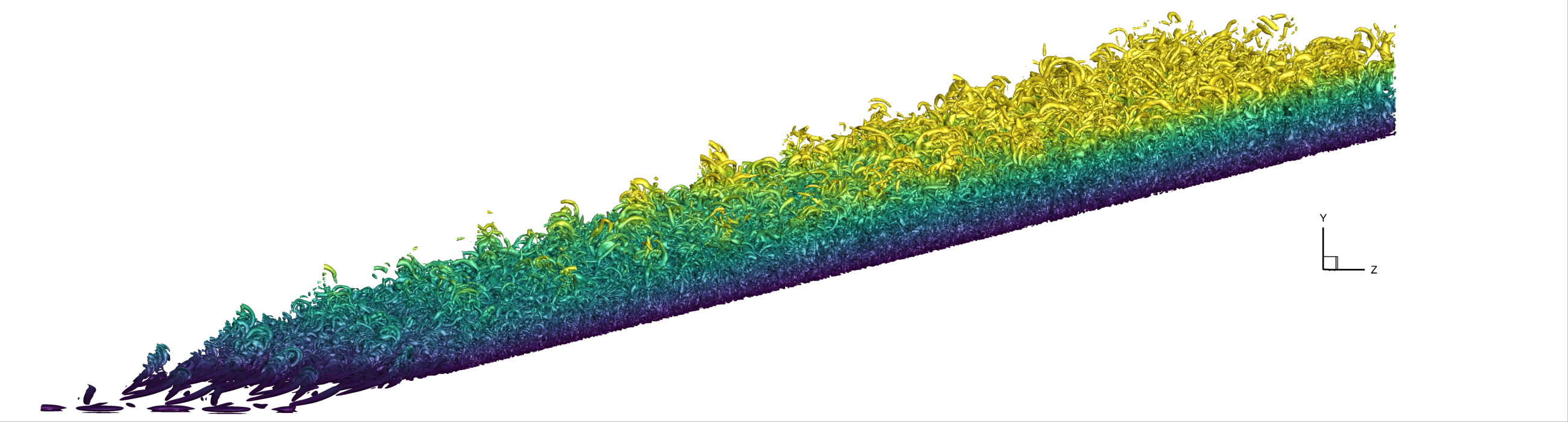}}
    \path [use as bounding box] rectangle (\wd\mysavebox, \ht\mysavebox);
    \path [clip] (0.0,1.0) -- (15:100) -- (-90:100) -- cycle;
    \node [anchor=south west, inner sep=0pt, outer sep=0pt] at (0,0) {\usebox\mysavebox};
    \begin{scope}[x={(a.south east)},y={(a.north west)}]
        \node [align=center] at (0.05,0.2) {(b)};
    \end{scope}
\end{tikzpicture}

\vspace{-2.2cm}
\begin{tikzpicture}
    \savebox\mysavebox{\includegraphics[width=\textwidth, trim={5 5 150 5}, clip]{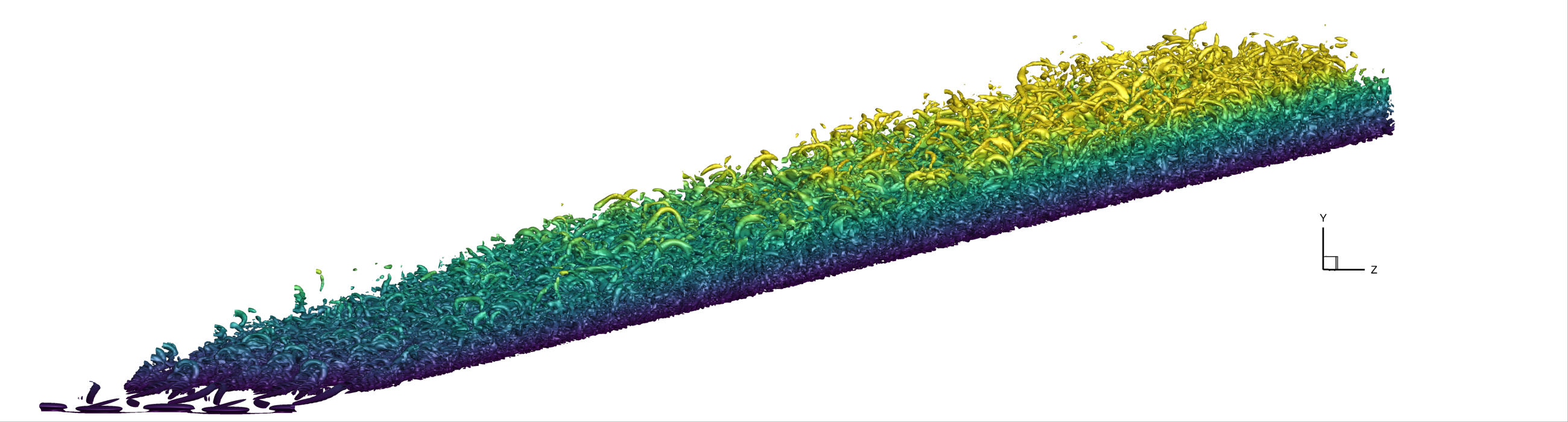}}
    \path [use as bounding box] rectangle (\wd\mysavebox, \ht\mysavebox);
    \path [clip] (0.0,1.0) -- (15:100) -- (-90:100) -- cycle;
    \node [anchor=south west, inner sep=0pt, outer sep=0pt] at (0,0) {\usebox\mysavebox};
    \begin{scope}[x={(a.south east)},y={(a.north west)}]
        \node [align=center] at (0.05,0.2) {(c)};
    \end{scope}
\end{tikzpicture}

% \vspace{-2.2cm}
% \begin{tikzpicture}
%     \savebox\mysavebox{\includegraphics[width=\textwidth, trim={5 5 150 5}, clip]{Q2_ILES-C.png}}
%     \path [use as bounding box] rectangle (\wd\mysavebox, \ht\mysavebox);
%     \path [clip] (0.0,1.0) -- (15:100) -- (-90:100) -- cycle;
%     \node [anchor=south west, inner sep=0pt, outer sep=0pt] at (0,0) {\usebox\mysavebox};
% \end{tikzpicture}

\vspace{-2.2cm}
\begin{tikzpicture}
    \savebox\mysavebox{\includegraphics[width=\textwidth, trim={5 5 150 5}, clip]{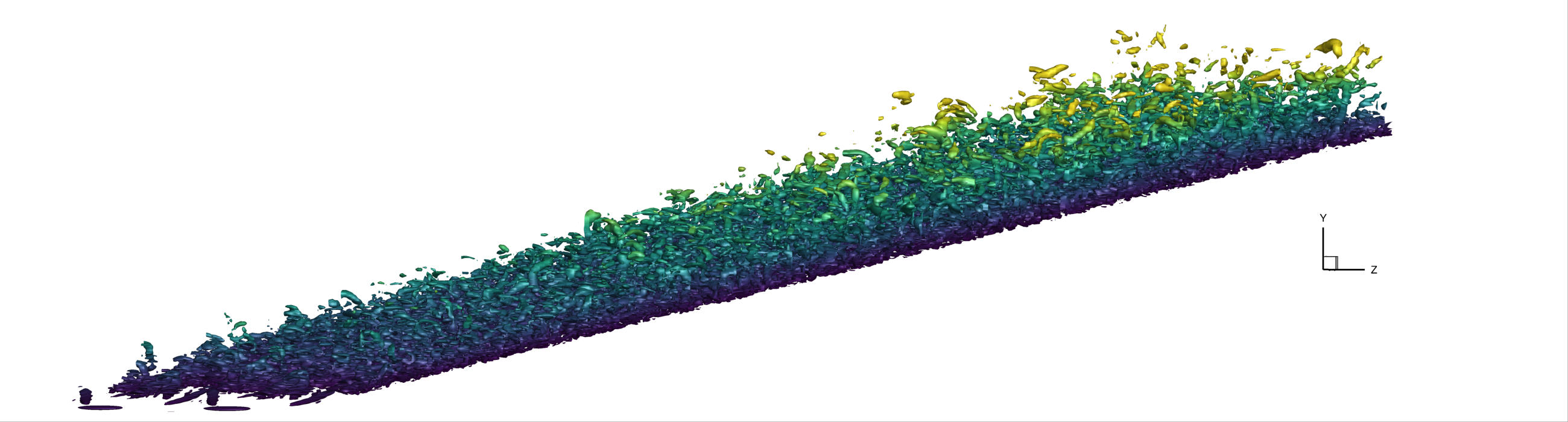}}
    \path [use as bounding box] rectangle (\wd\mysavebox, \ht\mysavebox);
    \path [clip] (0.0,1.0) -- (15:100) -- (-90:100) -- cycle;
    \node [anchor=south west, inner sep=0pt, outer sep=0pt] at (0,0) {\usebox\mysavebox};
    \begin{scope}[x={(a.south east)},y={(a.north west)}]
        \node [align=center] at (0.05,0.2) {(d)};
    \end{scope}
\end{tikzpicture}

\begin{tikzpicture}
    \node[anchor=south west,inner sep=0] (a) at (0,0) {\includegraphics[width=0.49\textwidth, trim={310 280 320 400}, clip]{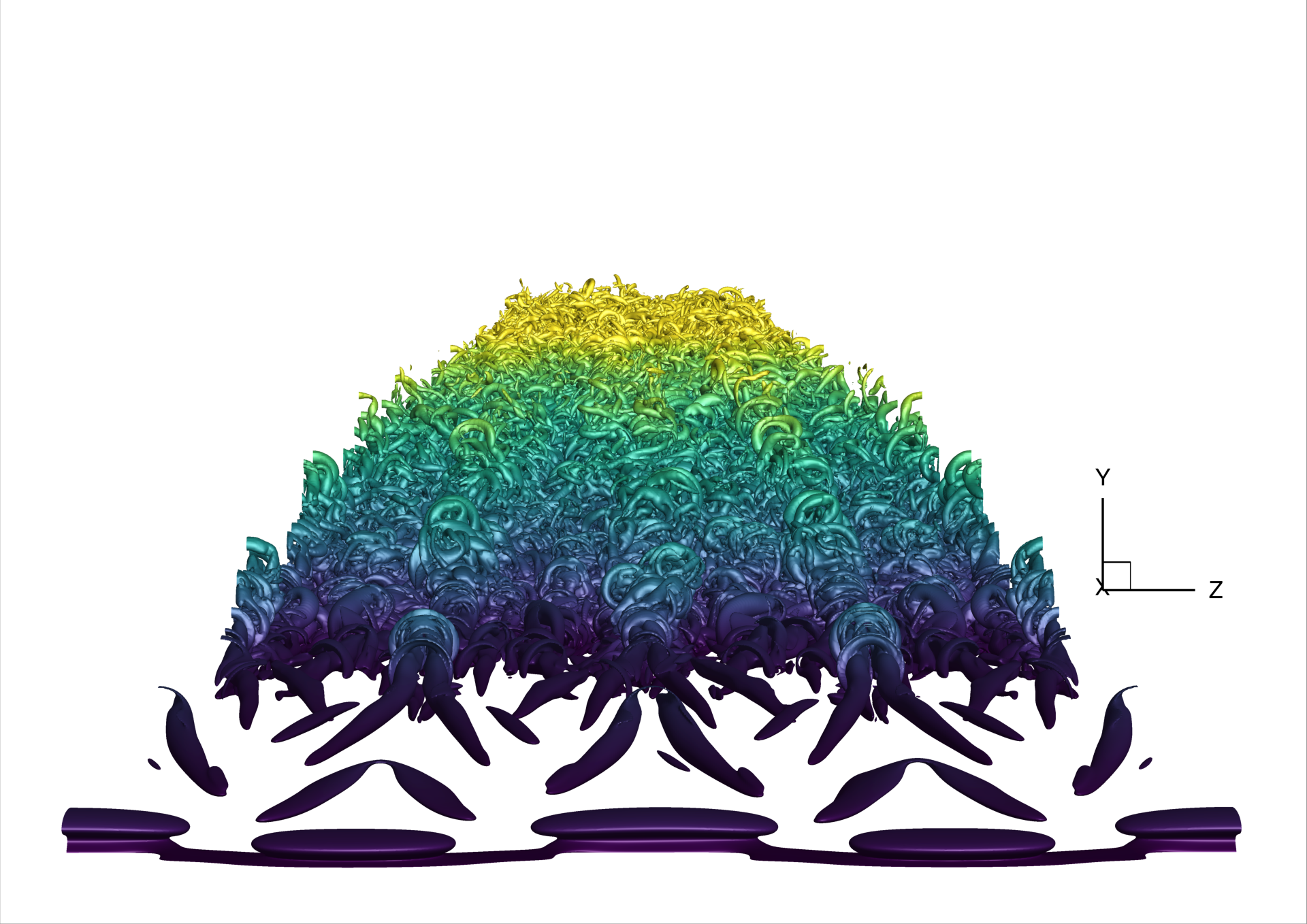}};
    \begin{scope}[x={(a.south east)},y={(a.north west)}]
        \node [align=center] at (0.05,0.9) {(e)};
    \end{scope}
\end{tikzpicture}
\begin{tikzpicture}
    \node[anchor=south west,inner sep=0] (a) at (0,0) {\includegraphics[width=0.49\textwidth, trim={310 280 320 400}, clip]{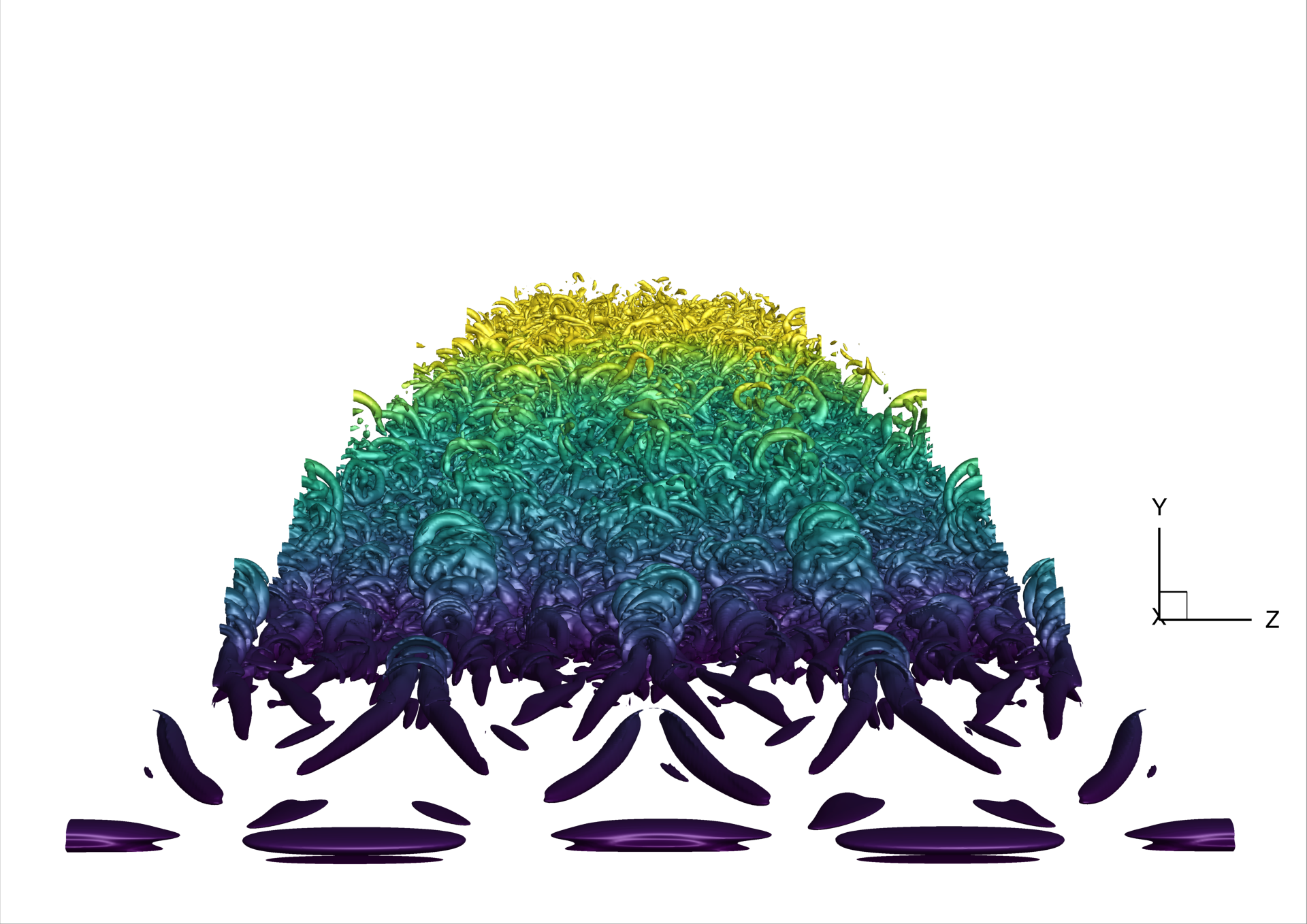}};
    \begin{scope}[x={(a.south east)},y={(a.north west)}]
        \node [align=center] at (0.05,0.9) {(f)};
    \end{scope}
\end{tikzpicture}

\begin{tikzpicture}
    \node[anchor=south west,inner sep=0] (a) at (0,0) {\includegraphics[width=0.49\textwidth, trim={310 280 320 400}, clip]{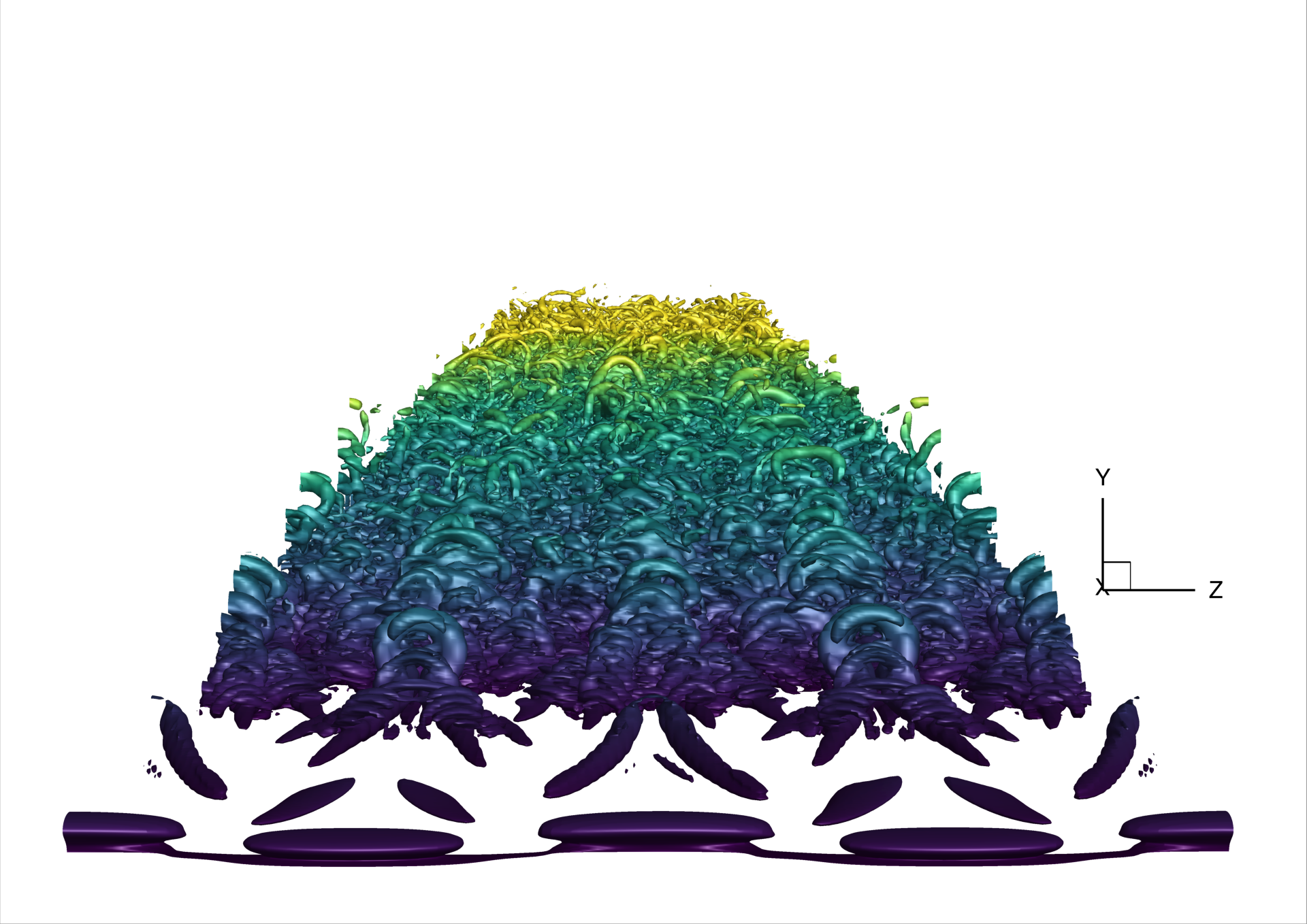}};
    \begin{scope}[x={(a.south east)},y={(a.north west)}]
        \node [align=center] at (0.05,0.9) {(g)};
    \end{scope}
\end{tikzpicture}
\begin{tikzpicture}
    \node[anchor=south west,inner sep=0] (a) at (0,0) {\includegraphics[width=0.49\textwidth, trim={310 280 320 400}, clip]{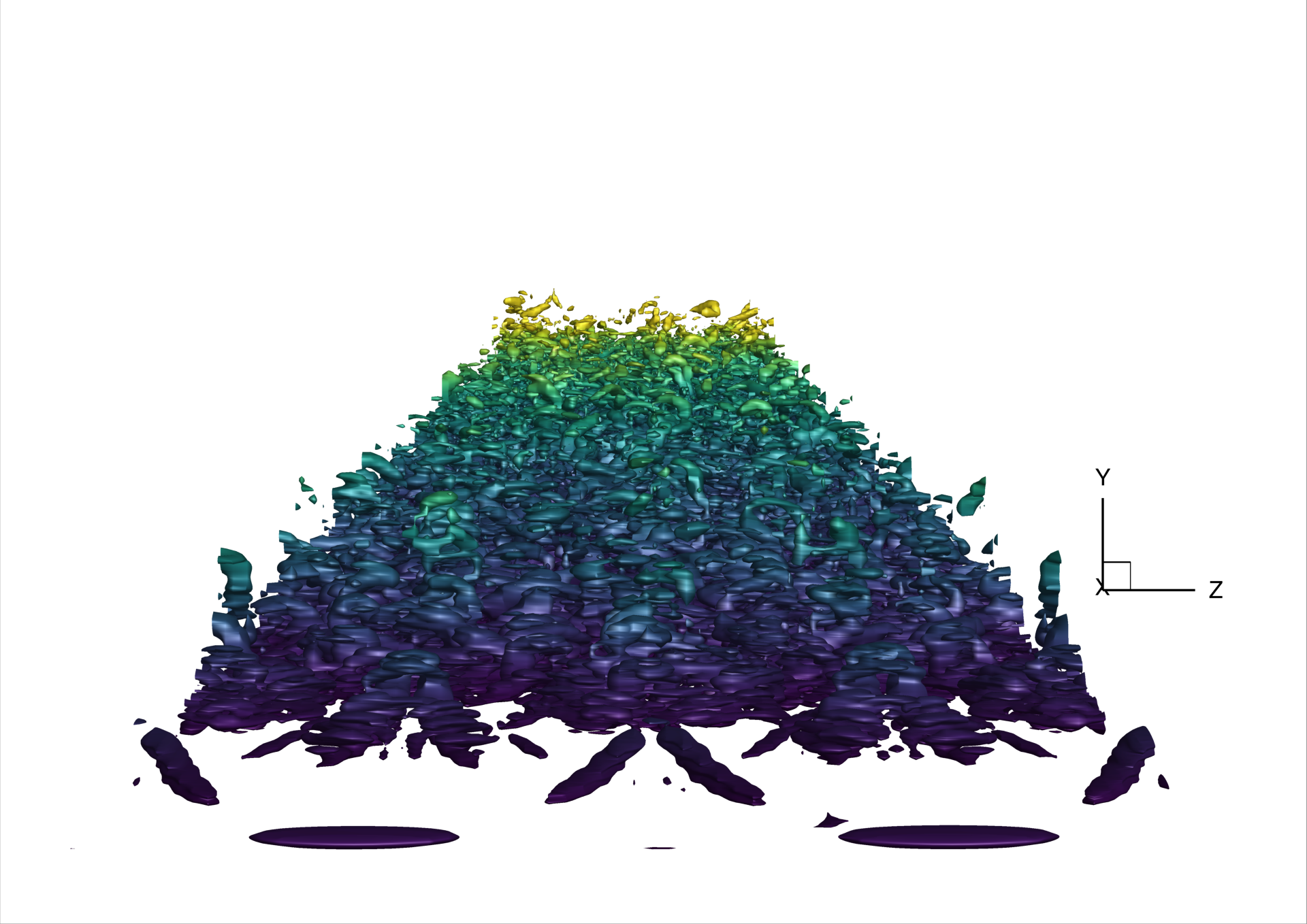}};
    \begin{scope}[x={(a.south east)},y={(a.north west)}]
        \node [align=center] at (0.05,0.9) {(h)};
    \end{scope}
\end{tikzpicture}
\caption{Lateral and frontal views of instantaneous snapshots for the $M=2.25$ supersonic boundary layer. Isosurfaces of the $Q$-criterion coloured by the distance from the wall. (a, e): DNS; (b, f): ILES-A; (c, g): ILES-B; (d, h): ILES-D.}
\label{fig:QcritM2}
\end{figure}

Five different computational grids have been considered and are listed in table~\ref{tab:grid}.
A reference solution is generated by means of a well-resolved DNS and is compared to data obtained on coarser, ILES-like grids, built by progressively halving the total number of points in the streamwise and spanwise directions. On the contrary, the wall-normal grid point distribution is kept unaltered such that discrepancies with respect to DNS data can directly be attributed to excessively coarse discretizations in the other two directions. The least-refined grid has less than 20 million grid points; that is, it is 64 times coarser than the grid used for DNS, which counts more than 1.2 billion grid points. Note that the $(\bullet)^+$ quantities listed in table~\ref{tab:grid} have been evaluated at the point where the skin friction coefficient peaks, corresponding to the most stringent requirements in terms of resolution. The overall streamwise evolutions of $\Delta x^+$ and $\Delta z^+$ for the five computational meshes, reported in figure~\ref{fig:pfg0}, clearly show that their values rapidly grow during transition and become approximately constant in the fully turbulent region. The influence of resolution can be appreciated from the lateral and frontal instantaneous views of the $Q$-criterion shown in figure~\ref{fig:QcritM2} for all cases but ILES-C (not shown), which bears a strong resemblance with case ILES-B. DNS (panels a and e) and ILES-A (panels b and f) are almost indistinguishable; fine-scale structures are properly resolved and the streamwise development of the boundary layers is alike in terms of integral thicknesses.
Some visual discrepancies start to appear for case ILES-B, in which both the transition region and the smallest features of the flow are shown to be more grainy. Lastly, the coarser grid of case ILES-D (panels d and h) results in exessive damping of the turbulent motions: hairpin-like structures are smeared out and the occurrence of coherent structures becomes much more sporadic.

%===============================================================================
%
\begin{figure}[!tb]
\centering
\begin{tikzpicture}
      \node[anchor=south west,inner sep=0] (a) at (0,0) {\includegraphics[width=0.96\columnwidth,trim={10 0 0 0},clip]
      {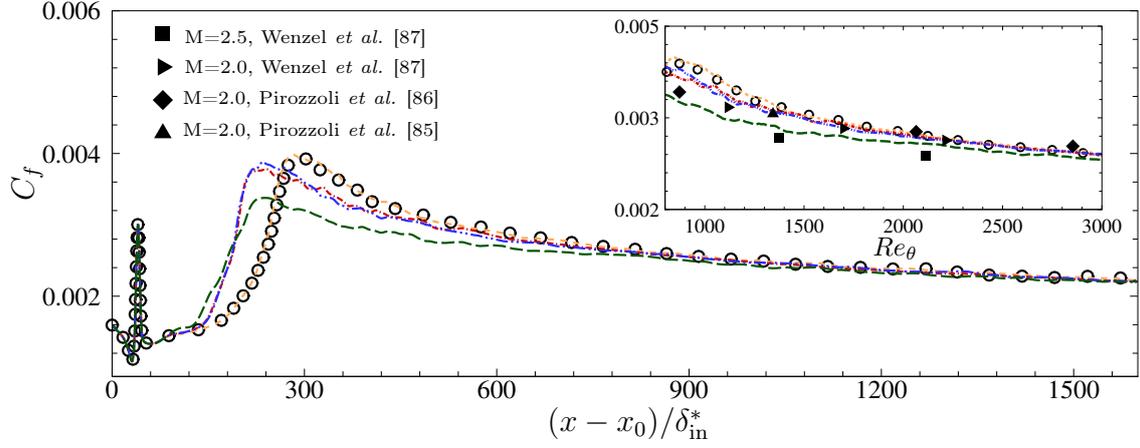}};
      \begin{scope}[x={(a.south east)},y={(a.north west)}]
        \node [align=center] at (0.53,0.00) {\large $(x-x_0)/\delta^*_\text{in}$};
        \node [align=center] at (0.77,0.40) {$Re_\theta$};
        \node [align=center,rotate=90] at (-0.01,0.55)  {\large $C_f$};
        \node [align=center] at (0.24 ,0.88) {\scriptsize M=2.5, Wenzel \emph{et al.} \cite{wenzel2018dns}};
        \node [align=center] at (0.24 ,0.81) {\scriptsize M=2.0, Wenzel \emph{et al.} \cite{wenzel2018dns}};
        \node [align=center] at (0.245,0.74) {\scriptsize M=2.0, Pirozzoli \emph{et al.} \cite{pirozzoli2011turbulence}};
        \node [align=center] at (0.245,0.668){\scriptsize M=2.0, Pirozzoli \emph{et al.} \cite{pirozzoli2008characterization}};
      \end{scope}
\end{tikzpicture}
\caption{Skin friction coefficient $C_f$ as a function of $(x-x_0)/\delta^*_\text{in}$. Sub-figure: $C_f$ as a function of $Re_\theta$ in the fully turbulent zone. Line legend as in table~\ref{tab:grid}.}
\label{fig:pfg1}
\end{figure}
\begin{figure}[!tb]
\centering
\begin{tikzpicture}
      \node[anchor=south west,inner sep=0] (a) at (0,0) {\includegraphics[width=0.47\columnwidth]
      {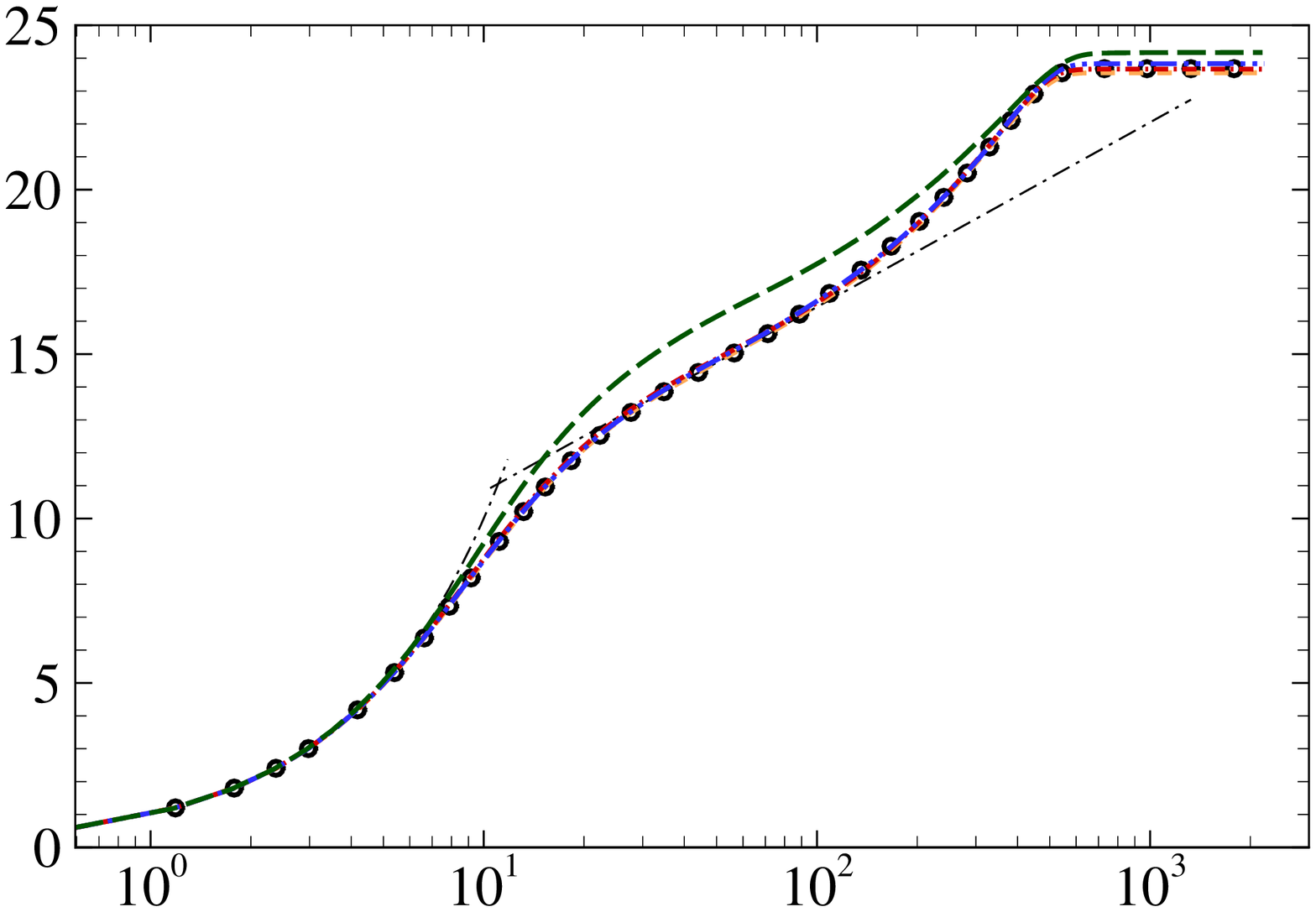}};
      \begin{scope}[x={(a.south east)},y={(a.north west)}]
        \node [align=center] at (0.17,0.87) {(a)};
        \node [align=center] at (0.55,0.03) {\large $y^+$};
        \node [align=center,rotate=90] at (-0.03,0.55)  {\large $u_{VD}^+$};
        \node [align=center] at (0.27,0.35) {\small $u^+=y^+$};
        \node [align=center,rotate=30] at (0.71,0.68) {\small $u^+=\frac{1}{0.41}\text{ln}(y^+)+5.2$};
      \end{scope}
\end{tikzpicture}
\hspace{-0.5cm}
\begin{tikzpicture}
      \node[anchor=south west,inner sep=0] (a) at (0,0) {\includegraphics[width=0.47\columnwidth]
      {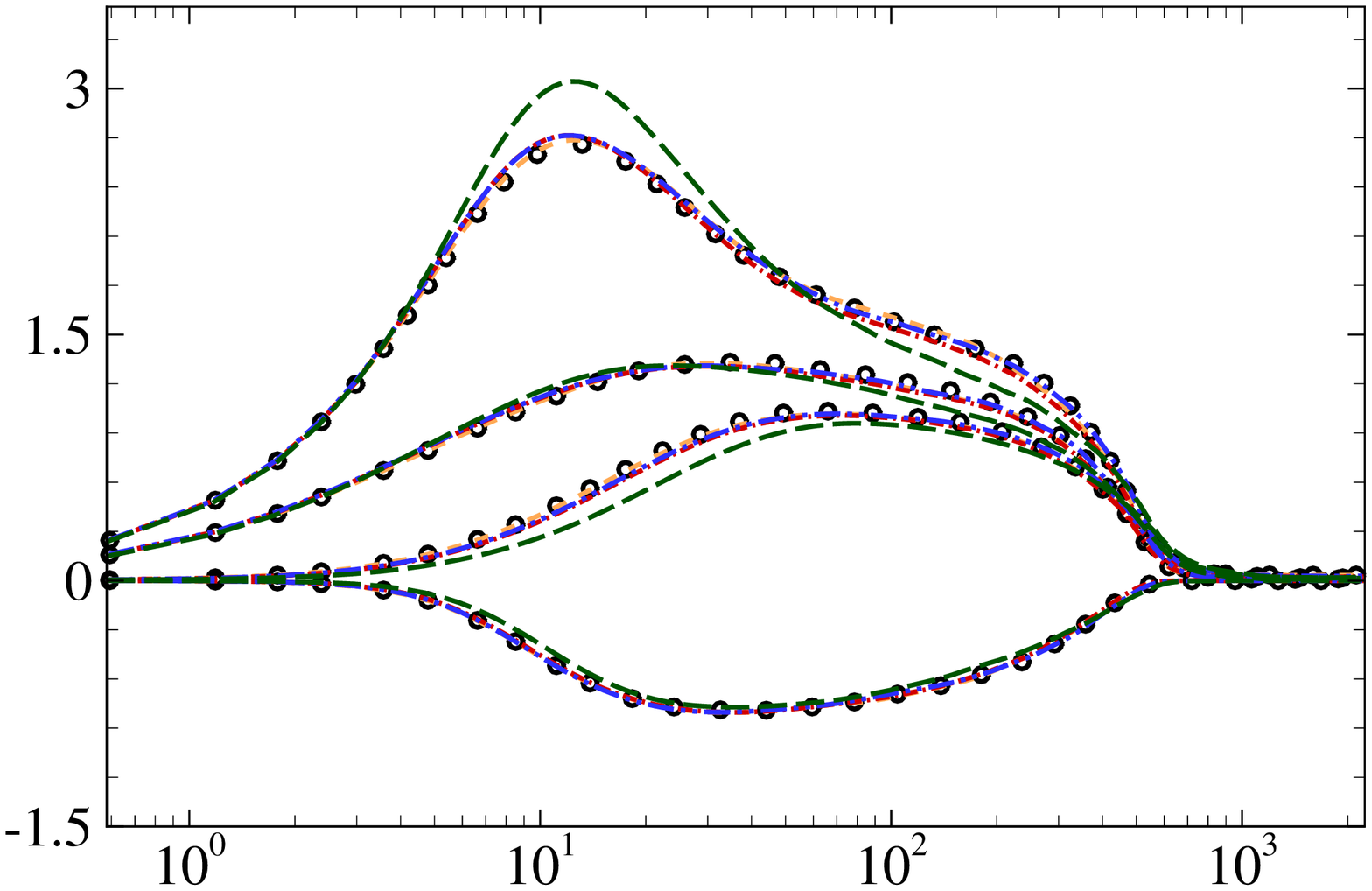}};
      \begin{scope}[x={(a.south east)},y={(a.north west)}]
        \node [align=center] at (0.17,0.87) {(b)};
        \node [align=center] at (0.55,0.03)  {\large $y^+$};
        \node [align=center] at (0.3,0.82)   {\scriptsize $u_\text{rms}^+$};
        \node [align=center] at (0.45,0.64)  {\scriptsize $w_\text{rms}^+$};
        \node [align=center] at (0.50,0.42)  {\scriptsize $v_\text{rms}^+$};
        \node [align=center] at (0.40,0.20)  {\scriptsize $-uv^+$};
        \node [align=center,rotate=90] at (-0.03,0.55)  {\large $u_{i,\text{rms}}^+$};
      \end{scope}
\end{tikzpicture}
\begin{tikzpicture}
      \node[anchor=south west,inner sep=0] (a) at (0,0) {\includegraphics[width=0.47\columnwidth]
      {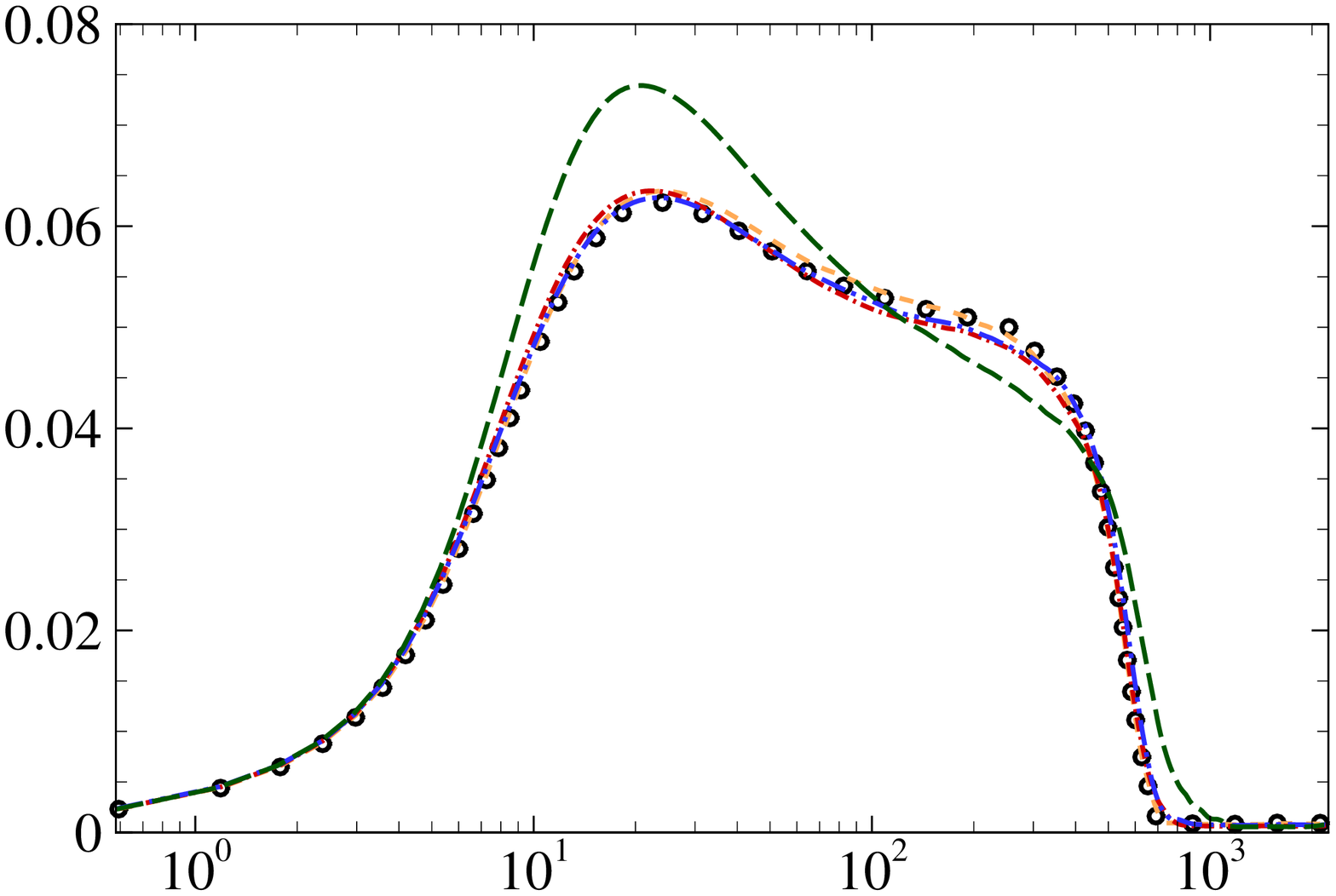}};
      \begin{scope}[x={(a.south east)},y={(a.north west)}]
        \node [align=center] at (0.17,0.87) {(c)};
        \node [align=center] at (0.55,0.01)  {\large $y^+$};
        \node [align=center,rotate=90] at (-0.03,0.55)  {\large $T_\text{rms}/T_w$};
      \end{scope}
\end{tikzpicture}
\hspace{-0.5cm}
\begin{tikzpicture}
      \node[anchor=south west,inner sep=0] (a) at (0,0) {\includegraphics[width=0.47\columnwidth]
      {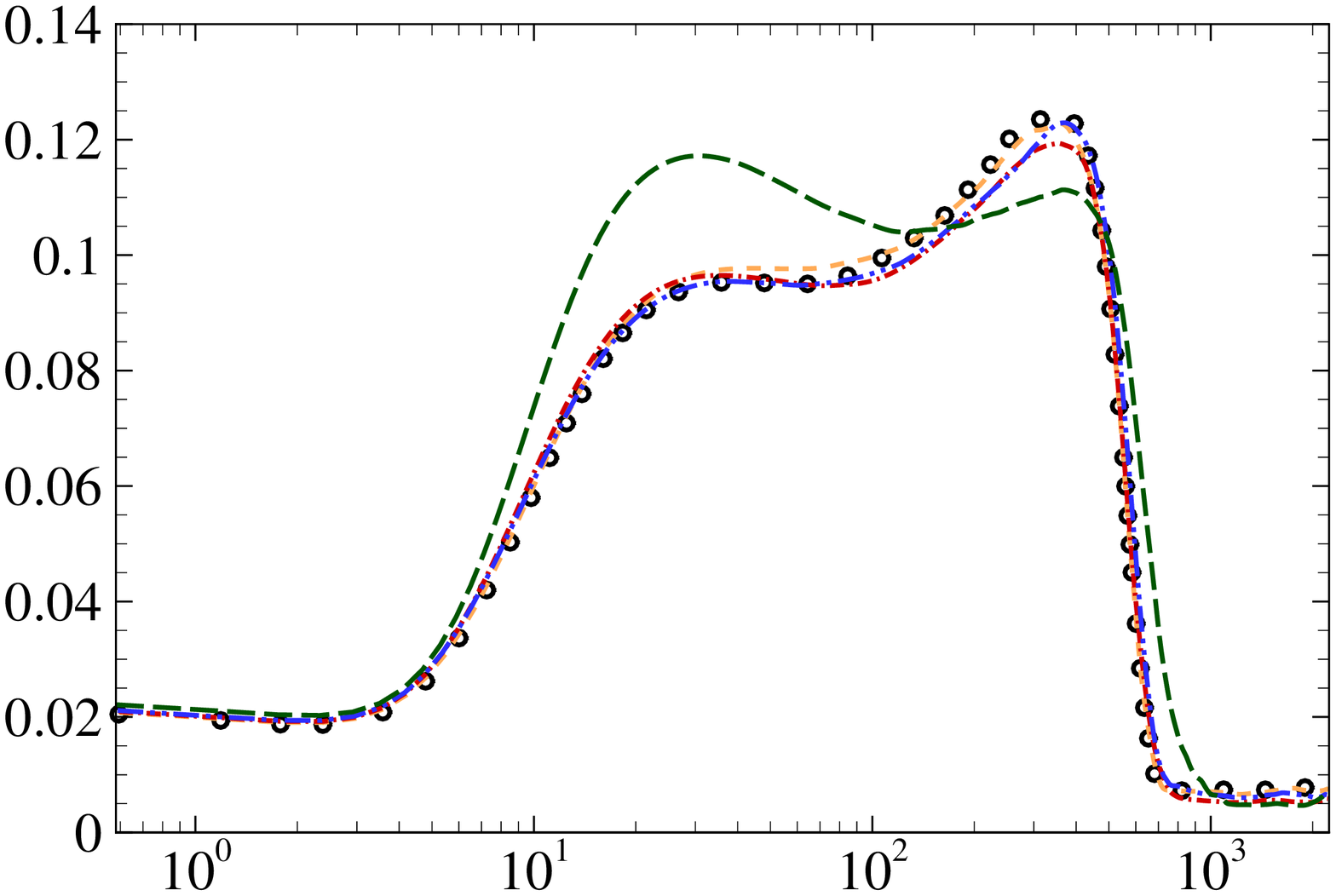}};
      \begin{scope}[x={(a.south east)},y={(a.north west)}]
        \node [align=center] at (0.17,0.87) {(d)};
        \node [align=center] at (0.55,0.01)  {\large $y^+$};
        \node [align=center,rotate=90] at (-0.03,0.55)  {\large $\rho_\text{rms}/\rho_w$};
      \end{scope}
\end{tikzpicture}
\caption{Wall-normal profiles of the Van-Driest-transformed longitudinal velocity (a), r.m.s. of velocity (b), temperature (c) and density (d), extracted at $Re_\theta=3500$. Line legend as in table~\ref{tab:grid}.}
\label{fig:pfg2}
\end{figure}

The streamwise profile of the skin friction coefficient $C_f$ is displayed in figure~\ref{fig:pfg1}, along with a close-up in the fully-turbulent region of the same quantity as a function of the momentum-thickness-based Reynolds number $Re_\theta$. DNS and ILES-A results are almost superposed both in the laminar-to-turbulent transition and fully-developed turbulent regions, confirming that grid resolutions of $\Delta x^+ < 15 $ and $\Delta z^+ < 10$ are fine enough to capture the main flow features. On the contrary, ILES-B and ILES-C cases present some discrepancies in the breakdown to turbulence, the $C_f$ peak being slightly smaller and moved towards the domain inlet. A comparison with ILES-A underlines the importance of keeping $\Delta z^+ < 10$ for the spanwise resolution, whereas the influence of doubling the streamwise mesh size (from $\Delta x^+ \approx 11.5$ to $\Delta x^+ \approx 22.5$) is shown to be much less pronounced. Lastly, the ILES-D grid results in wrong predictions of both transitional and fully turbulent regions.
Despite the much coarser resolutions with respect to the DNS, the fully-turbulent regions tend to match DNS predictions as $Re_\theta$ increases, mainly due to boundary-layer thickening. The $C_f$ evolution is indeed in good agreement with values extracted from numerical simulations of similar configurations available in literature \cite{pirozzoli2008characterization,pirozzoli2011turbulence,wenzel2018dns}, the slight discrepancies being related to different free-stream Mach numbers and forcing locations.
In figure~\ref{fig:pfg2}a, the Van-Driest-transformed velocity profiles collapse reasonably well for all the computational meshes except for ILES-D, in which the log-region values are shown to be largely overestimated (in accordance with observations of previous studies about the resolution limits for turbulent boundary layers \cite{sayadi2012large,poggie2015resolution}). Root-main-square values for each component of velocity, temperature and density are displayed in figures~\ref{fig:pfg2}b, c and d, respectively; here, the same conclusions hold: all the computational grids predict wall-normal profiles very close (or superposed) to the DNS solution, the only exception being ILES-D case which registers large mismatches in proximity of the inner peak.
The current grid resolution assessment confirms then the reliability of the numerical strategy in a wall-resolved ILES framework, pointing out that i) $\Delta x^+ < 15 $ and $\Delta z^+ < 10$ should be used to obtain DNS-like results, and ii) slightly coarser resolutions may be considered when focusing the analysis only on fully-turbulent regions.

%===============================================================================
\subsubsection{Hypersonic turbulent boundary layer with finite-rate chemistry}
\label{sec:bl_reac}
The last test case here considered points out the capabilities of the numerical code in reproducing turbulent configurations of high-$T$ flows, by considering a hypersonic, chemically out-of-equilibrium boundary layer undergoing laminar-to-turbulent transition.
The thermodynamic conditions are similar to those adopted in several stability studies (see Refs. \cite{malik1991real, hudson1997linear, perraud1999studies, franko2010effects, marxen2014direct, miro2018diffusion}); specifically, the imposed free-stream values are $M_\infty=10$, $T_\infty=\SI{350}{K}$ and $p_\infty=\SI{3596}{Pa}$, in conjunction with an adiabatic wall. These extreme conditions lead to large temperature values at the wall (of the order of $\approx \SI{5000}{K}$), thereby promoting a strong chemical activity inside the boundary layer.
The extent of the computational domain is $L_x \times L_y \times L_z = 8000 \delta^*_\text{in} \times 320 \delta^*_\text{in} \times 100 \pi\delta_\text{in}^*$, discretized with $N_x \times N_y \times N_z = 5520 \times 256 \times 240$ points. The same stretching function shown in equation~\eqref{eq:gridy} is used to generate the wall-normal grid distribution, with $\alpha=0.13$. Outside the boundary layer, air in equilibrium at its free-stream conditions is prescribed (namely, $Y_{\text{N}_2}=0.767082$ and $Y_{\text{O}_2}=0.232918$, similar to Marxen \textit{et al.} \cite{ marxen2014direct}), whereas a non-catalytic boundary condition is imposed at the wall. The locally self-similar profile, computed under finite-rate chemistry assumption, is prescribed as inlet condition at a distance corresponding to Re$_{ \delta^*}=6375$. Further details about the computation of the local self-similar solution are reported in~\ref{app:blasius}.\\
Differently from the supersonic case presented in section~\ref{sec:bl_pfg}, the forcing is located at Re$_{ \delta^*}=13880$ and the suction-and-blowing forcing function reads:
\begin{equation}
\label{eq:forcing_cne}
\frac{v_w}{u_\infty} =f(x)g(z) \sum_{m=1}^{N_\text{mode}} A_m \sin(\omega_m t + \beta_m z),
\end{equation}
the collection of $A_m$, $\omega_m$ and $\beta_m$ for each mode being listed in table~\ref{tab:forcing}. Lastly, a sponge layer is applied near the inlet to prevent abrupt distortions caused by the high Mach number. Statistics have been collected for approximately one turnover time after having reached a statistically steady state; the sampling time interval is equal to $\Delta t_\text{s} c_\infty /\delta^*_\text{in}= \num{3e-2}$.

\begin{table}[!tb]
 \caption{Parameters of the modes excited by the forcing function in equation~\eqref{eq:forcing_cne}: non-dimensional amplitude, frequency and spanwise wave number. \label{tab:forcing}}
 \centering
 \begin{tabular}{|c|ccc|cc|cc|cc|cc|cc|}
  \hline
  Mode     & 1  & 2 & 3 & 4 & 5 & 6 & 7 & 8 & 9 & 10 & 11 & 12 & 13  \\
  \hline
  $A_m \,\,(\times 10^{-3})$ & 50& 2.5 & 2.5 & 2.5 & 2.5 & \multicolumn{2}{c|}{2.5} & \multicolumn{2}{c|}{2.5} & \multicolumn{2}{c|}{2.5} & \multicolumn{2}{c|}{2.5} \\
  $\omega_m$ & 1.71 & 0.855 & 0 & 0 & 0 & \multicolumn{2}{c|}{1.71} & \multicolumn{2}{c|}{0.855} & \multicolumn{2}{c|}{1.71} & \multicolumn{2}{c|}{1.71} \\
  $\beta_m$  & 0 & 0 & $0.2$ & $0.4$ & $0.6$ & \multicolumn{2}{c|}{$\pm 0.2$} & \multicolumn{2}{c|}{$\pm 0.2$} & \multicolumn{2}{c|}{$\pm 0.4$} & \multicolumn{2}{c|}{$\pm 0.6$}  \\
  %  & 1.71   & 0.05                & 0    \\
  %  & 0.855  & $2.50\times 10^{-3}$ & 0    \\
  %  & 0      & $2.50\times 10^{-3}$ & +0.2 \\
  %  & 0      & $2.50\times 10^{-3}$ & +0.4 \\
  %  & 0      & $2.50\times 10^{-3}$ & +0.6 \\
  %  & 1.71   & $2.50\times 10^{-3}$ & -0.2 \\
  %  & 1.71   & $2.50\times 10^{-3}$ & +0.2 \\
  %  & 0.855  & $2.50\times 10^{-3}$ & -0.2 \\
  %  & 0.855  & $2.50\times 10^{-3}$ & +0.2 \\
  %  & 1.71   & $2.50\times 10^{-3}$ & -0.4 \\
  %  & 1.71   & $2.50\times 10^{-3}$ & +0.4 \\
  %  & 1.71   & $2.50\times 10^{-3}$ & -0.6 \\
  %  & 1.71   & $2.50\times 10^{-3}$ & +0.6 \\
  \hline
 \end{tabular}
\end{table}

\begin{figure}[!tb]
\centering
\begin{tikzpicture}
      \node[anchor=south west,inner sep=0] (a) at (0,0) {\includegraphics[width=0.46\columnwidth]
      {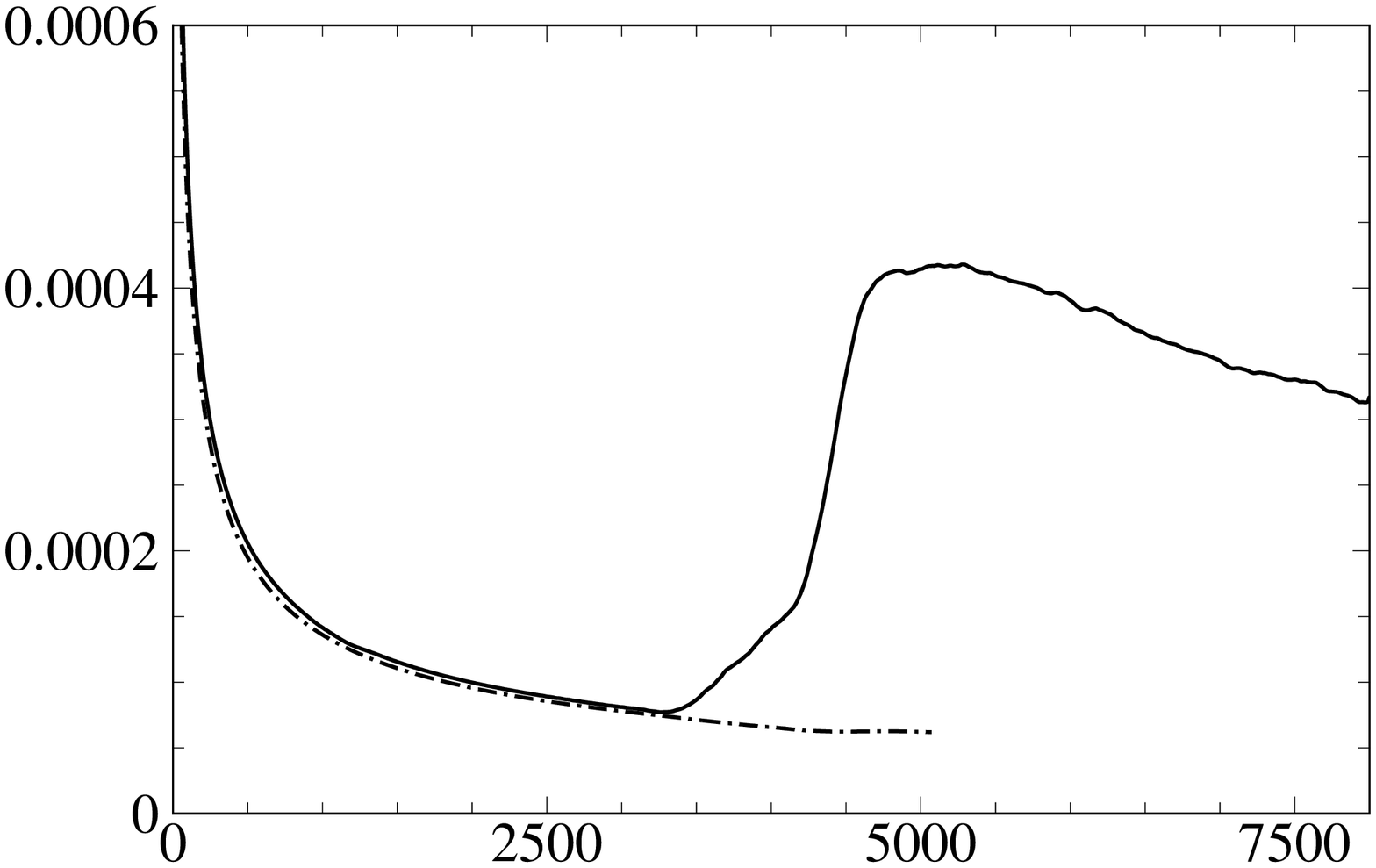}};
      \begin{scope}[x={(a.south east)},y={(a.north west)}]
        \node [align=center] at (-0.02,0.9) {(a)};
        \node [align=center] at (0.55,-0.02) {$(x-x_0)/\delta^*_\text{in}$};
        \node [align=center,rotate=90] at (-0.01,0.55)  {\large $C_f$};
        \node [align=center] at (0.68,0.25)  {$C_{f,\text{lam}}$};
      \end{scope}
\end{tikzpicture}
\hspace{-0.3cm}
\begin{tikzpicture}
      \node[anchor=south west,inner sep=0] (a) at (0,0) {\includegraphics[width=0.46\columnwidth]
      {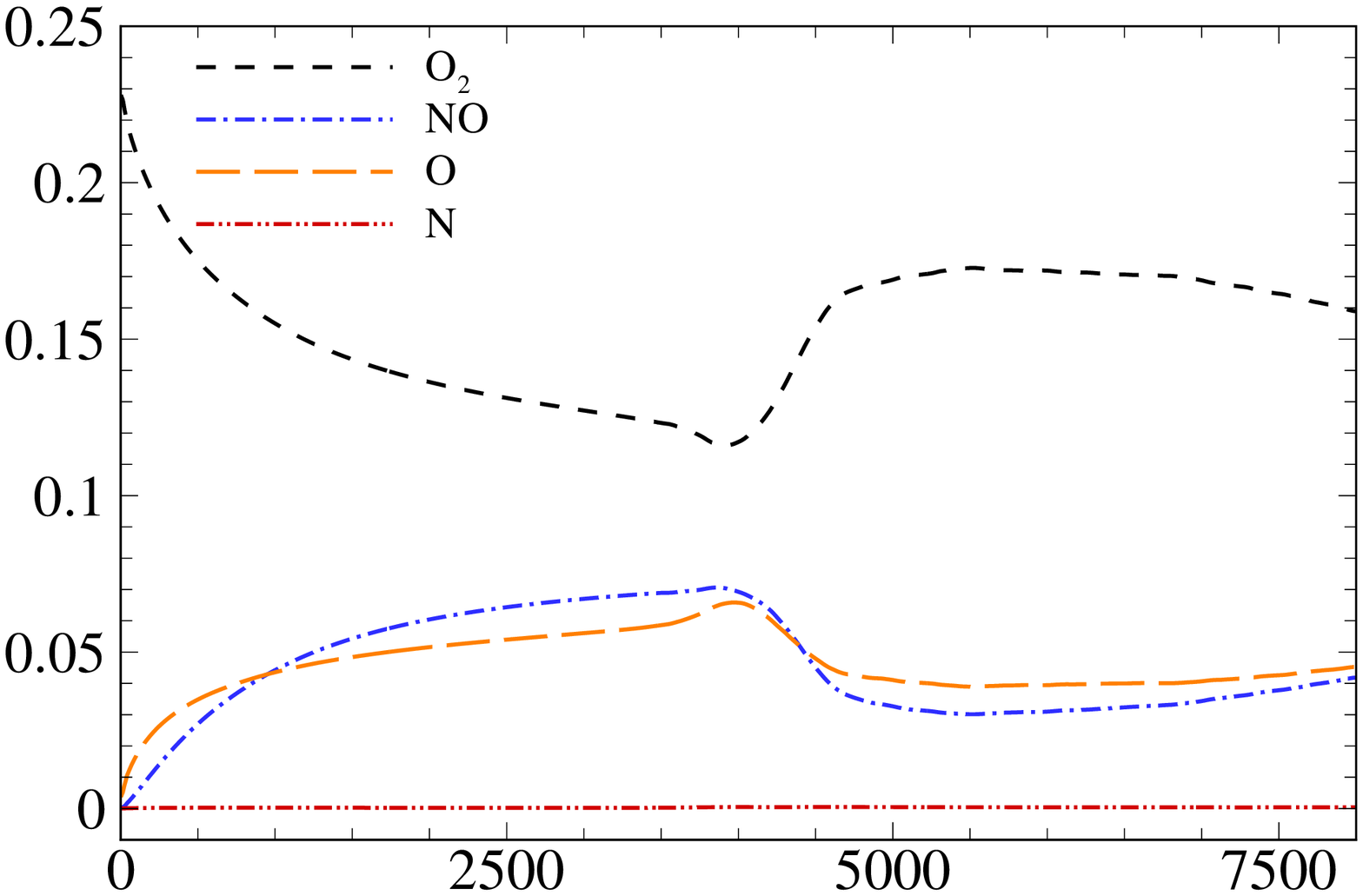}};
      \begin{scope}[x={(a.south east)},y={(a.north west)}]
        \node [align=center] at (-0.02,0.9) {(b)};
        \node [align=center] at (0.55,-0.02) {$(x-x_0)/\delta^*_\text{in}$};
        \node [align=center,rotate=90] at (-0.01,0.55)  {\large $\overline{Y_n}$};
      \end{scope}
\end{tikzpicture}
\vspace{-0.5cm}
\caption{Streamwise evolutions of (a) skin friction coefficient $C_f$ and laminar correlation $C_{f,\text{lam}}$, and (b) mean mass fractions $\overline{Y}_n$ along the wall along compared with the results of the locally self-similar solution at different stations (symbols). In panel b, $\overline{Y}_{\text{N}_2}$ is not shown being outside of the prescribed range.}
\label{fig:streamwise_reac}
\end{figure}

\begin{figure}[!tb]
\centering
\begin{tikzpicture}
      \node[anchor=south west,inner sep=0] (a) at (0,0) {\includegraphics[width=0.455\columnwidth]
      {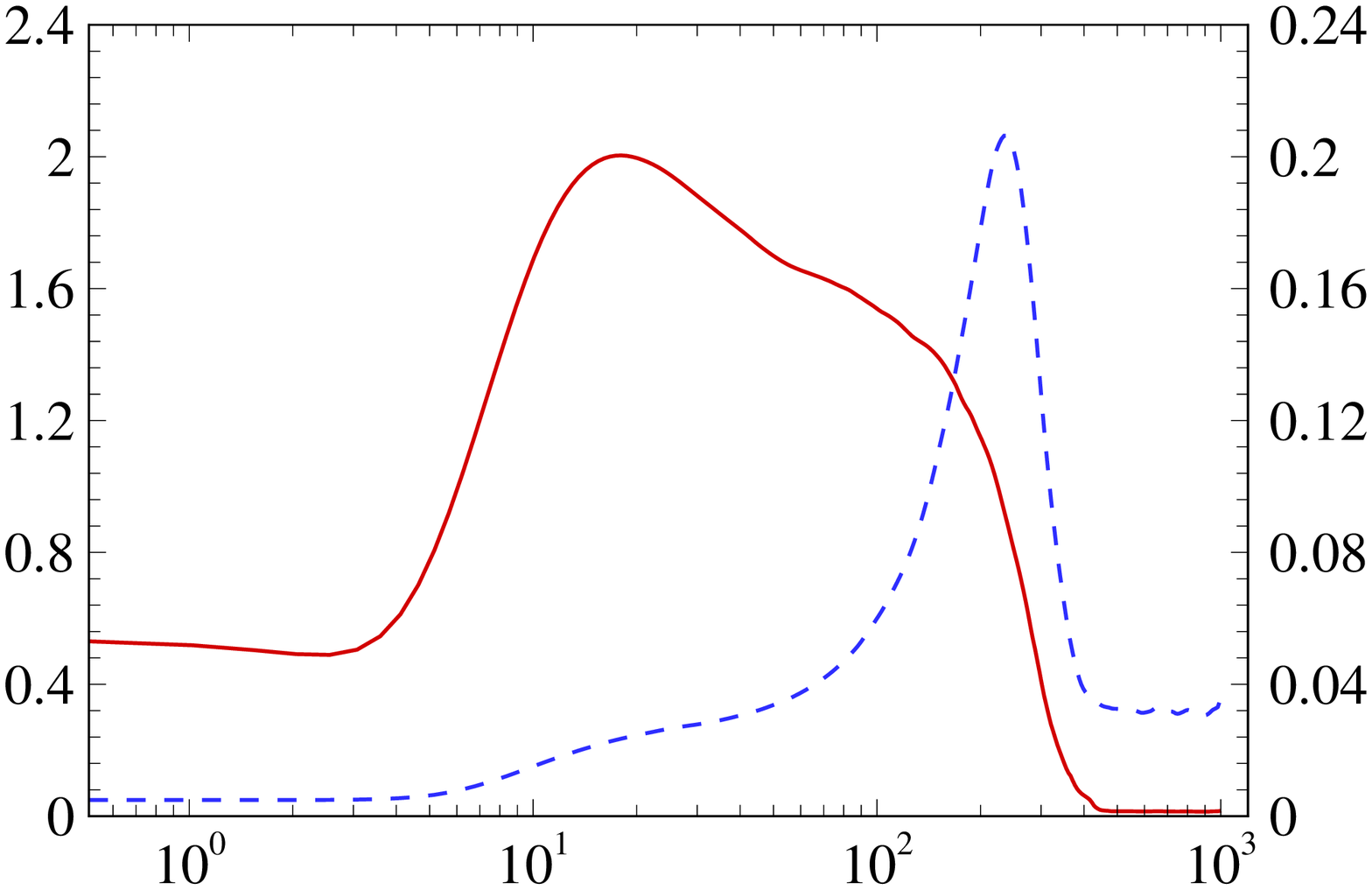}};
      \begin{scope}[x={(a.south east)},y={(a.north west)}]
        \node [align=center] at (-0.02,0.9) {(a)};
        \node [align=center] at (0.55,0.00) {\large $y^+$};
        \node [align=center,rotate=90] at (-0.01,0.55)  {$T_{rms}/T_\infty$};
        \node [align=center,rotate=270] at (1.04,0.53)  {$\rho_{rms}/\rho_\infty$};
        \node [align=center] at (0.32,0.60) {$\longleftarrow$};
        \node [align=center] at (0.81,0.60) {$\longrightarrow$};
      \end{scope}
\end{tikzpicture}
\hspace{-0.45cm}
\begin{tikzpicture}
      \node[anchor=south west,inner sep=0] (a) at (0,0) {\includegraphics[width=0.455\columnwidth]
      {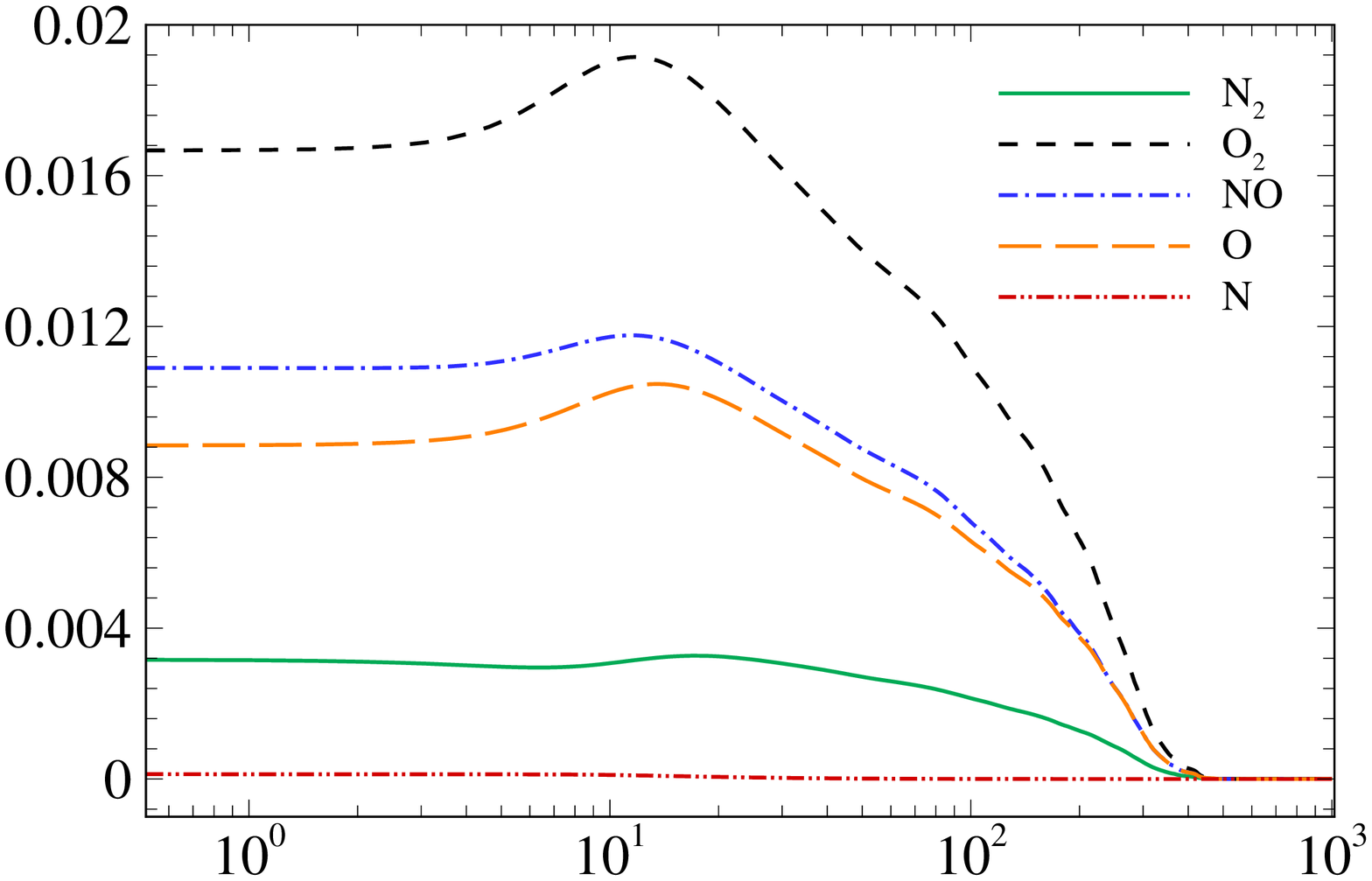}};
      \begin{scope}[x={(a.south east)},y={(a.north west)}]
        \node [align=center] at (-0.02,0.9) {(b)};
        \node [align=center] at (0.55,0.00) {\large $y^+$};
        \node [align=center,rotate=90] at (-0.02,0.55)  {$Y_{rms,n}$};
      \end{scope}
\end{tikzpicture}
\vspace{-1.cm}
\caption{Wall-normal profiles extracted at $Re_\theta=\num{11500}$ of (a) normalized r.m.s. temperature and density, and (b) r.m.s. species mass fractions.}
\label{fig:rms_reac}
\end{figure}

Figure~\ref{fig:streamwise_reac}a shows the streamwise evolution of the skin friction coefficient $C_f$ along with the compressible extension of Blasius' laminar correlation, $\DPS C_{f,lam}=\frac{0.664}{\sqrt{Re_x}}\sqrt{\frac{\overline{\rho}_w \overline{\mu}_w}{\rho_\infty \mu_\infty}}$. After the forcing strip (not visible in the figure being outside of $y$-range), the evolution of $C_f$ stays close to the laminar correlation up to $(x-x_0)/\delta^*_\text{in}\approx 3300$, where breakdown to turbulence (qualitatively similar to figure~\ref{fig:pfg1}) occurs. The averaged distribution of species mass fractions along the wall, displayed in figure~\ref{fig:streamwise_reac}b, shows that the amount of O$_2$ decreases (and contrariwise, O and NO increase) as the flows evolves in the laminar region. Afterwards, transition to turbulence enhances the mixing with external layers, inverting the previous trend; lastly, in the fully-developed turbulent region, dissociation of molecular oxygen continues and the mixture composition tends towards the equilibrium value. Additionally, the results obtained by the locally self-similar theory are also displayed and the discrepancies are acceptable, since the pressure gradient present in the full numerical solver is neglected in self-similar boundary layer equations. \\
Wall-normal distributions of selected second-order statistics are also investigated in figure~\ref{fig:rms_reac}. The profiles are extracted at a streamwise station close to the domain outlet (at $(x-x_0)/\delta^*_\text{in}=7800$), corresponding to $Re_\theta=\num{11500}$. Panel (a) shows that root-main-square values of temperature double the free-stream value at the peak of production, whereas density fluctuations stay relatively small. Accordingly, the fluctuating distributions of species mass fractions (figure~\ref{fig:rms_reac}b) exhibit the largest values where $T_\text{rms}$ peaks, although the strongest chemical activity is concentrated at the wall where the temperature reach the largest values. Of note, the mesh resolution in inner units at this station is equal to $\Delta x^+=4.8$, $\Delta y_w^+=0.52$ and $\Delta z^+=4.0$; based on the grid-refinement study carried out in section~\ref{sec:bl_pfg}, the current simulation may be then classified as a well-resolved DNS.

%===============================================================================

\section{Conclusions}
\label{sec:conclusions}
In this study, an efficient methodology for the numerical simulation of turbulent flows at high speeds has been presented. The main difficulties associated to such flows are related to the amount of intrinsic numerical dissipation of the scheme and the concurrent needs of obtaining a reliable picture of small-scale motions, while being able to handle strong discontinuities.

The proposed numerical scheme represents an high-order extension of the original artificial-viscosity-based method of Jameson \cite{jameson1981numerical}. It relies on a blending of second- and tenth-order derivatives equipped with a suitable shock detector, which results in a ninth-order-accurate scheme in smooth flow regions.
Several test cases have been considered to assess the capability of such technique to deal with shocks, shock/entropy perturbations and turbulence, both for calorically-perfect gases and high-temperature multicomponent flows. Different grid refinements (representative of typical resolutions used to carry out implicit Large-Eddy simulations) were analysed for multi-scale configurations; results were systematically compared to DNS data to quantify the error introduced by the adoption of under-resolved meshes.

A parametric study on the isentropic vortex advection was carried out and the formal orders of accuracy of the spatial discretization and temporal integration were correctly retrieved. The three shock tube configurations considered (Sod, Lax and Grossman-Cinnella) confirm that the scheme handles properly strong shocks and contact discontinuities, even in presence of chemical nonequilibrium processes.
Furthermore, the two-dimensional N$_2$-O$_2$ under-expanded jet, the compressible Tayor-Green Vortex and the supersonic boundary layer were used to prove the reliability and robustness of the numerical method when applied on coarse grids for ILES. It was shown that results tend seamlessly to DNS predictions as the resolution increases, and computations were stable also for severely under-resolved simulations. The last configuration investigated, a turbulent boundary layer at hypersonic speeds, demonstrates the good behaviour of the numerical scheme also with the concomitant occurrence of shocklets, broadband turbulence and chemical nonequilibrium processes.

Future works have been planned concerning the extension of the current scheme to curvilinear meshes, the coupling with skew-symmetric formulations, and the numerical investigations of high-speed configurations encompassing thermal relaxation phenomena by means of multi-temperature models.

\section*{Acknowledgments}
This work was granted access to the HPC resources of IDRIS and TGCC under the allocation 2019-2B10947 made by GENCI (Grand Equipement National de Calcul Intensif).

\bibliography{bibref.bib}

\appendix
\section{Locally self-similar solution for laminar boundary layers}
\label{app:blasius}
\begin{figure}[!tb]
   \centering
     \begin{tikzpicture}
      \node[anchor=south west,inner sep=0] (a) at (0,0) {\includegraphics[width=0.44\columnwidth, trim={5 5 5 5}, clip]{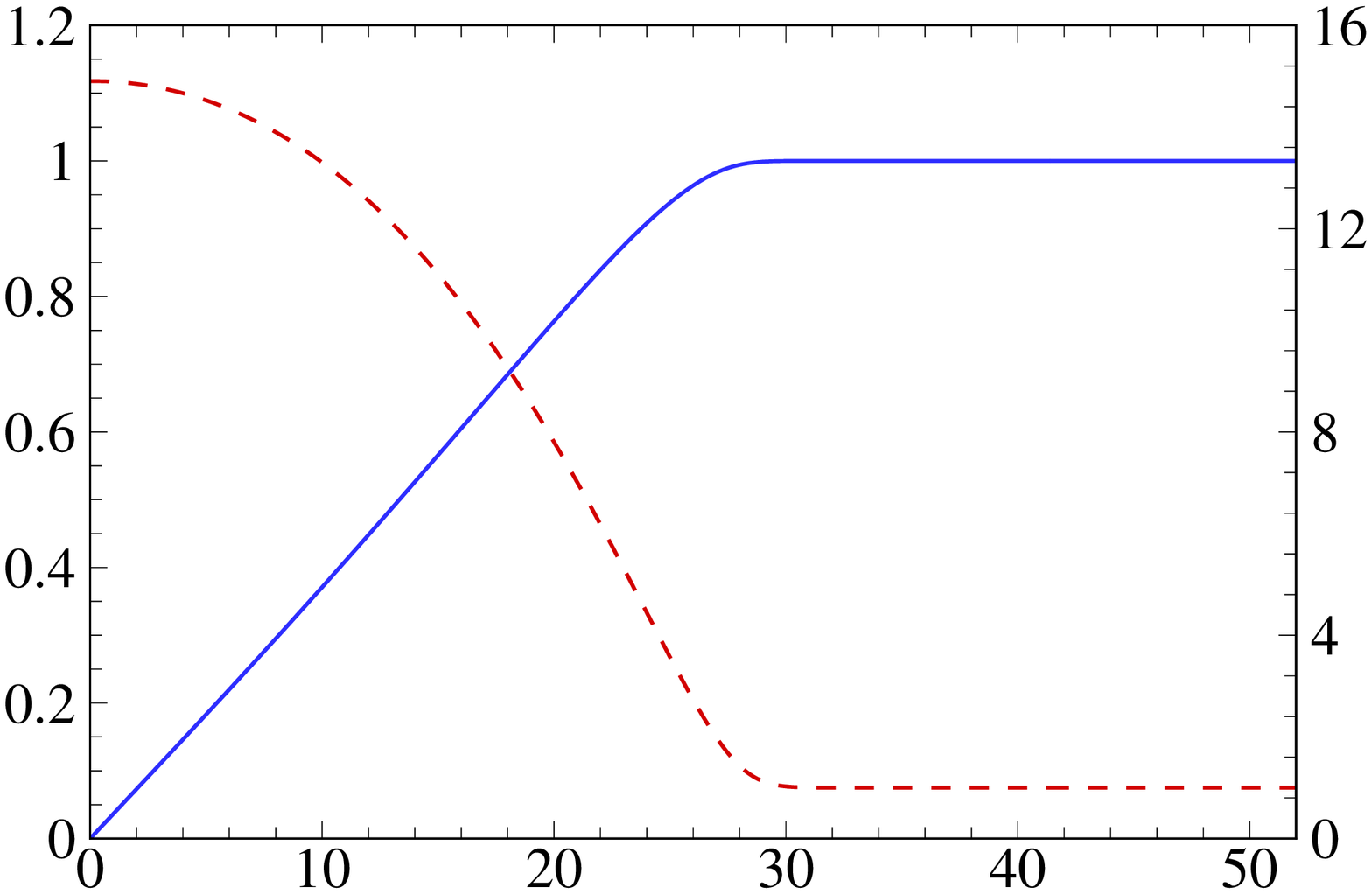}};
      \begin{scope}[x={(a.south east)},y={(a.north west)}]
        \node [align=center] at (0.56,0.0) {$\eta_i$};
        \node [align=center,rotate=90] at (-0.01,0.54) {$u/U_e$};
        \node [align=center,rotate=270] at (1.01,0.53) {$T/T_e$};
        \node [align=center] at (0.25,0.44) {$\longleftarrow$};
        \node [align=center] at (0.50,0.48) {$\longrightarrow$};
      \end{scope}
  \end{tikzpicture}
     \begin{tikzpicture}
      \node[anchor=south west,inner sep=0] (a) at (0,0) {\includegraphics[width=0.44\columnwidth, trim={5 5 5 5}, clip]{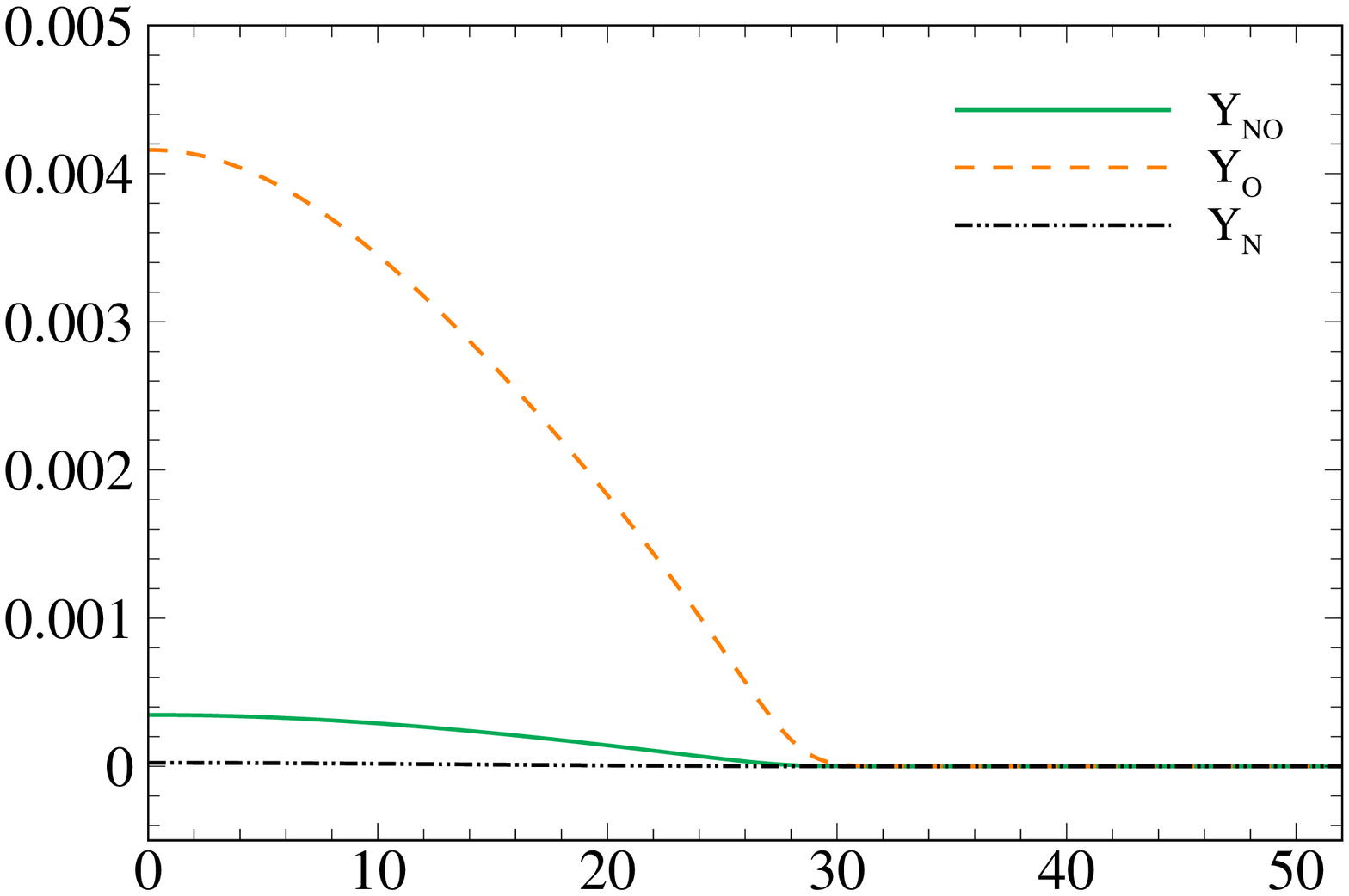}};
      \begin{scope}[x={(a.south east)},y={(a.north west)}]
        \node [align=center] at (0.56,0.0) {$\eta_i$};
      \end{scope}
  \end{tikzpicture}
\caption{Evolution of normalized velocity and normalized temperature (left) and species mass fractions (right), at the inflow. $Y_{\text{N}_2}$ and $Y_{\text{O}_2}$ are not shown being outside the range.}
\label{fig:blasius1}
\end{figure}

\noindent The boundary-layer equations for a steady, compressible, multicomponent, reacting, two-dimensional flow (without pressure gradient) write:

\begin{align}
 \frac{\partial (\rho u)}{\partial x} + \frac{\partial (\rho v)}{\partial y} & = 0 \\
 \rho u \frac{\partial Y_n}{\partial x} + \rho v \frac{\partial Y_n}{\partial y} & = \frac{\partial}{ \partial y}\left(\rho_n u^D_n\right) + \dot{\omega}_n \\
 \rho u \frac{\partial u}{\partial x} + \rho v \frac{\partial u}{\partial y} & = \frac{\partial}{\partial y}\left(\mu \frac{\partial u }{\partial y} \right)
\end{align}
% \noindent \textit{y-Momentum:}
% \begin{equation}
%  \frac{\partial p }{\partial y} = 0
% \end{equation}
% \noindent \textit{Total enthalpy conservation:}
\begin{equation}
 \rho u \frac{\partial h}{\partial x} + \rho v \frac{\partial h}{\partial y} = \frac{\partial}{\partial y }\left(\lambda \frac{\partial T}{\partial y} \right) + \frac{\partial}{\partial y}\left( \sum_n \rho_n u^D_n h_n \right) + \mu \left( \frac{\partial u}{\partial y}\right)^2
\end{equation}
Independent variable transformations are introduced as follows (see Ref. \cite{lees1956laminar}):
\begin{equation}
\xi = \rho_e \mu_e U_e x = \xi(x), \qquad \eta = \frac{u_e}{\sqrt{2 \xi}} \int_0^y \rho dy= \eta (x,y).
\end{equation}
Considering the definition of the stream-function:
\begin{equation}
\frac{ \partial \psi}{ \partial y} = \rho u \qquad  \frac{ \partial \psi}{ \partial x} = - \rho v
\end{equation}
and the equivalent expression in terms of transformed variables
\begin{equation}
\frac{\partial \psi}{\partial \xi} = \frac{1}{\sqrt{2 \xi}}f(\eta) \qquad \frac{\partial \psi}{\partial \eta} = \sqrt{2 \xi} f'(\eta),
\end{equation}

\noindent with $f'=u/U_e$, one can manipulate the boundary-layer equations and obtain their formulation in the self-similar coordinate system:

\begin{equation}
\left(Cf''\right)'+ff'' = 0
\end{equation}
\begin{equation}
\left[\frac{\rho^2 D_n}{\rho_e \mu_e}\left(Y'_n -Y_n\sum_{i=1}^{NS}\frac{D_i}{D_n} Y_i'\right)\right]' + fY'_n+ \frac{2 \xi \dot{\omega}_n/\rho}{\rho_e \mu_e u_e^2 } = 0
\label{eq:blasius_species}
\end{equation}
\begin{equation}
\frac{T_e}{h_{0,e}}\left(\frac{C \lambda}{\mu}\theta'\right)' + fg' + \frac{u_e^2}{h_{0,e}}ff'f'' + \frac{u_e^2}{h_{0,e}}\left( Cf'f''\right)'+\left[\sum_{n=1}^{NS} \frac{h_n}{h_{0,e}} \frac{\rho^2 D_n}{\rho_e \mu_e}\left(Y'_n -Y_n\sum_{i=1}^{NS}\frac{D_i}{D_n}Y_i'\right)\right]'= 0
\end{equation}
Here, $g=h/h_{0,e}$ is the self-similar parameter for the enthalpy, $h_{0,e}$ being the edge stagnation enthalpy, $\theta = T/T_e$ and $C=\rho \mu/\rho_e \mu_e$. For the sake of clarity, in case of frozen chemistry or chemical equilibrium, the only $\xi$-dependent term in equation~\eqref{eq:blasius_species} vanishes and the resulting equations become globally self-similar.
The previous system of equations can be integrated numerically, subjected to the following boundary conditions:
\begin{align}
f'=f=0, \qquad Y'_n=0, \qquad g'=0 \qquad & \text{for} \quad \eta=0\\
f'=1, \qquad Y_n=Y_{n,e}, \qquad g=1-\frac{1}{2}\frac{u_e^2}{h_{0,e}} \qquad & \text{for} \quad \eta \rightarrow \infty
\end{align}
In the specific configuration of section~\ref{sec:bl_reac}, the perscribed edge variables are $U_e=\SI{3754}{m/s}$, $T_e=\SI{350}{K}$ and $h_{0,e}=\SI{7.1}{MJ/kg}$. Considering a distance from the leading edge equal to $x=\SI{0.01}{m}$, we obtain the inflow profiles of normalized temperature, normalized velocity and species mass fractions, shown in figure~\ref{fig:blasius1} as a function of the incompressible similitude variable $\eta_i=y \sqrt{Re_x}/x$.
% Moreover, a comparison between the locally self-similar solution and the solver numerical solution is provided; figure~\ref{fig:blasius2} displays the skin friction coefficient distribution (left) and the wall-normal evolution of dimensionless velocity and temperature at $x=\SI{0.05}{m}$.

\end{document}